\documentclass[14pt]{article}
\usepackage{epsf}
\usepackage{color}
\usepackage{framed}
\usepackage{amsmath,amsfonts,amsbsy,amssymb}

\numberwithin{equation}{section}

\usepackage{indentfirst}
\usepackage{mathtools}
\usepackage{arydshln} 
\usepackage{framed}
\usepackage{mathrsfs}

\newcommand{\nn}{\nonumber}

\def\be{\begin{equation}}
\def\ee{\end{equation}}
\def\bse{\begin{subequations}}
\def\ese{\end{subequations}}
\def\bal{\begin{align}}
\def\ealn{\end{align}}
\def\tr{\text{tr}}
\def\bs{\boldsymbol}

\begin{document}

\begin{titlepage}

\def\slash#1{{\rlap{$#1$} \thinspace/}}

\begin{flushright} 

\end{flushright} 

\vspace{0.1cm}

\begin{Large}

\begin{center}

{\bf  
Perfectly  
Spherical Bloch Hyper-spheres  \\  from \\ Quantum Matrix Geometry
 }\end{center}
\end{Large}


\vspace{1cm}

\begin{center}
{\bf Kazuki Hasebe}   \\ 
\vspace{0.5cm} 
\it{
National Institute of Technology, Sendai College,  
Ayashi, Sendai, 989-3128, Japan} \\

\vspace{0.5cm} 

{\sf
khasebe@sendai-nct.ac.jp} 

\vspace{0.8cm} 

{\today} 

\end{center}

\vspace{0.8cm}

\begin{abstract}
\noindent

\baselineskip=20pt

Exploiting analogies between the precessing quantum spin system and the charge-monopole system, 
 we construct  Bloch hyper-spheres with $\it{exact}$ spherical symmetries in arbitrary dimensions. Such  Bloch hyper-spheres are  realized  as a collection of the orbits of a precessing quantum spin. The geometry of Bloch hyper-spheres is exactly equal to the quantum Nambu geometry of  higher dimensional fuzzy spheres.  
The stabilizer group symmetry of the Bloch hyper-sphere necessarily  introduces degenerate spin-coherent states, giving rise  to the Wilczek-Zee geometric phase  of  non-Abelian  monopoles associated with the hyper-sphere holonomy.  The degenerate spin-coherent states induce matrix-valued  quantum geometric tensors.  
 While the minimal spin Bloch hyper-spheres  exhibit similar properties in even and odd dimensions, their  large spin counterparts differ qualitatively depending on the parity of the dimensions.   Exact correspondences between  spin-coherent states and   monopole harmonics  in higher dimensions are established. 
We also investigate density matrices described by Bloch hyper-balls  and elucidate  their corresponding statistical and geometric properties, such as von Neumann entropies and Bures quantum  metrics.  

\end{abstract}

\end{titlepage}

\newpage 

\tableofcontents

\newpage 

\section{Introduction}

The geometry of quantum states provides an indispensable perspective for a deeper understanding of both quantum mechanics and quantum information \cite{Bengtsson-Zyczkowski-2006, Nielsen-Chuang-2005, Chruscinski-Jamiolkowski-2004, Bohm-et-al-2003}. Its importance has also grown rapidly in recent advances in material science \cite{Torma-2023, Lambert-Sorensen-2023}. 
 The Bloch sphere \cite{Bloch-1946}  
serves as the fundamental geometry of  two-level quantum mechanics. 
In  two-level quantum mechanics with  conical  degeneracy, 
Berry's geometric phase \cite{Berry-1984} was  recognized in the adiabatic evolution of the energy eigenstate  
 \cite{Herzberg-LonguetHiggins-1963}.    Soon after Berry's work, 
Wilczek and Zee introduced 
 a non-Abelian  version of the geometric phase for degenerate energy levels  \cite{Wilczek-Zee-1984}. 
As the Bloch sphere illustrates the underlying geometry of such a two-level quantum mechanics and Berry's geometric   phase, higher dimensional Bloch spheres  (Bloch  hyper-spheres)  realize  a paradigmatic example of  the geometry of multi-level quantum mechanics and the Wilczek-Zee non-Abelian phase.  
In the recent developments of quantum matter  \cite{Wilczek-2012},  
higher dimensional topological phases can be accessed through the concept of synthetic dimensions \cite{Price-et-al-2015, Price-et-al-2016, Ozawa-Price-et-al-2016, Wang-Price-Zhang-Chong-2020}, and 
higher dimensional topologies have attracted increasing attention. In particular, the non-Abelian geometric phases  have recently been observed  by state-of-the-art tabletop experiments  \cite{Ma-Jia-Bi-et-al-2023, Zhang-Chen-Zhang-et-al-2022, Ma-Bi-et-al-2021, Sugawa-et-al-2018, Li-Duca-et-al-2016}.

A two-level Hamiltonian  for qubit is introduced as 
\be 
 H =\sum_{i=1}^3 x_i \cdot \frac{1}{2}\sigma_i. \label{zeediracso3}
\ee
Its eigenstates are referred to as the spin-coherent states or Bloch coherent states \cite{Klauder-1960,Radcliffe-1971,Perelomov-1972, Arecchi-Courtens-Gilmore-Thomas-1972}. 
 In the context of  quantum information, the qubit state is first given, and then the Bloch vector $x_i$ is determined  to   visualize the geometry of the qubit.    Meanwhile,  in quantum physics, a quantum mechanical Hamiltonian is usually given first, and quantum states  follow as its eigenstates. 
The Hamiltonian (\ref{zeediracso3}) is ubiquitous in the quantum world and plays a crucial role in various contexts of physics:   
When $x_{i}$ represent the direction of an applied  magnetic field, the Hamiltonian 
(\ref{zeediracso3})  is called the Zeeman magnetic interaction term.  Meanwhile, if $x_i$ are taken to be a crystal momentum, the Hamiltonian  is known in material science  as the Dirac (or Weyl) Hamiltonian  where the spin index of the Pauli matrices signifies  the two-band index.\footnote{For the real spin $\frac{1}{2}\sigma_i$ and momentum $x_i$, (\ref{zeediracso3}) simply stands for the helicity.} For these reasons, we refer to the Hamiltonian (\ref{zeediracso3}) as the ($SO(3)$) Zeeman-Dirac Hamiltonian in this paper. The Bloch sphere emerges as the underlying geometry behind all of the physical systems described by the Zeeman-Dirac Hamiltonian. 
For a large spin model, such as nuclear spin, we employ the Zeeman-Dirac Hamiltonian of $SU(2)$  spin matrices with  spin magnitude $S(=1/2, 1, 3/2, \cdots)$: 
\be 
 H =\sum_{i=1}^3 x_i \cdot S_i, 
\label{zeediracso3large}
\ee
which accommodates   $\it{equally}$ spaced $2S+1$ energy levels. 
 As demonstrated by Berry \cite{Berry-1984}, the  geometric phase associated with the adiabatic evolution of the spin-coherent state of the Hamiltonian (\ref{zeediracso3large}) is identical to the $U(1)$ phase accounted for by the Dirac magnetic monopole with magnetic charge $S$  \cite{Dirac-1931, Wu-Yang-1976}.   For a  
 general $N$ level system with $\it{arbitrary}$ level spacing or an $N$-qudit, the corresponding  Hamiltonian  is  represented by the  $N\times N$ Hermitian matrix  expanded by the $SU(N)$ matrix generators (apart from the trivial $U(1)$ unit matrix corresponding to an overall energy shift). 
 The $SO(3)$ Zeeman-Dirac Hamiltonian of   $S=\frac{N-1}{2}$ (\ref{zeediracso3large}) is realized as  a special case of such  $SU(N)$ Hamiltonians.    
The study of the $SU(N)$ generalization of the Zeeman-Dirac model has a rather long history \cite{Hioe-Eberly-1981, Kimura-2003, Byrd-Khaneja-2003,  Graf-Piechon-2021, Kemp-Cooper-Unal-2022}, and  the $SU(N)$ spin-coherent states have been constructed in   
 Refs.\cite{Anadan-Stodolsky-1987,Gitman-Shelepin-1993,Gnutzmanny-Kus-1998}. 
 The  $SU(N)$ spin magnetism  is crucial for quantum information processing using alkaline-earth atoms 
\cite{Gorshkov-Hermele-et-al-2010}. 
The underlying geometry of such $SU(N)$ models is described by an $SU(N)$ generalized  Bloch ``sphere'', $i.e.$, $\mathbb{C}P^{N-1}$ geometry  \cite{Kemp-Cooper-Unal-2022, Byrd-Boya-Mims-Sudarshan-2006, Uskov-Rau-2008, Rau-2021}, as it reproduces the Bloch sphere in the special  $N=2$ case, $\mathbb{C}P^1\simeq S^2$.  The $SU(4)$ generalization of the Bloch sphere has recently been implemented   experimentally in a photonic device \cite{Zhang-Zhao-et-al-2022}.  
However, it should  be noted that this $SU(N)$ generalization  leads to  unitarily symmetric manifolds rather than spherically symmetric ones as just mentioned.  

Another extension  of the $SO(3)$ Zeeman-Dirac Hamiltonian, and perhaps even more interesting  in some sense, is the time-reversal symmetric $S=3/2$ quadrupole Hamiltonian \cite{Mead-1987}.  
This $S=3/2$ quadrupole Hamiltonian is equivalent to   the $SO(5)$ Zeeman-Dirac Hamiltonian made of the $SO(5)$  gamma matrices\footnote{Recall that the Pauli matrices are equivalent to the gamma matrices of  $SO(3)$. } $\gamma_a$ \cite{Avron-Sadun-Segert-Simon-1988, Avron-Sadun-Segert-Simon-1989}: 
\be
H =\sum_{a=1}^5 x_a \cdot \frac{1}{2}\gamma_a. \label{so5simpzd} 
\ee
While this Hamiltonian  is a special case of the $SU(4)$ Hamiltonian,  it is important in its own right.   The $SO(5)$ model  is closely related to  special Jahn-Teller systems \cite{Mead-1992,Apsel-Chancey-OBrien-1992} and an  
ultra-cold atom system of spin $3/2$ fermions \cite{Wu-Hu-Zhang-2003}\footnote{See \cite{Larson-Sjoqvist-Ohberg-book} for conical singularities  in various contexts of physics.}. The Hamiltonian (\ref{so5simpzd}) also plays a  crucial role as the parent Hamiltonian of topological insulator \cite{Ryu-S-F-L-2010}.   This $SO(5)$  Hamiltonian exhibits two energy levels of equal magnitude but opposite sign, similar to the $SO(3)$ Hamiltonian (\ref{zeediracso3}). 
Each of the energy levels accommodates double degeneracy attributed to  the existence of time-reversal symmetry (the Kramers theorem).    
 The adiabatic evolution of the doubly degenerate $SO(5)$ spin-coherent states  in each energy level naturally induces the Wilczek-Zee non-Abelian connection \cite{Levay-1990,Levay-1991,Johnsson-Aitchison-1997}, which is identified as the gauge field of      Yang's $SU(2)$ monopole \cite{Yang-1978-1, Yang-1978-2} or  the BPST instanton  \cite{Belavin-Polyakov-Schwartz-Tyupkin-1975}.  
 Very recently, the  $SO(5)$ Zeeman-Dirac Hamiltonian has been  implemented in  cold atom systems and meta-materials, in which  the physical consequences peculiar to  the $SU(2)$ monopole  have been  observed experimentally \cite{Sugawa-et-al-2018, Ma-Bi-et-al-2021}. 

\begin{figure}[tbph]
\center 
\includegraphics*[width=140mm]{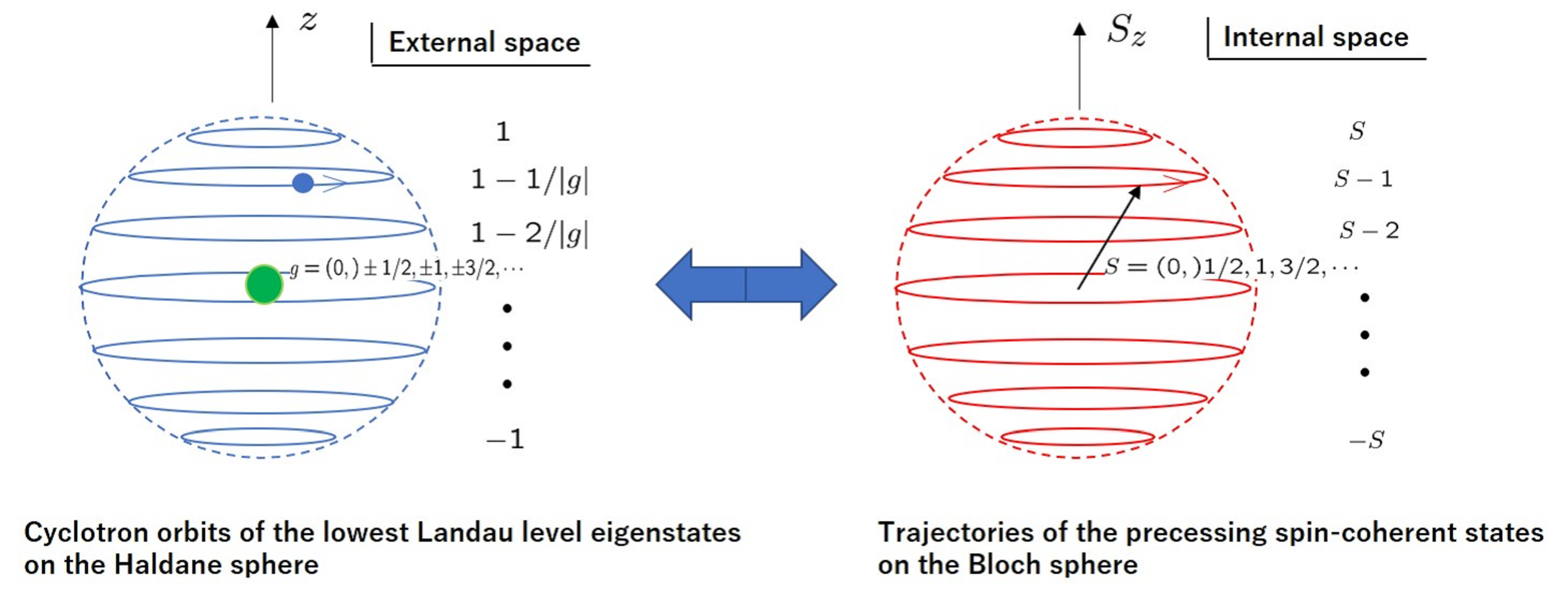}
\caption{Analogies between the electron cyclotron orbits of the Landau model \cite{Haldane-1983} (left) and the  orbits of the quantum spin precession (right). 
}
\label{mono.fig}
\end{figure}

The $SO(3)$ Zeeman-Dirac Hamiltonians of large spins were readily constructed by replacing the Pauli matrices with general $SU(2)$ spin matrices.  
However, it is  far from obvious  how to generalize the $SO(5)$  Hamiltonian  
for arbitrarily large spins. This is because    
the gamma matrices themselves are $\it{not}$ generators of the $SO(5)$ group (but their commutators are), and we cannot adopt $SO(5)$ generators of large spins for this purpose. 
To construct the gamma matrices of large spins, we utilize the  analogy  between  the charge-monopole system on a sphere (known as the Landau model)  and the precession of the quantum spin (Fig.\ref{mono.fig}): 
The trajectories of the precessing spin on the Bloch sphere can be interpreted as the cyclotron orbits of a charged particle on a two-sphere in the Dirac monopole background \cite{Wu-Yang-1976, Haldane-1983}  (Fig.\ref{mono.fig}). 
We exploit this observation to construct the generalized gamma matrices of large spins. 
This idea is in line with recent developments of  non-commutative geometry  \cite{Hasebe-2015, Ishiki-Matsumoto-Muraki-2016, Hasebe-2018,  Ishiki-Matsumoto-Muraki-2018, Matsuura-Tsuchiya-2020, Matsuura-Tsuchiya-2020, Nair-2020, Hasebe-2020-1, Adachi-Ishiki-Matsumoto-Saito-2020, Hasebe-2021, Steinacker-2021, Adachi-Ishiki-Kanno-2022, Hasebe-2023-1}, in particular from the quantum matrix geometry of the higher dimensional fuzzy spheres \cite{Hasebe-2023-1, Hasebe-2021, 
Hasebe-2020-1,   Hasebe-2018, Ishiki-Matsumoto-Muraki-2016}.\footnote{
The quantum geometry of the fuzzy sphere is now being applied to various branches of physics \cite{Zhu-Han-Huffman-Hofmann-He-2022, Fan-Fischler-Kubischta-2022, Cuomo-Komargodski-Mezei-RavivMoshe-2022, Cuomo-Fuente-Monin-Pirskhalava-Rattazzi-2018, Cuomo-Delacretaz-Mehta-2021}. 
}   
We  present a  systematic construction of exactly spherical Bloch hyper-spheres  and investigate  their unique properties.  We will see that  higher dimensional Zeeman-Dirac models necessarily exhibit  energy level degeneracies and realize   the  Wilczek-Zee connections of non-Abelian monopoles. We also construct  Bloch hyper-balls and address their implications in  quantum statistics and quantum statistical geometry.

This paper is organized as follows. In Sec.\ref{sec:blochfourthreetwo}, we review the original Bloch sphere and the spin-coherent states. Section \ref{sec:blochfour} discusses the $SO(5)$  Zeeman-Dirac models and their geometric structures. In Sec.\ref{sec:geneso5so4}, we  construct  $SO(4)$ Zeeman-Dirac models and clarify their properties.  We analyze the  Zeeman-Dirac models  in arbitrary dimensions  in Sec.\ref{sec:quansphigh}. In Sec.\ref{sec:blochball}, we introduce the density matrices associated with Bloch hyper-balls and exploit their statistical properties,  such as von Neumann entropy and Bures information metric. Sec.\ref{sec:summary} is devoted to summary and discussion.

\section{Bloch sphere and the $SO(3)$ Zeeman-Dirac model}\label{sec:blochfourthreetwo}

As a warm-up, we  review the Bloch sphere and the spin-coherent states with emphasis on their relation to the $SO(3)$ Zeeman-Dirac model. We will clarify the relationship between the  spin-coherent states and the Landau level eigenstates.

\subsection{Minimal spin model}\label{append:so3minimal}

\begin{figure}[tbph]
\center
\includegraphics*[width=150mm]{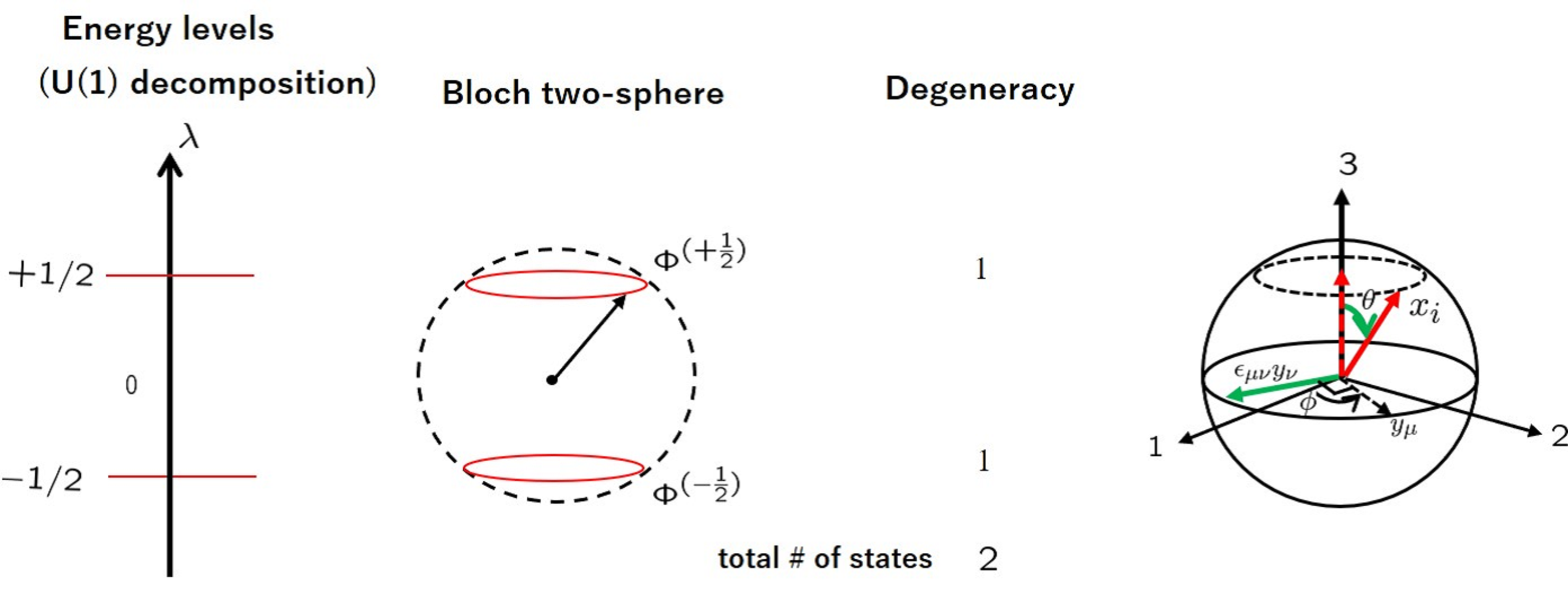}
\caption{The eigenvalues and the eigenstates of the $SO(3)$ Zeeman-Dirac model (left and middle) and the rotation of the spin (right).  }
\label{spins2.fig}
\end{figure}

We introduce the $SO(3)$ minimal Zeeman-Dirac model: 
\be
H=\sum_{i=1}^3 x_i\cdot \frac{1}{2}\sigma_i, \label{so3spinmodemin}
\ee
where $x_i$ are the coordinates on $S^2$ or the components of the Bloch vector: 
\be
x_1=\cos\phi\sin\theta, ~~x_2=\sin\phi\sin\theta, ~~x_3=\cos\theta. \label{blochvec3d}
\ee
It is easy  to solve the eigenvalue problem of this $2\times 2$ matrix Hamiltonian (\ref{so3spinmodemin}):
\be
H\Phi^{(\lambda)} =\lambda\cdot \Phi^{(\lambda)},  
\ee
where the eigenvalues are  given by 
\be
\lambda=+{1}/{2}, -{1}/{2}. \label{eigenso3hasimp}
\ee
The corresponding eigenstates are known as the spin-coherent states  
\be
\Phi^{(+\frac{1}{2})} = \frac{1}{\sqrt{2(1+x_3)}}\begin{pmatrix}
1+x_3  \\
x_1+ix_2  
\end{pmatrix}=\begin{pmatrix}
\cos(\frac{\theta}{2}) \\ 
\sin(\frac{\theta}{2}) e^{i\phi}
\end{pmatrix},~~\Phi^{(-\frac{1}{2})} = \frac{1}{\sqrt{2(1+x_3)}}\begin{pmatrix}
-x_1+ix_2 \\
 1+x_3
\end{pmatrix}=\begin{pmatrix}
-\sin(\frac{\theta}{2}) e^{-i\phi}\\
\cos(\frac{\theta}{2} )
\end{pmatrix}, \label{spinso3coherentstates}
\ee
which are normalized as 
\be
{\Phi^{(+\frac{1}{2})}}^{\dagger}\Phi^{(+\frac{1}{2})}= {\Phi^{(-\frac{1}{2})}}^{\dagger}\Phi^{(-\frac{1}{2})}= 1, ~~~~{\Phi^{(+\frac{1}{2})}}^{\dagger}\Phi^{(-\frac{1}{2})}=0.  \label{orthonospinso3coh}
\ee
Notice that the eigenvalues (\ref{eigenso3hasimp}) are the diagonal components of $\frac{1}{2}\sigma_3$. In other words, the eigenstates (\ref{spinso3coherentstates}) carry  the quantum numbers of the   $U(1)$ sub-algebra of  $SU(2)$.  
The eigenvalues $\lambda=\pm 1/2$ have a nice geometric meaning as  the  latitudes on the Bloch sphere where the spin-coherent states are oriented (see the left of Fig.\ref{spins2.fig}).    
 We can generate the spin-coherent states by the following well-known geometric manipulation. 
The projection of the Bloch vector $x_i$ onto the $xy$-plane is given by 
\be
y_1=\cos\phi,~~y_2=\sin\phi. 
\ee
The spin-coherent state with $\lambda=+1/2$ can be obtained by rotating  the north-pole oriented spin-coherent state about the $\epsilon_{\mu\nu}y_{\nu}$-axis by  $\theta$ (see the right of Fig.\ref{spins2.fig}). 

Such a manipulation is demonstrated by the non-linear realization matrix 
\be
\Phi=e^{-i{\theta}\sum_{\mu,\nu=1}^2 \epsilon_{\mu\nu}y_{\mu}\frac{1}{2}\sigma_{\nu}}, \label{defmso3nl}
\ee
which  is expanded as 
\begin{align}
 \Phi&=\cos(\frac{\theta}{2})1_2 -i\sin(\frac{\theta}{2})\sum_{\mu,\nu=1}^2 \epsilon_{\mu\nu}y_{\mu}\sigma_{\nu}=\begin{pmatrix}
\cos\frac{\theta}{2} & -\sin\frac{\theta}{2} e^{-i\phi} \\
\sin\frac{\theta}{2} e^{i\phi}  & \cos\frac{\theta}{2}
\end{pmatrix}\nn\\
&=\frac{1}{\sqrt{2(1+x_3)}}((1+x_3)1_2 -i\epsilon_{\mu\nu}x_{\mu}\sigma_{\nu}) =
\frac{1}{\sqrt{2(1+x_3)}} 
\begin{pmatrix}
1+x_3 & -x_1+ix_2 \\
x_1+ix_2  & 1+x_3
\end{pmatrix}.
\end{align}
The spin-coherent states (\ref{spinso3coherentstates}) are indeed obtained from $\Phi$ as 
\be
\Phi^{(+\frac{1}{2})} =  \Phi\begin{pmatrix}
1 \\
0
\end{pmatrix}, ~~~\Phi^{(-\frac{1}{2})} =  \Phi\begin{pmatrix}
0 \\
1
\end{pmatrix} \label{mfromspinco}
\ee
or 
\be
\Phi =\begin{pmatrix}
\Phi^{(+\frac{1}{2})} & \Phi^{(-\frac{1}{2})}
\end{pmatrix}. 
\ee
Since $\Phi$ has a clear geometrical meaning and simultaneously contains the two spin-coherent states as its columns, we will use the non-linear realization matrix (\ref{defmso3nl}) instead of  the spin-coherent states themselves.  
Obviously, $\Phi$ denotes a unitary matrix that diagonalizes the Zeeman-Dirac Hamiltonian: 
\be
\Phi^{\dagger}H \Phi=\frac{1}{2}\sigma_3. \label{diagpau}
\ee
It is important to note that the diagonalization can be justified solely from the group theoretical properties of the $SU(2)$. 
Solving eigenvalue problems for large-sized matrix Hamiltonians is generally laborious. However, the geometric method makes it feasible, since the properties of the $SU(2)$ group are universal regardless of the magnitude of spin.
Non-linear realization matrix $\Phi$ (\ref{defmso3nl}) is factorized as\footnote{The factorization (\ref{facphisu2}) implies that $\Phi$ is a special case of  Wigner's $D$ function (see Chap.3 of Ref.\cite{Sakurai-book}, for instance), $\Phi=e^{-i\frac{\phi}{2}\sigma_3} e^{-i\frac{\theta}{2}\sigma_2} e^{-i\frac{\chi}{2}\sigma_3}|_{\chi=-\phi}$. 
  }  
\be
\Phi 
=e^{-i\frac{\phi}{2}\sigma_3 } e^{-i\frac{\theta}{2}\sigma_2 }e^{i\frac{\phi}{2}\sigma_3 }. \label{facphisu2}
\ee
A similar factorization holds for non-linear realizations of  arbitrary spin matrices. This factorization significantly reduces numerical  computation time in using   $\Phi$, especially for large spin matrices.  
As noted in (\ref{diagpau}), $\Phi$ enjoys the  $U(1)$ degrees of freedom (apart from the overall $U(1)$) 
\be
\Phi ~\rightarrow~\Phi\cdot e^{i\frac{\chi}{2} \sigma_3}, \label{u1trans}
\ee
which corresponds to the degrees of freedom for the relative phase of two  spin-coherent states. For the original  Hamiltonian (\ref{so3spinmodemin}),  this $U(1)$ symmetry acts as  
\be
e^{-i\frac{\chi}{2}\tilde{\sigma}_3}~ H ~e^{i\frac{\chi}{2}\tilde{\sigma}_3} =H, 
\ee
where 
\be
\tilde{\sigma}_3 \equiv \Phi \sigma_3 \Phi^{\dagger}=\sum_{i=1}^3 x_i \sigma_i ~(=2H). 
\ee
The $U(1)$ transformation, $e^{i\frac{\chi}{2}\tilde{\sigma}_3 } =e^{i\chi \sum_{i=1}^3 x_i \frac{1}{2}\sigma_i}$,  stands for the $SO(2)$ rotation of $\chi$ about  the direction pointed by the Bloch vector, and so the geometric origin of the $U(1)$  symmetry is understood as the $SO(2)$ stabilizer group of the two-sphere, $S^2 \simeq SO(3)/SO(2)$. It is  obvious that any rotations about the direction pointed by the Bloch vector do not change the $SO(3)$ Hamiltonian (\ref{so3spinmodemin}).    
An invariant quantity under the $U(1)$ transformation (\ref{u1trans}) is given by 
\be
 {\Phi^{(\pm 1/2)}}^{\dagger}\sigma_i \Phi^{(\pm 1/2)} =\pm x_i, 
\ee
which is nothing but  the Bloch vector (\ref{blochvec3d}).   
The Berry connections of the spin-coherent states  are derived as \cite{Berry-1984}
\be
A^{(\pm \frac{1}{2})} =-i{{\Phi}^{(\pm \frac{1}{2})}}^{\dagger}d{\Phi}^{(\pm \frac{1}{2})} =\pm \frac{1}{2}(1-\cos\theta)d\phi=\mp \frac{1}{2(1+x_3)}\epsilon_{ij 3}x_jdx_i , \label{u1monogfun}
\ee
which are realized as the diagonal components of the pure $SU(2)$ gauge field: 
\be
-i\Phi^{\dagger}d\Phi =\begin{pmatrix}
A^{(+\frac{1}{2})} & * \\
* & A^{(-\frac{1}{2})}
\end{pmatrix}. \label{aacomd}
\ee
The $U(1)$ degrees of freedom (\ref{diagpau}) formally correspond to the $U(1)$ gauge transformations through (\ref{aacomd}): 
\be
A^{(\pm \frac{1}{2})} ~~\rightarrow~~ A^{(\pm \frac{1}{2})}\pm \frac{1}{2}d \chi. 
\ee
The Berry connection (\ref{u1monogfun}) is exactly equal to the monopole gauge field with magnetic charge $\lambda=\pm 1/2$.  A natural question may arise about the relationship between the Zeeman-Dirac model and the Landau model.  
 Let us recall  the $SO(3)$ Landau model in the $U(1)$ monopole background (see \cite{Hasebe-2015} for instance).    
The degenerate lowest Landau level eigenstates of monopole charge $\pm 1/2$ are given by the monopole harmonics \cite{Wu-Yang-1976}
\footnote{
The monopole harmonics are defined on a two-sphere and their orthonormal relations  are given by 
\be
\int_{S^2} d\theta d\phi~\sin\theta ~{\phi_{\alpha}^{(\lambda)}}^* \phi_{\beta}^{(\lambda')} =2\pi \delta_{\alpha\beta}\delta_{\lambda \lambda'}. 
\ee
}
\begin{subequations}
\begin{align}
&\lambda=+\frac{1}{2}~~:~~\phi_1^{(+\frac{1}{2})}=\cos(\frac{\theta}{2}), ~~~\phi_2^{(+\frac{1}{2})} = \sin(\frac{\theta}{2})~e^{-i\phi},\\
&\lambda=-\frac{1}{2}~~:~~\phi_1^{(-\frac{1}{2})}=-\sin(\frac{\theta}{2})~e^{i\phi},~~~ \phi_2^{(-\frac{1}{2})} =\cos(\frac{\theta}{2}).
\end{align}\label{llldoublets}
\end{subequations}
Interestingly,  these lowest Landau level eigenstates  constitute the  spin-coherent states (\ref{spinso3coherentstates}): 
\be
\Phi^{(+\frac{1}{2})} =\begin{pmatrix}
{\phi_1^{(+\frac{1}{2})}}^* \\
{\phi_2^{(+\frac{1}{2})}}^* \\
\end{pmatrix},~~~\Phi^{(-\frac{1}{2})} =\begin{pmatrix}
{\phi_1^{(-\frac{1}{2})}}^* \\
{\phi_2^{(-\frac{1}{2})}}^* \\
\end{pmatrix}. \label{spinlllreso3}
\ee

\subsection{Large spin model}

\begin{figure}[tbph]
\center 
\includegraphics*[width=170mm]{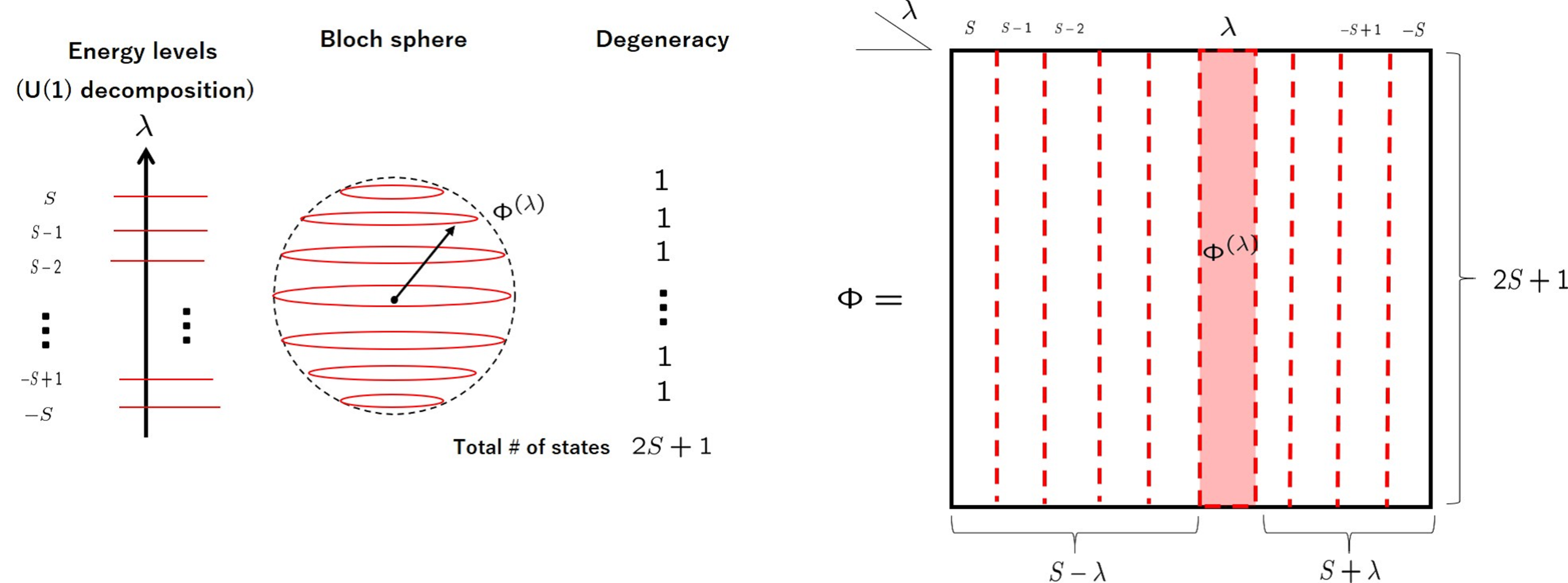}
\caption{The Bloch sphere with large spin $S$ and the $SO(3)$ spin-coherent state ${\Phi}^{(\lambda)}$ in $\Phi$. }
\label{b2comp.fig}
\end{figure}

We extend the previous discussion to  arbitrary $SU(2)$   spin matrices
 $(S=0,1/2, 1, 3/2, \cdots)$, which satisfy $[S_i, S_j]=i\epsilon_{ijk}S_k$ and 
\be
\sum_{i=1}^3 S_i S_i =S(S+1)\bs{1}_{2S+1}.  \label{subsu2}
\ee
The matrix components of the spin matrices are given by   
\begin{align}
&(S_x)_{mn} =\frac{1}{2}(\sqrt{(S+m)(S-n)} ~\delta_{m-1,n} +\sqrt{(S-m)(S+n)} ~\delta_{m,n-1}), \nn\\
&(S_y)_{mn} =i\frac{1}{2}(\sqrt{(S+m)(S-n)} ~\delta_{m-1,n} -\sqrt{(S-m)(S+n)} ~\delta_{m,n-1}), \nn\\
&(S_z)_{mn} = m\delta_{m,n}~~~~(m,n=S,S-1, S-2, \cdots, -S). \label{genelargespinmat}
\end{align}
The $S_z$ takes a diagonal form,  
\be
S_z =\begin{pmatrix}
S & 0 & 0 & 0 & 0  \\
0 & S-1 & 0 & 0 & 0 \\
0 & 0 &  S-2 & 0 & 0 \\
0 & 0 & 0 & \ddots & 0 \\
0 & 0 & 0 & 0 & -S
\end{pmatrix}. \label{matrixszdia}
\ee
The $SO(3)$   Hamiltonian (\ref{so3spinmodemin}) is simply generalized as 
\be
H=\sum_{i=1}^3  x_i S_i. \label{quantumspinlar}
\ee
As indicated before, we apply the geometric method to solve the eigenvalue problem of (\ref{quantumspinlar}): 
\be
\Phi^{\dagger}H \Phi=S_3, \label{diags3}
\ee
where $\Phi$  denotes the  non-linear realization matrix
\be
\Phi=e^{-i\theta \sum_{\mu,\nu =1}^2 \epsilon_{\mu, \nu}y_{\mu}S_{\nu}}=e^{-i\phi S_3 } e^{-i\theta S_2} e^{i\phi S_3}. \label{matsu2nonlin}
\ee 
In the notation 
\be
\Phi \equiv (\Phi^{(S)}~\Phi^{(S-1)}~\Phi^{(S-2)}~\cdots ~\Phi^{(-S)}), \label{phicolumn}
\ee
(\ref{diags3}) is restated as  
\be
H \Phi^{(\lambda)} =\lambda \cdot \Phi^{(\lambda)},
\ee
where 
\be
\lambda=S, ~S-1,~S-2, ~\cdots, -S.
\ee
The $SO(3)$ spin-coherent state\footnote{Since $S$ takes both half-integer and  integer values, $\Phi^{(\lambda)}$ may be more appropriately  called the $SU(2)$ spin-coherent states  rather than the $SO(3)$.   
}  $\Phi^{(\lambda)}$ is realized as  the $\lambda$th column of (\ref{phicolumn})  and  
 denotes the spin coherent state oriented to the latitude $\lambda$ on the Bloch sphere. Note that the spectra of $H$ are nicely illustrated as the latitudes on the Bloch sphere (Fig.\ref{b2comp.fig}).  
As $\Phi$ is a unitary matrix,  the $\Phi^{(\lambda)}$ apparently satisfy the ortho-normal relations  
\be
{\Phi^{(\lambda)}}^{\dagger}\Phi^{(\lambda')}=\delta_{\lambda \lambda'}.
\ee
Equation (\ref{diags3}) is invariant under the $U(1)$ transformation   
\be
\Phi ~~\rightarrow~~\Phi\cdot e^{i\chi S_3} 
\ee
or 
\be
\Phi^{(\lambda)} ~~\rightarrow~~\Phi^{(\lambda)} e^{i\lambda \chi}. 
\ee
In the following, we discuss several important $U(1)$ invariant quantities. 
The Bloch vector is a $U(1)$-invariant quantity, as implied by the equation   
\be
{\Phi^{(\lambda)}}^{\dagger}S_i \Phi^{(\lambda)}=\lambda \cdot x_i. 
\ee
Another important $U(1)$ invariant quantity is the quantum geometric tensor \cite{Provost-Vallee-1980}\footnote{
The quantum geometric tensor appears as the second order term of the expansion of  overlapped wave-functions. For even higher order geometric tensors, see Refs.\cite{Hetenyi-Levay-2023, Avdoshkin-Popov-2023}.} 
\be
\chi_{\mu\nu}^{(\lambda)} =\partial_{\theta_{\mu}}{\Phi^{(\lambda)}}^{\dagger}  \partial_{\theta_{\nu}}{\Phi^{(\lambda)}} -    \partial_{\theta_{\mu}}{\Phi^{(\lambda)}}^{\dagger}  {\Phi^{(\lambda)}}~ {\Phi^{(\lambda)}}^{\dagger}\partial_{\theta_{\nu}}{\Phi^{(\lambda)}}
~~~(\theta_{\mu}=\theta, \phi). \label{u1geoten}
\ee
Using the spin matrices (\ref{genelargespinmat}) and  mathematical software,\footnote{We checked the validity of (\ref{fishermetge2}) up to $S=7/2$.} 
we can easily show that the symmetric part of $\chi_{\mu\nu}^{(\lambda)}$  provides the  two-sphere metric:  
\be
g^{(\lambda)}_{\theta_{\mu}\theta_{\nu}}=\frac{1}{2}(\chi^{(\lambda)}_{\theta_{\mu}\theta_{\nu}}+\chi^{(\lambda)}_{\theta_{\nu}\theta_{\mu}}) =\frac{1}{2}(S(S+1) -\lambda^2)~ g_{\theta_{\mu}\theta_{\nu}}^{(S_2)}  \label{fishermetge2}
\ee
with 
\be
g_{\theta_{\mu}\theta_{\nu}}^{(S^2)} =\text{diag}(g^{S^2}_{\theta \theta}, g_{\phi\phi}^{S^2}) =\text{diag}(1, ~\sin^2\theta).  
\ee
The Berry phase associated with the spin-coherent state $\Phi^{(\lambda)}$ can be derived as 
\be
-i\Phi^{\dagger}d\Phi =\begin{pmatrix}
A^{(S)} & * & * & * \\
* & A^{(S-1)} & * & * \\
* & * & \ddots & * \\
* & * & * & A^{(-S)}
\end{pmatrix}
\ee
or 
\be
A^{(\lambda)}=-i{\Phi^{(\lambda)}}^{\dagger}d\Phi^{(\lambda)}=-\lambda \frac{1}{1+x_3}\epsilon_{ij3}x_j dx_i =\lambda(1-\cos\theta)d\phi.
\label{arbgaugeu1}
\ee
Note that the energy eigenvalue $\lambda$ appears as the monopole charge in (\ref{arbgaugeu1}). 
The corresponding field strength $F^{(\lambda)}=dA^{(\lambda)}=\frac{1}{2}F^{(\lambda)}_{\theta_{\mu}\theta_{\nu}}d\theta_{\mu}\wedge d\theta_{\nu}$ is equal to the anti-symmetric part of the quantum geometric tensor:  
\be
F^{(\lambda)}_{\theta_{\mu}\theta_{\nu}} =-i(\chi^{(\lambda)}_{\theta_{\mu}\theta_{\nu}}-\chi^{(\lambda)}_{\theta_{\nu}\theta_{\mu}})=\lambda\sin(\theta)\epsilon_{\mu\nu}, 
\ee
which is also a $U(1)$ gauge invariant quantity.   
The corresponding first Chern number is evaluated as 
\be
\text{ch}_1^{(\lambda)}=\frac{1}{2\pi}\int F^{(\lambda)} =2\lambda =\text{sgn}(\lambda)\cdot D_{SO(3)}(|\lambda| -\frac{1}{2})  =-\text{ch}_1^{(-\lambda)}, 
\ee
where 
\be
D_{SO(3)}(S) \equiv 2S+1. 
\ee

\begin{figure}[tbph]
\center 
\includegraphics*[width=170mm]{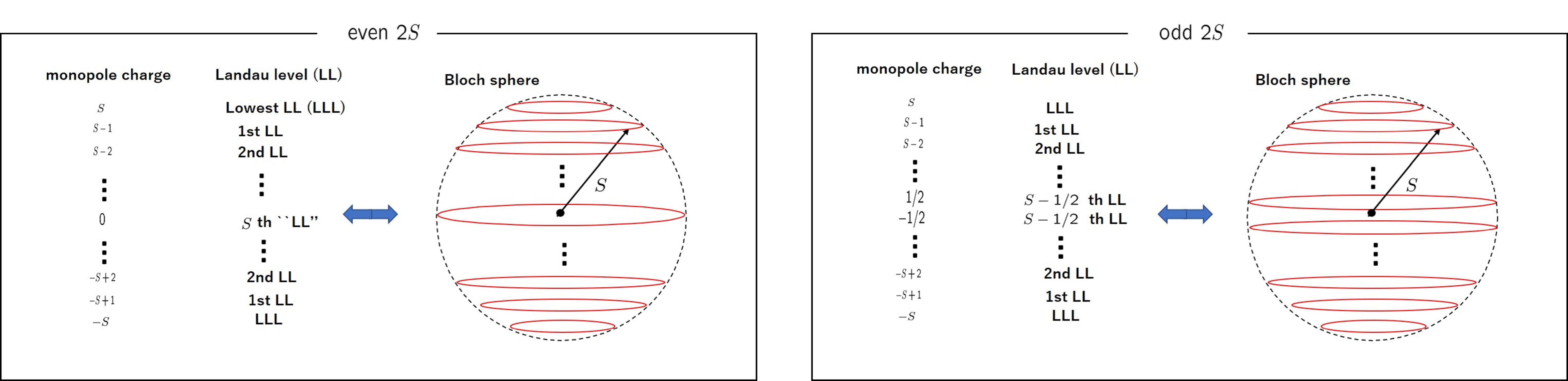}
\caption{Correspondence between the  monopole harmonics and the $SO(3)$ spin-coherent states. }
\label{so3ll.fig}
\end{figure}

It is known that  the Landau level eigenstates are also embedded in the non-linear realization matrix $\Phi$  \cite{Hasebe-2021}. 
Assume that $g$ denotes the monopole charge and  $N$ signifies the Landau level index. 
For the $SU(2)$ spin index, we have  the identification  
\be
S =N+|g|, \label{rel1so3}
\ee
and for the $U(1)$ index,  
\be
S-\lambda =N-g+|g|. \label{rel2so3}
\ee
The quantities on the left-hand sides of (\ref{rel1so3}) and (\ref{rel2so3}) come from the $SO(3)$ Zeeman-Dirac model,  while  those on the right-hand sides come from the $SO(3)$ Landau model.  
From (\ref{rel1so3}) and (\ref{rel2so3}), we have 
\be
N=S-|\lambda|, ~~~g=\lambda.
\ee
Take $\phi_{1}^{(g)}, \phi_2^{(g)}, \cdots, \phi_{2S+1}^{(g)}$  for the $N=(S-|g|)$th Landau level eigenstates in the $U(1)$ monopole background with magnetic charge $g$ (Fig.\ref{so3ll.fig})\footnote{The monopole harmonics satisfy 
\be
\int_{S^2} d\Omega_2~ {\phi_{\alpha}^{(\lambda)}}^* \phi_{\beta}^{(\lambda)} =A(S^2)\frac{1}{D_{SO(3)}(S)}\delta_{\alpha\beta}=\frac{4\pi}{2S+1}\delta_{\alpha\beta},
\ee
with $d\Omega_2=\sin\theta d\theta d\phi$, $D_{SO(3)}(S)=2S+1$ and $A(S^2)=\int_{S^2} d\Omega_2~=4\pi$. The monopole configuration (\ref{arbgaugeu1}) is represented as 
\be
A^{(\lambda)}=-i\sum_{\alpha=1}^{2S+1}\phi_{\alpha}^{(\lambda)}d{\phi_{\alpha}^{(\lambda)}}^*. 
\ee
}, then  the $SO(3)$ spin-coherent state  
is represented as 
\be
\Phi^{(\lambda)}=\begin{pmatrix}
{\phi_1^{(\lambda)*}} \\
{\phi_2^{(\lambda)*}} \\
\vdots \\
{\phi_{2S+1}^{(\lambda)*}} \\
\end{pmatrix}, \label{genso3spinco}
\ee
which signifies the exact relation between the spin-coherent states and the monopole harmonics: The spin-coherent states of large spin $S$  consist of the $(2S+1)$-fold degenerate Landau level eigenstates of $N=S-|\lambda|$ in the monopole background with magnetic charge $\lambda$ (Fig.\ref{so3ll.fig}).  

In the above discussions, we started from the Zeeman-Dirac model and later addressed the relationship to the Landau model. However, we can reverse the flow of the argument.  Suppose that the $SO(3)$ Landau model is first given and  the Landau level eigenstates are known.  
We can  generate the large spin  matrices $S_i$ by the following formula   
\be
\int_{S^2} d\Omega_2 ~{\phi^{(\lambda)}_{\alpha}}^*~ x_i ~\phi^{(\lambda)}_{\beta} =\frac{4\pi \lambda}{S(S+1)(2S+1)}(S_i)_{\alpha\beta}. \label{matelefromxa}
\ee
In the present $SO(3)$ case, since arbitrary spin matrices were already known,  the construction of  spin matrices from the Landau level eigenstates was  unnecessary. However, in the case of $SO(5)$ and other higher dimensional groups, 
this procedure is crucial when constructing  large spin gamma matrices.

\section{Bloch four-sphere and the $SO(5)$ Zeeman-Dirac model }\label{sec:blochfour}

Here, we extend the results of Sec.\ref{sec:blochfourthreetwo} to the  $SO(5)$ Zeeman-Dirac model.  The basic idea comes from  the   analogy between  the cyclotron motion  on a four-sphere and  
the  $SO(5)$  spin precession in internal space.  

\subsection{Minimal spin model}

The geometric phase of the minimal $SO(5)$ Zeeman-Dirac model \cite{Avron-Sadun-Segert-Simon-1988, Avron-Sadun-Segert-Simon-1989} has been investigated in Refs.\cite{Levay-1990, Levay-1991, Johnsson-Aitchison-1997}.  Here, we reproduce the previous results using the group theoretical method.

\begin{figure}[tbph]
\center 
\includegraphics*[width=130mm]{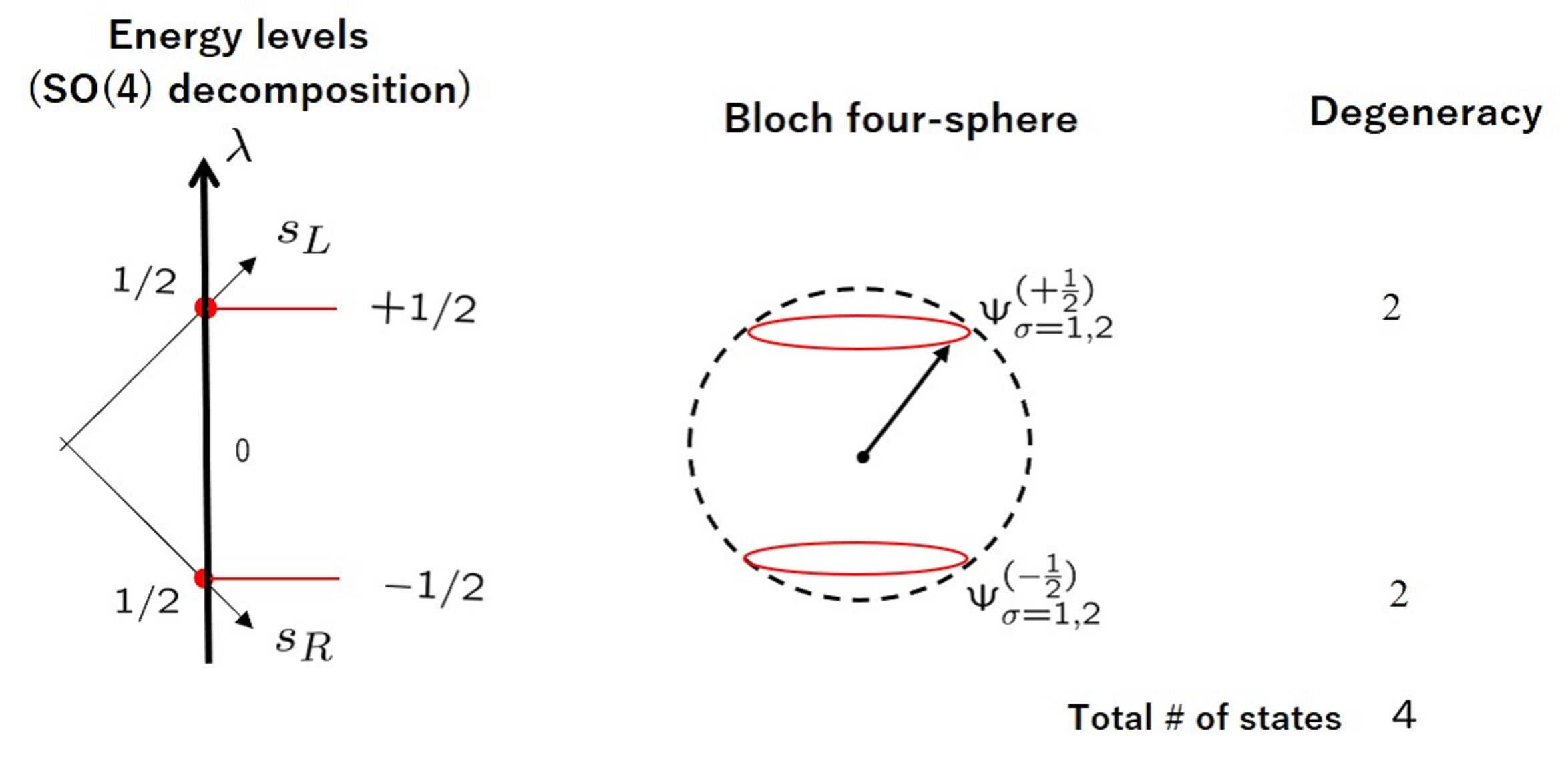}
\caption{The eigenvalues and the eigenstates of the minimal $SO(5)$ Zeeman-Dirac model.  }
\label{bloch.fig}
\end{figure}

We adopt  the following $SO(5)$ gamma matrices 
\be
\gamma_{\mu=1,2,3,4} =  \begin{pmatrix}
0 & \bar{q}_{\mu} \\
q_{\mu} & 0 
\end{pmatrix}, ~~\gamma_5= 
\begin{pmatrix}
\bs{1}_2 & 0 \\
0 & -\bs{1}_2
\end{pmatrix} ~~~~~~(q_\mu =\{-i\sigma_i, \bs{1}_2\}, ~~~\bar{q}_{\mu} =\{i\sigma_i, \bs{1}_2\}.   
\label{oriso5gam}
\ee
These satisfy 
\be
\{\gamma_a, \gamma_b\}=2\delta_{ab}\bs{1}_4 ~~~~(a,b=1,2,3, 4,5) \label{so5prgam}
\ee
and  yield the $SO(5)$ generators as 
\begin{equation}
{\sigma}_{ab}=-i\frac{1}{4}[\gamma_a,\gamma_b], \label{so5gammasigma}
\end{equation}
or 
\be
\sigma_{\mu\nu} =\frac{1}{2}\begin{pmatrix}
\eta_{\mu\nu}^{(+) i}\sigma_i & 0 \\
0 & \eta_{\mu\nu}^{(-) i}\sigma_i
\end{pmatrix},~~~\sigma_{\mu 5} = i\frac{1}{2} \begin{pmatrix}
0 & \bar{q}_{\mu} \\
-q_{\mu} & 0 
\end{pmatrix}= -\sigma_{5\mu}.
\ee
Here, $\eta_{\mu\nu}^{(\pm) i}$ denote the 't Hooft tensors, 
\be
\eta_{\mu\nu}^{(\pm) i} \equiv \epsilon_{\mu\nu i 4} \pm \delta_{\mu i}\delta_{\nu 4} \mp \delta_{\nu i}\delta_{\mu 4}. \label{defthoof}
\ee
The minimal  $SO(5)$ Zeeman-Dirac Hamiltonian is given by the following $4\times 4$ matrix\footnote{
A four-level matrix Hamiltonian is generally given by 
\be
H=\sum_{A=1}^{15}n_A \cdot \frac{1}{2}\lambda_A,
\ee
where $\lambda_A$ are $SU(4)$ Gell-Mann matrices. 
The minimal $SO(5)$ Hamiltonian (\ref{so5diham}) is realized in the special case 
\be
n_A=\sum_{a=1}^5\eta_{a 6}^A x_a 
\ee
where $\eta_{a b}^A$ denote the $SU(4)$  generalized 't Hooft symbol \cite{Hasebe-2020-3}.}
\be
H=\sum_{a=1}^5 x_a \cdot \frac{1}{2}\gamma_a ~~~~(\sum_{a=1}^5 x_ax_a =1), \label{so5diham}
\ee
where $x_a$ denote the coordinates of a four-sphere: 
\begin{align}
&x_1 =\cos\phi\sin\theta\sin\chi\sin\xi,~~x_2=\sin\phi\sin\theta\sin\chi\sin\xi,~~x_3=\cos\theta\sin\chi\sin\xi,\nn\\
&x_4=\cos\chi\sin\xi,~~x_5=\cos\xi.
\end{align}
The parameter $\xi$ signifies the azimuthal angle on $S^4$.  
Due to the property  (\ref{so5prgam}), the square of $H$ (\ref{so5diham}) becomes  
\be
H^2 =\frac{1}{4}\sum_{a=1}^5 x_a x_a \bs{1}_4 =\frac{1}{4}\bs{1}_4 ,
\ee
which implies that the eigenvalues of $H$ are 
\be
\lambda=\pm \frac{1}{2}.
\ee
Each eigenvalue is doubly degenerate. 
In the above diagonalization, we utilized the specific properties of the gamma matrices (\ref{so5prgam}) that  $SO(5)$ gamma matrices of large spin do not respect.  
For later convenience, we develop a geometric method for the present   case.  To orient the  $SO(5)$ spin-coherent state to the direction $x_a$,  we introduce the $SO(5)$  non-linear realization matrix  \cite{Hasebe-2020-1, Hasebe-2021}:
\be
\Psi=e^{i\xi \sum_{\mu=1}^4 y_\mu \sigma_{\mu 5}}, \label{non-lineso5}
\ee
where $y_{\mu}$ denote the coordinates of the $S^3$-latitude on the four-sphere at the azimuthal angle $\xi$: 
\be
y_1 =\cos\phi\sin\theta\sin\chi,~~y_2=\sin\phi\sin\theta\sin\chi,~~y_3=\cos\theta\sin\chi,
~~y_4=\cos\chi.
\ee
Note the resemblance between (\ref{defmso3nl}) and (\ref{non-lineso5}). 
The matrix $\Psi$ is represented by the $S^4$ coordinates as 
\be
\Psi= \cos(\frac{\xi}{2})\bs{1}_4 +2i\sin(\frac{\xi}{2})~\sum_{\mu=1}^4 y_{\mu}\sigma_{\mu 5} = \frac{1}{\sqrt{2(1+x_5)}}\begin{pmatrix}
(1+x_5) \bs{1}_2 & -x_{\mu}\bar{q}_{\mu} \\
x_{\mu}{q}_{\mu} & (1+x_5) \bs{1}_2
\end{pmatrix}, \label{psiminiso5}
\ee
which is factorized as    
\be
\Psi= N(\chi, \theta, \phi)^{\dagger}\cdot  e^{i\xi\sigma_{4 5}}\cdot  N(\chi, \theta, \phi) 
\label{hdmh}
\ee
where 
\be
N(\chi, \theta, \phi)\equiv e^{i\chi \sigma_{4 3}} e^{ i\theta \sigma_{3 1}} e^{i\phi \sigma_{1 2}} . 
\ee
It is not difficult to check that (\ref{psiminiso5}) diagonalizes the $SO(5)$  Hamiltonian,  
\be
\Psi^{\dagger} H \Psi =\frac{1}{2}\gamma_5, \label{fromhtogamma5}
\ee
or 
\be
H 
\Psi = \Psi \frac{1}{2}\gamma_5. \label{spinsso5cohmat1}
\ee
In the notation 
\be
\Psi=\biggl({\Psi}^{(+\frac{1}{2})} ~\vdots~ {\Psi}^{(-\frac{1}{2})}  \biggr)=
\biggl({\Psi}^{(+\frac{1}{2})}_1 ~ {\Psi}^{(+\frac{1}{2})}_2 ~\vdots~ {\Psi}^{(-\frac{1}{2})}_1 ~ {\Psi}^{(-\frac{1}{2})}_2  \biggr),
\label{mpsifours}
\ee
the eigenvalue equation (\ref{spinsso5cohmat1}) is restated as 
\be
H{\Psi}_\sigma^{({\lambda})} = \lambda ~{\Psi}_\sigma^{({\lambda})}, \label{so5cohespin1}
\ee
where $\sigma=1,2$ for each of $\lambda=+1/2, -1/2$. 
The identification (\ref{mpsifours}) indeed reproduces the $SO(5)$ spin-coherent states in the previous literature \cite{Levay-1990, Levay-1991, Johnsson-Aitchison-1997}: 
\begin{subequations}
\begin{align}
&{\Psi}^{(+\frac{1}{2})}_1
=\frac{1}{\sqrt{2(1+x_5)}} \begin{pmatrix}
1+x_5  \\
0 \\
x_4-ix_3\\
x_2-ix_1 
\end{pmatrix},~ ~ {\Psi}^{(+\frac{1}{2})}_2
=\frac{1}{\sqrt{2(1+x_5)}}  \begin{pmatrix}
0 \\
1+x_5  \\
-x_2-ix_1 \\
x_4 +ix_3 
\end{pmatrix},\\
&{\Psi}^{(-\frac{1}{2})}_1 
=\frac{1}{\sqrt{2(1+x_5)}}\begin{pmatrix}
-x_4-ix_3\\
x_2-ix_1 \\
1+x_5 \\
0 
\end{pmatrix},~ ~ {\Psi}^{(-\frac{1}{2})}_2 
 =\frac{1}{\sqrt{2(1+x_5)}}\begin{pmatrix}
-x_2-ix_1\\
-x_4+ix_3  \\
0 \\
1+x_5 
\end{pmatrix}.
\end{align}
\end{subequations}
See Fig.\ref{bloch.fig} also.  
Since $\gamma_5$ is immune to the $SO(4)$ rotations generated by $\sigma_{\mu\nu}$, 
Eq.(\ref{fromhtogamma5}) implies   the existence of the $SO(4)$  symmetry: 
\be
\Psi ~~\rightarrow~~\Psi\cdot e^{i\frac{1}{2}\omega_{\mu\nu}\sigma_{\mu\nu}}
\ee
or 
\be
\Psi^{(\pm 1/2)} ~~\rightarrow~~\Psi^{(\pm 1/2)} \cdot e^{i\frac{1}{4}\eta_{\mu\nu}^{(\pm )i}  \omega_{\mu\nu}\sigma_i}.
\ee
For the original Hamiltonian (\ref{so5diham}), the $SO(4)$ symmetry is represented as    
\be
e^{-i\frac{1}{2}\omega_{\mu\nu}\tilde{\sigma}_{\mu\nu}}~ H ~e^{i\frac{1}{2}\omega_{\mu\nu}\tilde{\sigma}_{\mu\nu}} =H, 
\ee
where $\tilde{\sigma}_{\mu\nu}$ denote the $SO(4)$ matrix generators of the form 
\be
\tilde{\sigma}_{\mu\nu} \equiv \Psi \sigma_{\mu\nu} \Psi^{\dagger}.
\ee
Such an $SO(4)$ symmetry is considered to be the ``internal'' symmetry of the $SO(5)$ Zeeman-Dirac Hamiltonian in the sense that the $SO(4)$ transformation 
does not change the direction of  the Bloch vector $x_a$, and   the double degeneracy in each energy level is a consequence of such an $SO(4)$ symmetry.  
The Bloch vector represents an $SO(4)$ invariant quantity:  
\be
{\Psi_{\sigma}^{(\pm \frac{1}{2})}}^{\dagger}\gamma_a \Psi^{(\pm \frac{1}{2})}_{\tau} = \pm x_a \delta_{\sigma\tau}. 
\ee
The Wilczek-Zee connections associated with  the $SO(5)$ spin-coherent states are derived as  
\begin{subequations}
\begin{align}
&A^{(+ \frac{1}{2})} =-i{{\Psi}^{(+\frac{1}{2})}}^{\dagger}d{\Psi}^{(+\frac{1}{2})}
=-\frac{1}{2(1+x_5)}\eta_{\mu\nu}^{(+) i}  \sigma_i x_\nu dx_\mu,  
\\
&A^{(-\frac{1}{2})} =-i{{\Psi}^{(-\frac{1}{2})}}^{\dagger}d{\Psi}^{(-\frac{1}{2})}
=-\frac{1}{2(1+x_5)}{\eta}_{\mu\nu}^{(-)i}  \sigma_i x_\nu dx_\mu, 
\end{align}\label{so4gaugemonobi}
\end{subequations}
which are exactly equal to the gauge field configuration of Yang's $SU(2)$ monopoles \cite{Yang-1978-1, Zhang-Hu-2001}. 
This implies a close relation to  the $SO(5)$ Landau model \cite{Hasebe-2020-1, Hasebe-2021}.    
Take  $\bs{\psi}_{\alpha=1,2,3,4}^{(\pm 1/2)}$ for the lowest Landau level eigenstates in the $SU(2)$ monopole/anti-monopole  with the second Chern number $+1/-1$.\footnote{The lowest Landau level eigenstates  are explicitly given by 
\begin{align}
&\!\!\!\!\!\!\!\bs{\psi}^{(+\frac{1}{2})}_1\!=\!\sqrt{\frac{1+x_5}{2}} \begin{pmatrix}
1 \\
0 
\end{pmatrix},~ \bs{\psi}^{(+\frac{1}{2})}_2 \!=\!\sqrt{\frac{1+x_5}{2}} \begin{pmatrix}
0 \\
1 
\end{pmatrix},~\bs{\psi}^{(+\frac{1}{2})}_3\!=\!\frac{1}{\sqrt{2(1+x_5)}}\begin{pmatrix}
x_4+ix_3\\
-x_2+ix_1 
\end{pmatrix},~\bs{\psi}^{(+\frac{1}{2})}_4 \!=\!\frac{1}{\sqrt{2(1+x_5)}}\begin{pmatrix}
x_2+ix_1\\
x_4-ix_3  
\end{pmatrix}, \nn\\
&\!\!\!\!\!\!\!\bs{\psi}^{(-\frac{1}{2})}_1\!=\!\frac{1}{\sqrt{2(1+x_5)}} \begin{pmatrix}
-x_4+ix_3\\
-x_2+ix_1 
\end{pmatrix},~ \bs{\psi}^{(-\frac{1}{2})}_2 \!=\!\frac{1}{\sqrt{2(1+x_5)}}  \begin{pmatrix}
x_2+ix_1 \\
-x_4 -ix_3 
\end{pmatrix},~\bs{\psi}^{(- \frac{1}{2})}_3\!=\!\sqrt{\frac{1+x_5}{2}} \begin{pmatrix}
1 \\
0 
\end{pmatrix},~ \bs{\psi}^{(-\frac{1}{2})}_4 \!=\!\sqrt{\frac{1+x_5}{2}} \begin{pmatrix}
0 \\
1 
\end{pmatrix}.
\end{align}
} They are  embedded in $\Psi$ (\ref{mpsifours}) as 
\be
\Psi^{\dagger} =\begin{pmatrix}
\bs{\psi}_1^{(+\frac{1}{2})} & \bs{\psi}_2^{(+\frac{1}{2})} & \bs{\psi}_3^{(+\frac{1}{2})} & \bs{\psi}_4^{(+\frac{1}{2})} \\
\bs{\psi}_1^{(-\frac{1}{2})} & \bs{\psi}_2^{(-\frac{1}{2})} & \bs{\psi}_3^{(-\frac{1}{2})} & \bs{\psi}_4^{(-\frac{1}{2})} 
\end{pmatrix}
\ee
or 
\be
\Psi^{(+\frac{1}{2})}=(\Psi_1^{(+\frac{1}{2})} ~\Psi_2^{(+\frac{1}{2})}) 
=\begin{pmatrix}
{\bs{\psi}_1^{(+\frac{1}{2})}}^{\dagger} \\
{\bs{\psi}_2^{(+\frac{1}{2})}}^{\dagger} \\
{\bs{\psi}_3^{(+\frac{1}{2})}}^{\dagger} \\
{\bs{\psi}_4^{(+\frac{1}{2})}}^{\dagger} 
\end{pmatrix} ,~~~~~\Psi^{(-\frac{1}{2})}=(\Psi_1^{(-\frac{1}{2})} ~\Psi_2^{(-\frac{1}{2})})
=\begin{pmatrix}
{\bs{\psi}_1^{(-\frac{1}{2})}}^{\dagger} \\
{\bs{\psi}_2^{(-\frac{1}{2})}}^{\dagger} \\
{\bs{\psi}_3^{(-\frac{1}{2})}}^{\dagger} \\
{\bs{\psi}_4^{(-\frac{1}{2})}}^{\dagger} 
\end{pmatrix}. 
\ee

\subsection{Large spin  model}\label{subsec:so5largespin}

Now we explore  $SO(5)$ Zeeman-Dirac models with  large spin. To construct  large-spin $SO(5)$ gamma matrices, we utilize the Landau level eigenstates of the $SO(5)$  Landau model \cite{Hasebe-2023-1}.  
We  take the matrix elements of the four-sphere coordinates with the  (lowest) Landau level eigenstates 
\be
(\Gamma_a)_{\alpha\beta} =
2(S+2)
 \int_{S^4}d\Omega_4 ~\psi_{\alpha}^{\dagger}~x_a ~\psi_{\beta}, \label{biggamamaso5}
\ee
where $\alpha$ runs from 1 to  
\be
D_{SO(5)}(S) =\frac{1}{3}(S+1)(2S+1)(2S+3). \label{dso5sss}
\ee
The explicit matrix forms of $\Gamma_a$ are given by  
\begin{subequations}
\begin{align}
(\Gamma_{\mu})_{(s_L', m'_L,  s_R', m'_R;~ s_L, m_L,  s_R, m_R  )}&=-2~\biggl(\sqrt{(S-\lambda+1)(S+\lambda+2)}~Y_{\mu}^{(+,-)}(s_L, s_R)_{( m'_L,  m'_R; ~m_L,  m_R  )}\delta_{s_L', s_L+\frac{1}{2}} \delta_{s_R', s_R-\frac{1}{2}}\nn\\
&~~~~~~~+\sqrt{(S+\lambda+1)(S-\lambda+2)}~Y_{\mu}^{(-,+)}(s_L, s_R)_{( m'_L,  m'_R;~ m_L,  m_R  )}\delta_{s_L', s_L-\frac{1}{2}} \delta_{s_R', s_R+\frac{1}{2}}\biggr), \\
(\Gamma_5)_{(s_L', m'_L, s_R', m'_R;~s_L, m_L, s_R, m_R  )}&=2\lambda\delta_{s_L's_L}\delta_{s_R's_R}\delta_{m'_L m_L}\delta_{m'_R, m_R}, \label{geneexga5}
\end{align}\label{gamma5ge}
\end{subequations}
where $s_L$, $s_R$, $s_L'$ and $s_R'$ are non-negative integers or half-integers subject to  $s_L'+s_R'=s_L+s_R=S$ and $\lambda \equiv s_L-s_R$. The quantities, $Y_{\mu}^{(+,-)}(s_L, s_R)$ and $Y_{\mu}^{(-,+)}(s_L, s_R)$, are defined in \cite{Hasebe-2020-1}.  
For $S=1/2$, $\Gamma_a$ (\ref{gamma5ge}) are reduced to the original $SO(5)$ gamma matrices (\ref{oriso5gam}).
 For $S=1$, see Appendix \ref{appendix:genegamm}. 

The matrices $\Gamma_a$ (\ref{gamma5ge}) can be  regarded as a natural generalization of the gamma matrices, as they 
satisfy
\footnote{While  (\ref{so5gamagenebas}) is  a natural generalization of the basic properties of the gamma matrices 
\be
\sum_{a=1}^5 \gamma_a \gamma_a =5\cdot \bs{1}_4, ~~~[\gamma_a, \gamma_b, \gamma_c, \gamma_d]=- 4!\epsilon_{abcde} \gamma_e, 
\ee
$\Gamma_a$ $(S\ge 1)$ fail to have a similar property to (\ref{so5prgam}): 
\be
\Gamma_a\Gamma_a\not \propto \bs{1}~~(\text{no sum for }a), ~~~\Gamma_a\Gamma_b\neq -\Gamma_b\Gamma_a ~~(a\neq b) . 
\ee
}      
\begin{subequations}
\begin{align}
&\sum_{a=1}^5 \Gamma_a \Gamma_a =4S(S+2)\bs{1}_{D_{SO(5)}(S)}, \label{s4quno} \\
&[\Gamma_a, \Gamma_b, \Gamma_c, \Gamma_d] =-16(S+1)\epsilon_{abcde} \Gamma_e, \label{so5gamagenebas2}
\end{align}\label{so5gamagenebas}
\end{subequations}
where   $[~,~,~,~]$ represents the Nambu four-bracket, which denotes the total antisymmetric combination of the four entities inside the bracket. These relations (\ref{so5gamagenebas}) are exactly equal to the definition of the fuzzy four-sphere    \cite{Castelino-Lee-Taylor-1997, Grosse-Klimcik-Presnajder-1996}.  
The $SO(5)$ matrix generators $\Sigma_{ab}$ with  matrix dimension (\ref{dso5sss}) can be obtained from the commutators of the  $\Gamma_a$s: 
\be
\Sigma_{ab} =-i\frac{1}{4}[\Gamma_a, \Gamma_b]. \label{so5genelargspin}
\ee
The matrices $\Gamma_a$ transform as an $SO(5)$ vector, 
\be
[\Sigma_{ab}, \Gamma_c] =i\delta_{ac}\Gamma_b- i\delta_{bc}\Gamma_a,
\ee
or 
\be
\Gamma_a~~\rightarrow~~R_{ab}\Gamma_{b},
\ee
where $R_{ab}\equiv e^{i\frac{1}{2}\omega_{ab}\Sigma_{ab}^\text{(vec)}}~~((\Sigma_{ab}^\text{(vec)})_{cd}\equiv -i\delta_{ac}\delta_{bd}+i\delta_{ad}\delta_{bc})$ denote   $SO(5)$ group elements,   
\be
R_{ac}R_{bc}=\delta_{ab}, ~~~~~\epsilon_{abcde}R_{aa'}R_{bb'}R_{cc'}R_{dd'}=\epsilon_{a'b'c'd'e'}R_{ee'}. 
\ee
It is obvious that (\ref{so5gamagenebas}) are $SO(5)$ covariant equations,   demonstrating the $SO(5)$ spherical symmetry of the present system.  
   In the large $S$ limit, Eq.(\ref{s4quno}) becomes  $\sum_{a=1}^5 \frac{1}{2}\Gamma_a \cdot \frac{1}{2}\Gamma_a\sim S^2\bs{1}_{D_{SO(5)}(S)}$, which implies that  $\frac{1}{2}\Gamma_a$ represent  quantum spin matrices of spin $S$. 
The diagonal matrix $\frac{1}{2}\Gamma_5$ (\ref{geneexga5}) is given by  
\be
\frac{1}{2}\Gamma_5 =
\begin{pmatrix}
S \bs{1}_{2S+1} & 0  & 0 & 0  & 0 \\
0 &  (S-1)\bs{1}_{4S} & 0  & 0 & 0 \\
0 &  0 & (S-2)\bs{1}_{3(2S-1)} & 0  & 0  \\
0 &  0 & 0 & \ddots  & 0  \\
0 &  0 & 0 & 0  & -S\bs{1}_{2S+1}  \\
\end{pmatrix} =\bigoplus_{\lambda=-S}^{S} \lambda~ \bs{1}_{D_{SO(4)}(s_L, s_R)},\label{gammafivei}
\ee
where 
\be
D_{SO(4)}(s_L, s_R)=(2s_L+1)(2s_R+1)=(S+\lambda+1)(S-\lambda+1)
\ee
with bi-spin index of $SU(2)_L\otimes SU(2)_R\simeq SO(4)$: 
\be
s_L \equiv \frac{S}{2}+\frac{\lambda}{2}, ~~~~s_R \equiv \frac{S}{2}-\frac{\lambda}{2}. 
\label{bispinindexso4}
\ee
Note that   $\frac{1}{2}\Gamma_5$ (\ref{gammafivei}) is  exactly equal to   $S_z$ (\ref{matrixszdia})  up to the degeneracies.   

\begin{figure}[tbph]
\center 
\includegraphics*[width=160mm]{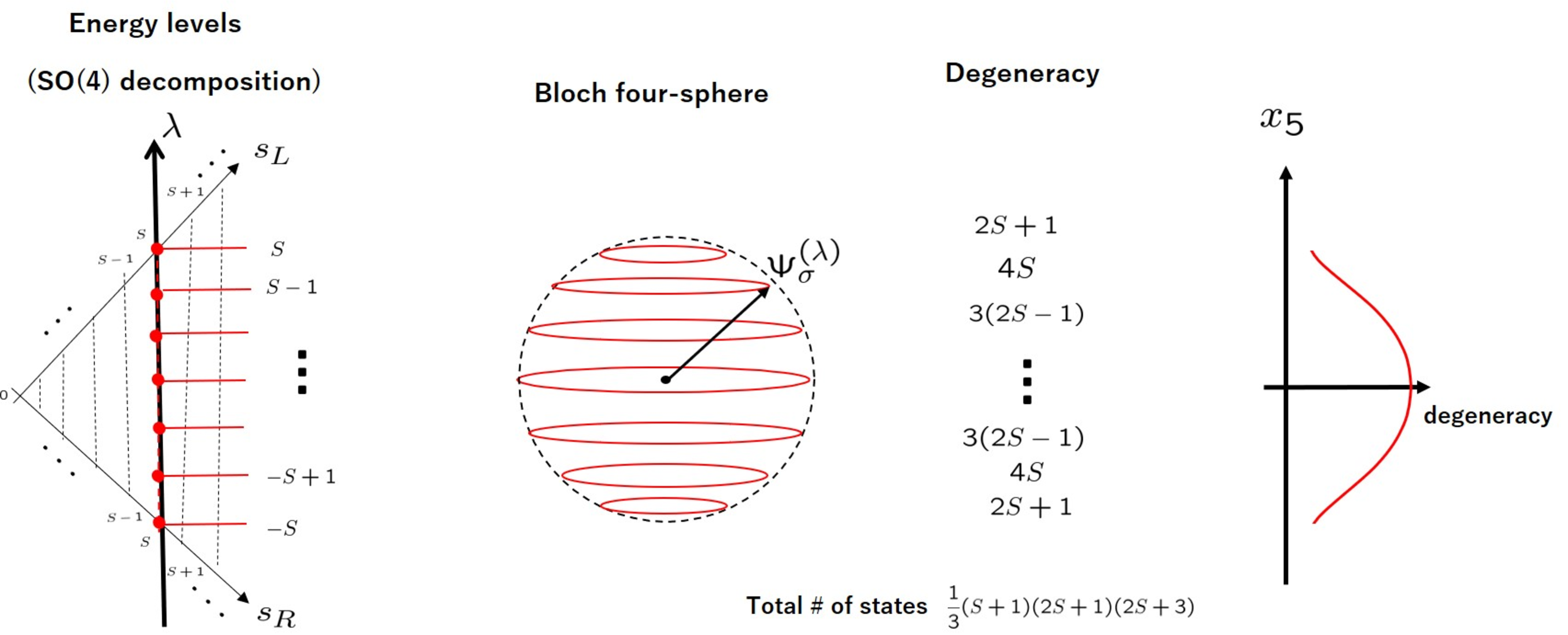}
\caption{The $SO(5)$ Zeeman-Dirac model with large spin $S$.   }
\label{lll.fig}
\end{figure}

\begin{figure}[tbph]
\center 
\includegraphics*[width=110mm]{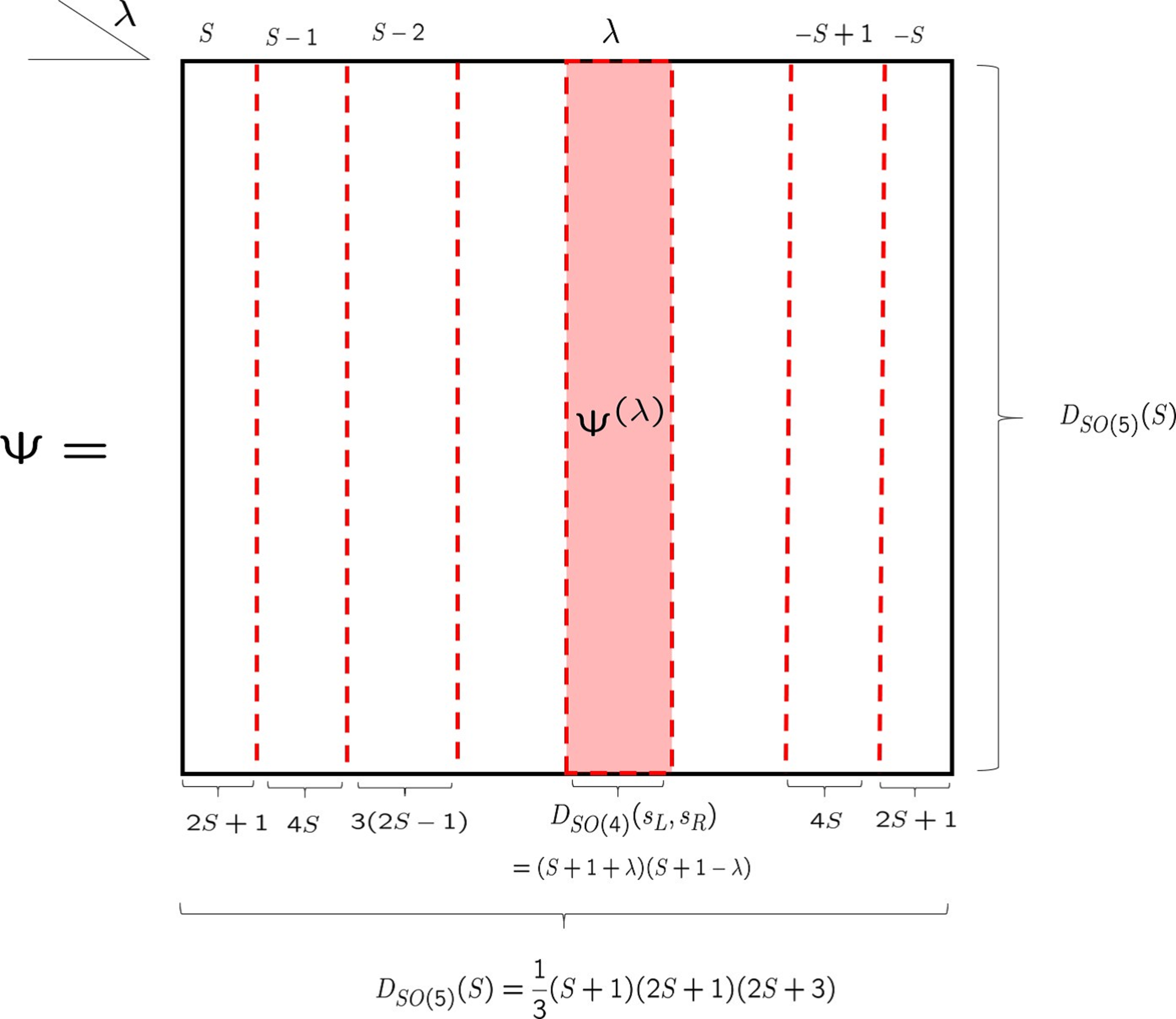}
\caption{The $SO(5)$ spin-coherent state matrix ${\Psi}^{(\lambda)}$ in  $\Psi$. }
\label{mat.fig}
\end{figure}

We now introduce the $SO(5)$ large-spin Zeeman-Dirac Hamiltonian as 
\be
H=\sum_{a=1}^5 x_a \cdot \frac{1}{2}\Gamma_a  ~~~~(\sum_{a=1}^5 x_ax_a =1). \label{geneso5ham}
\ee
Since the $\Gamma_a$ behave as an $SO(5)$ vector, we can safely use the group-theoretic method to diagonalize this Hamiltonian.  
Replacing $\sigma_{ab}$ with $\Sigma_{ab}$ (\ref{so5genelargspin}), we readily obtain 
\be
\Psi=e^{i\xi \sum_{\mu=1}^4 y_\mu \Sigma_{\mu 5}}= {N}(\chi, \theta, \phi)^{\dagger}\cdot  e^{ -i\xi \Sigma_{45 }}\cdot  N(\chi, \theta, \phi)~~~~~(N(\chi, \theta, \phi)\equiv e^{i\chi \Sigma_{4 3}} e^{ i\theta \Sigma_{3 1}} e^{i\phi \Sigma_{1 2}}), \label{lllnonre}
\ee 
which diagonalizes the Hamiltonian, 
\be
\Psi^{\dagger}H \Psi= \frac{1}{2}\Gamma_5.  \label{genediagso5gam}
\ee
The eigenvalues of the $SO(5)$ Hamiltonian range from $-S$ to $S$, with each interval between the  adjacent eigenvalues being  $1$. The degeneracy $D_{SO(4)}(s_L, s_R)$ takes a convex form with a peak at the equation $x_5=0$ (Fig.\ref{lll.fig}). 
The explicit degenerate eigenstates can be identified from the non-linear realization matrix (Fig.\ref{mat.fig}): 
\begin{align}
\Psi
&=\biggl(\Psi^{(S)} ~\vdots ~\Psi^{(S-1)}) ~\vdots ~ \Psi^{(S-2)}) ~\vdots~  \cdots~ \vdots~ \Psi^{(-S)} \biggr)\nn\\
&=\biggl( 
{\Psi}^{(S)}_1    \cdots   {\Psi}^{(S)}_{2S+1} ~\vdots~  {\Psi}^{(S-1)}_1  \cdots   {\Psi}^{(S-1)}_{4S} 
   ~\vdots~ {\Psi}^{(S-2)}_1    \cdots   {\Psi}^{(S-2)}_{3(2S-1)} ~\vdots~  \cdots   \cdots  \cdots \! ~\vdots~ {\Psi}^{(-S)}_1   \cdots    {\Psi}^{(-S)}_{2S+1} 
\biggr). \label{eigeninm}
\end{align}
The columns ${\Psi}_\sigma^{(\lambda)}$ $(\lambda=S, S-1, \cdots, -S, ~\sigma=1,2,\cdots, D_{SO(4)}(s_L, s_R))$ denote the  $SO(5)$ spin-coherent states that satisfy  
\be
\sum_{a=1}^5 (x_a \cdot \frac{1}{2}\Gamma_a) \Psi^{(\lambda)}_{\sigma} = \lambda\Psi^{(\lambda)}_{\sigma} ~~~~(\sigma=1,2,\cdots, D_{SO(4)}(s_L, s_R)). 
\ee
Their ortho-normal relations are given by  
\be
{\Psi^{(\lambda)}_{\sigma}}^{\dagger}\Psi^{(\lambda')}_{\tau} =\delta_{\sigma\tau}\delta_{\lambda\lambda'}.
\ee
Since $\Gamma_5$ is immune to the $SO(4)$ transformations, $[\Gamma_5, \Sigma_{\mu\nu}]=0$,   there exist $SO(4)$ degrees of freedom in (\ref{genediagso5gam}): 
\be
\Psi ~~\rightarrow ~~\Psi \cdot e^{i\frac{1}{2}\omega_{\mu\nu} \Sigma_{\mu\nu}}. 
\ee
The Bloch vector  is an $SO(4)$ invariant quantity: 
\be
 {\Psi^{(\lambda)}}^{\dagger}\Gamma_a \Psi^{(\lambda)}=  2\lambda \cdot x_a \bs{1}_{D_{SO(4)}(s_L, s_R)} .
\ee
Unlike the previous $SO(3)$ case (\ref{u1geoten}), the quantum geometric tensor becomes a matrix-valued tensor,  which transforms as an $SO(4)$ covariant quantity (not an $SO(4)$ invariant quantity): 
\be
\chi_{\theta_{\mu}\theta_{\nu}}^{(\lambda)}=\partial_{\theta_{\mu}}{\Psi^{(\lambda)}}^{\dagger}  \partial_{\theta_{\nu}}{\Psi^{(\lambda)}} -   \partial_{\theta_{\mu}}{\Psi^{(\lambda)}}^{\dagger}  {\Psi^{(\lambda)}}~ {\Psi^{(\lambda)}}^{\dagger}\partial_{\theta_{\nu}}{\Psi^{(\lambda)}} 
~~~(\theta_{\mu}, \theta_{\nu}=\xi, \chi, \theta, \phi).  \label{chiso4}
\ee
(See Appendix \ref{append:matrixtensgeo} for more details on matrix-valued quantum geometric tensors.)   
The trace of its symmetric part gives rise to the four-sphere metric:\footnote{Using a mathematical software, we checked the validity of (\ref{fishermetge4}) up to $S=2$. 
Similar calculations have been performed in the context of the Landau models \cite{Hasebe-2021,  Ishiki-Matsumoto-Muraki-2018}.
}
\be
g^{(\lambda)}_{\theta_{\mu}\theta_{\nu}}
=\frac{1}{2}\text{tr}(\chi^{(\lambda)}_{\theta_{\mu}\theta_{\nu}}+\chi^{(\lambda)}_{\theta_{\nu}\theta_{\mu}})~\propto ~g_{\theta_\mu \theta_\nu}^{(S^4)} 
= \text{diag}(1, ~\sin^2\xi, ~\sin^2\xi\sin^2\chi, ~\sin^2\xi\sin^2\chi\sin^2\theta).    \label{fishermetge4}
\ee
The dependence of $S$ and $|\lambda|$ is accounted for by the proportionality  coefficient  omitted in (\ref{fishermetge4}).  

\begin{figure}[tbph]
\center 
\includegraphics*[width=140mm]{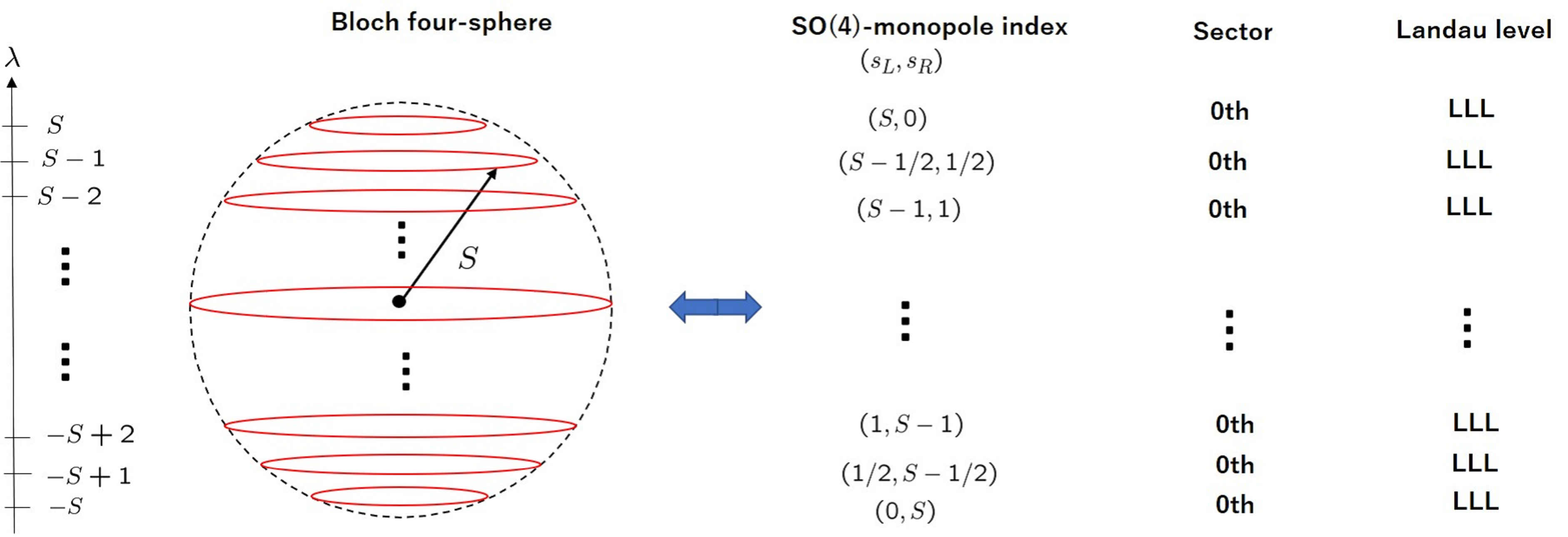}
\caption{ The  Bloch four-sphere and the $SO(5)$ Landau level eigenstates.  }
\label{so5ll.fig}
\end{figure}

Following a similar calculation in \cite{Hasebe-2021}, 
the Wilczek-Zee connections are derived as 
\be
 -i\Psi^{\dagger}d\Psi =\begin{pmatrix}
A^{(S)} & * &  * & *  & * \\
* & A^{(S-1)} & *  & * & * \\
* & * & \ddots  & * & * \\
 * & * & * & A^{(-S+1)} & * \\
  * & * & * & * & A^{(-S)} 
\end{pmatrix},
\ee
where 
\be
A^{(\lambda)}=-i{{\Psi}^{(\lambda)}}^{\dagger}d{\Psi}^{(\lambda)}
=-\frac{1}{1+x_5} \Sigma_{\mu\nu}^{(s_L, s_R)} x_\nu dx_\mu=\frac{1}{2}\omega_{\mu\nu \theta_{\rho}}\Sigma_{\mu\nu}^{(s_L, s_R)}d\theta_{\rho}. \label{so4bwzg}
\ee
Here, $\omega_{\mu\nu \theta_{\rho}}$ denote the spin-connection of $S^4$ \cite{Hasebe-2020-1} and  
$\Sigma_{\mu\nu}^{(s_L, s_R)}$  signify the $SO(4)$ matrix generators  
\be
\Sigma_{\mu\nu}^{(s_L, s_R)} \equiv \eta^{(+)i}_{\mu\nu}S_i^{(s_L)}\otimes \bs{1}_{2s_R+1} + \bs{1}_{2s_L +1}\otimes \eta^{(-)i}_{\mu\nu}S_i^{(s_R)}, 
\ee
with 't Hooft tensors $\eta_{\mu\nu}^{(\pm)i}$  (\ref{defthoof}).  
 The Wilczek-Zee connections  $A^{(\lambda)}$ in (\ref{so4bwzg}) coincide with the gauge fields of the $SO(4)$ monopoles \cite{Hasebe-2021}.\footnote{The stereographic projection of the $SO(4)$ monopole is given by the $SO(4)$ BPST instanton configuration on $\mathbb{R}^4$: 
\be
A_\mu=-\frac{2}{x^2+1}\Sigma_{\mu\nu}^{(s_L, s_R)}x_\nu, ~~~F_{\mu\nu} =-\frac{4}{(x^2+1)^2}\Sigma_{\mu\nu}^{(s_L, s_R)}, 
\ee
which does not satisfy either the self- or the anti-self dual equation, but realizes a solution of the pure Yang-Mills field equation.   
}  
The corresponding curvature, $F_{\theta_\mu \theta_\nu}=\partial_{\theta_\mu} A_{\theta_{\nu}} -\partial_{\theta_{\nu}} A_{\theta_{\mu}} +i[A_{\theta_{\mu}}, A_{\theta_{\nu}}]$, is equal to the antisymmetric part of (\ref{chiso4}): 
\be 
F_{\theta_\mu \theta_\nu}^{(\lambda)}  =-i(\chi^{(\lambda)}_{\theta_{\mu}\theta_{\nu}}-\chi^{(\lambda)}_{\theta_{\nu}\theta_{\mu}})
=\frac{1}{2}e^{\mu'}_{~\theta_{\mu}}\wedge e^{\nu'}_{~\theta_{\nu}} \Sigma_{\mu'\nu'}^{(s_L, s_R)}, \label{fieldstso4}
\ee
where $e^{\mu'}_{~\theta_{\mu}}$ denote the vierbein of $S^4$ \cite{Hasebe-2020-1}. 
The $SO(4)$ monopole is essentially the composite of the $SU(2)$ monopole and the $SU(2)$ anti-monopole to have   two topological invariants, the  second Chern number and a generalized Euler number \cite{Hasebe-2021}: 
\begin{subequations}
\begin{align}
&\text{ch}_2^{(\lambda)}\equiv \frac{1}{8\pi^2}\int_{S^4}\tr(F\wedge F) =\frac{1}{8\pi^2}\int_{S^4}\tr(\mathcal{F}\wedge \mathcal{F})=\frac{2}{3}(S+1) ~\lambda ~(S+1+\lambda)(S+1-\lambda), 
 \label{chernumbsecp}\\
&\tilde{c}_2^{(\lambda)}\equiv  \frac{1}{8\pi^2}\int_{S^4}\tr(F\wedge \mathcal{F}) = \frac{1}{8\pi^2}\int_{S^4}\tr(\mathcal{F} \wedge F) 
=\frac{1}{3}(S(S+2)+\lambda^2) ~(S+1+\lambda)(S+1-\lambda), 
\label{genereul}
\end{align}\label{twotopinv}
\end{subequations}
where $\mathcal{F}$ stands for the dual field strength of $F$ with the replacement of the $SO(4)$  matrix generators $\Sigma^{(s_L,  s_R)}_{\mu\nu}$ in (\ref{fieldstso4}) by $\frac{1}{2}\epsilon_{\mu\nu\rho\sigma}\Sigma_{\rho,\sigma}^{(s_L, s_R)}$. The topological numbers (\ref{twotopinv}) exhibit the reflection symmetry: 
\be
\text{ch}_2^{(\lambda)} =-\text{ch}_2^{(-\lambda)}, ~~~~\tilde{c}_2^{(\lambda)} =+\tilde{c}_2^{(-\lambda)}. 
\ee
The Atiyah-Singer index theorem tells that \cite{Hasebe-2021}
\be
\text{ch}_2^{(\lambda)} =\text{sgn}(\lambda)\cdot D_{SO(5)}(S-\frac{1}{2}, |\lambda|-\frac{1}{2})=-\text{ch}_2^{(-\lambda)} , \label{chernumbsec}
\ee
where $\text{sgn}(0)\equiv 0$ and
\be
D_{SO(5)}(S-\frac{1}{2}, |\lambda|-\frac{1}{2}) \equiv \frac{2}{3}(S+1)|\lambda|(S+|\lambda|+1)(S-|\lambda|+1). 
\ee
The  $SO(5)$ spin-coherent state matrices in (\ref{eigeninm}) are represented as  
\be
\Psi^{(\lambda)}=
\begin{pmatrix}
{\Psi}^{(\lambda)}_1 & {\Psi}^{(\lambda)}_2 & \cdots & {\Psi}^{(\lambda)}_{D_{SO(4)}(s_L, s_R)}   
\end{pmatrix}
=
\begin{pmatrix}
{\boldsymbol{\psi}_{1}^{(\lambda)}}^{\dagger}\\
{\boldsymbol{\psi}_{2}^{(\lambda)}}^{\dagger}\\
{\boldsymbol{\psi}_{3}^{(\lambda)}}^{\dagger}\\
\vdots \\
{\boldsymbol{\psi}_{D_{SO(5)}(S)}^{(\lambda)}}^{\dagger}\\
\end{pmatrix}, \label{corresppsiso5mon}
\ee
where   $\boldsymbol{\psi}_{\alpha}^ {(\lambda)}$ are 
 the $SO(5)$ Landau level eigenstates  of the $SO(4)$ monopole background with bi-spin index $(s_L, s_R)=(\frac{S}{2}+\frac{\lambda}{2}, \frac{S}{2}-\frac{\lambda}{2})$ (\ref{bispinindexso4}) (see Fig.\ref{so5ll.fig}).\footnote{
The orthonormal relations for the $SO(5)$ monopole harmonics are given by  
\be
\int_{S^4}d\Omega_4 ~{\bs{\psi}_{\alpha}^{(\lambda)}}^{\dagger}\bs{\psi}_{\beta}^{(\lambda)}=A(S^4)\frac{D_{SO(4)}(s_L, s_R)}{D_{SO(5)}(S)} =8\pi^2 \frac{(S+\lambda+1)(S-\lambda+1)}{(S+1)(2S+1)(2S+3)}~~~~(\alpha,\beta=1,2,\cdots, D_{SO(5)}(S)),
\ee
where $A(S^4)=\frac{8\pi^2}{3}$.  
The $SO(4)$ monopole gauge field (\ref{so4bwzg}) can also be represented as  
\be
A^{(\lambda)}=-i\sum_{\alpha=1}^{D_{SO(5)}(S)} \bs{\psi}^{(\lambda)}_{\alpha}d{\bs{\psi}_{\alpha}^{(\lambda)}}^{\dagger}.
\ee
} 
The correspondence between the spin-coherent states and the Landau level eigenstates is given as follows: 
\begin{align}
&D_{SO(5)}(S)~~~~~~~:~~ \text{Dimension of the spin-coherent states} ~~~=~~\text{Degeneracy of the Landau level eigenstates}  \nn\\
&D_{SO(4)}(s_L, s_R)~:~~\text{Degeneracy of the spin-coherent states}~~=~~\text{Dimension of the  Landau level eigenstates} \nn
\end{align}

\section{Bloch three-sphere and $SO(4)$ Zeeman-Dirac model}\label{sec:geneso5so4}

This section discusses the $SO(4)$ Zeeman-Dirac models. The properties of the large-spin $SO(4)$ Zeeman-Dirac models  are quite different from those of the  $SO(3)$ and $SO(5)$ models. 

\subsection{Minimal spin model}

\begin{figure}[tbph]
\center 
\includegraphics*[width=130mm]{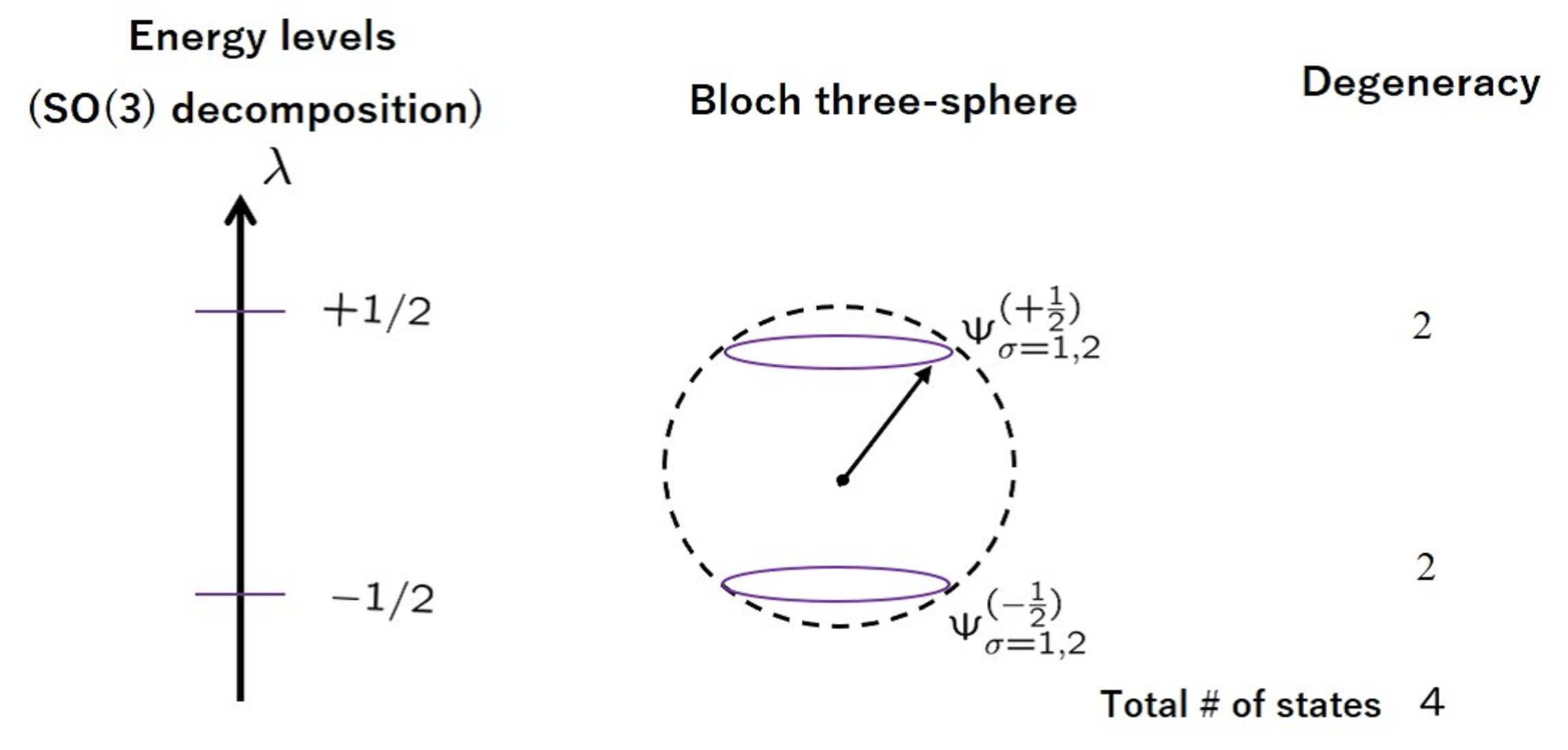}
\caption{$SO(4)$ minimal Zeeman-Dirac model.   }
\label{bl3.fig}
\end{figure}

 With the  $SO(4)$ gamma matrices $\gamma_{\mu}$ (\ref{oriso5gam}),  we construct the  minimal $SO(4)$ Zeeman-Dirac model, 
\be
H=\sum_{\mu=1}^4 x_{\mu}\cdot \frac{1}{2}\gamma_{\mu}=\frac{1}{2}\begin{pmatrix}
0 & \sum_{\mu=1}^4 x_{\mu}\bar{q}_{\mu} \\
\sum_{\mu=1}^4 x_{\mu} {q}_{\mu} & 0  
\end{pmatrix} ~~~~(\sum_{\mu=1}^4 x_{\mu}x_{\mu}=1). \label{so4diraczeeham}
\ee
As the $SO(5)$ minimal Hamiltonian (\ref{so5diham}) is reduced to (\ref{so4diraczeeham})  on the $S^3$-equator  ($\xi=\frac{\pi}{2}$) of the four-sphere,  they share similar properties, such as  $H^2 =\frac{1}{4}\bs{1}_4$. 
With the $S^3$-coordinates  
\be
x_1=\sin\theta\cos\phi\sin\chi,~~x_2=\sin\theta\sin\phi\sin\chi, ~~~x_3=\cos\theta\sin\chi, ~~~x_4=\cos\chi,  \label{xsso4}
\ee
we  introduce a unitary matrix in a similar manner to (\ref{non-lineso5})\footnote{
Using (\ref{hdmh}), we can factorize (\ref{so4matnon}) as  
\be
\Psi(\chi, \theta, \phi) =N(\theta, \phi)^{\dagger} \cdot  e^{i\chi\sigma_{34}} \cdot N(\theta, \phi)~~~~(N(\theta, \phi) \equiv e^{i\theta \sigma_{31} } e^{i\phi \sigma_{12}}).
\ee
}
\be
\Psi(\chi, \theta, \phi)=e^{i\chi\sum_{i=1}^3 y_i\sigma_{i4}}= \begin{pmatrix}
U(\chi, \theta, \phi)  & 0 \\
0 & U(\chi, \theta, \phi) ^{\dagger}
\end{pmatrix} ~~~~(y_{i=1,2,3}\equiv \frac{1}{\sin\chi}x_i), \label{so4matnon}
\ee
where 
\be
U(\chi, \theta, \phi) \equiv  e^{i\frac{\chi}{2}y_i\sigma_i} =\frac{1}{\sqrt{2(1+x_4)}}
((1+x_4)\bs{1}_2 +ix_i\sigma_i).
\ee
The unitary matrix $\Psi$  transforms the  $SO(4)$ minimal Hamiltonian  into the form     
\be
\Psi^{\dagger}H \Psi=\frac{1}{2}\gamma_4. \label{psihgamma4}
\ee
Applying another unitary transformation  
\be
V\equiv 
e^{i\frac{\pi}{2}\sigma_{45}} =\frac{1}{\sqrt{2}}\begin{pmatrix}
\bs{1}_2 & -\bs{1}_2 \\
\bs{1}_2 & \bs{1}_2
\end{pmatrix} ~~~~~(V^{\dagger}\gamma_4 V=\gamma_5),
\ee
we can diagonalize the $SO(4)$  Hamiltonian (\ref{so4diraczeeham}) as 
\be
\tilde{\Psi}^{\dagger}H \tilde{\Psi} =\frac{1}{2}\gamma_5,
\ee
where 
\be
\tilde{\Psi} \equiv \Psi ~V =\frac{1}{\sqrt{2}} \begin{pmatrix}
U & -U \\
U^{\dagger} & U^{\dagger}
\end{pmatrix}. 
\ee
Therefore, the $SO(4)$ spin-coherent states that satisfy   
\be
H\tilde{\Psi}_{\sigma}^{(\pm \frac{1}{2})} =\pm \frac{1}{2}\tilde{\Psi}_{\sigma}^{(\pm \frac{1}{2})} ~~~~(\sigma=1,2)
\ee
are obtained as  
\be
\tilde{\Psi}
=(~ \tilde{\Psi}_{1}^{(+\frac{1}{2})} ~ \tilde{\Psi}_{2}^{(+\frac{1}{2})} ~\vdots~ \tilde{\Psi}_{1}^{(-\frac{1}{2})} ~ \tilde{\Psi}_{2}^{(-\frac{1}{2})} ~)
\ee
where 
\begin{align}
&\tilde{\Psi}_{1}^{(+\frac{1}{2})} 
=\frac{1}{2\sqrt{1+x_4}} \begin{pmatrix}
1+x_4 +ix_3 \\
-x_2 +ix_1 \\
1+x_4 -ix_3 \\
x_2-ix_1
\end{pmatrix}, ~~\tilde{\Psi}_{2}^{(+\frac{1}{2})} =\frac{1}{2\sqrt{1+x_4}} \begin{pmatrix}
x_2 +ix_1 \\
1+x_4 -ix_3 \\
-x_2-ix_1 \\
1+x_4 +ix_3 
\end{pmatrix}, \nn\\
&\tilde{\Psi}_{1}^{(-\frac{1}{2})} =\frac{1}{2\sqrt{1+x_4}} \begin{pmatrix}
-1-x_4 -ix_3 \\
x_2 -ix_1 \\
1+x_4 -ix_3 \\
x_2-ix_1
\end{pmatrix}, ~~\tilde{\Psi}_{2}^{(-\frac{1}{2})} =\frac{1}{2\sqrt{1+x_4}} \begin{pmatrix}
-x_2 -ix_1 \\
-1-x_4 +ix_3 \\
-x_2-ix_1 \\
1+x_4 +ix_3 
\end{pmatrix}. 
\end{align}
See Fig.\ref{bl3.fig}.   The eigenvalues  and the 
degeneracies of the $SO(4)$ minimal model are equal to those of the $SO(5)$ minimal model.   
Equation (\ref{psihgamma4}) is invariant under the $SO(3)$ transformation 
\be
\Psi~~\rightarrow~~\Psi \cdot e^{i\frac{1}{2}\omega_{ij}\sigma_{ij}},
\ee
where $\sigma_{ij} =\frac{1}{2}\epsilon_{ijk}\begin{pmatrix}
\sigma_k & 0 \\
0 & \sigma_k
\end{pmatrix}$ are the $SO(3)$ matrix generators that commutate with $\gamma_4$. 
This  symmetry brings the $SO(3)$ degeneracy to each energy level. 
As an $SO(3)$ gauge invariant quantity,  the $SO(4)$ Bloch vector satisfies  
\be
({{\tilde{\Psi}}}^{(\pm \frac{1}{2})}_{\sigma})^{\dagger}
\gamma_\mu{\tilde{\Psi}}_{\tau}^{(\pm \frac{1}{2})} =\pm  x_\mu \delta_{\sigma\tau}. 
\ee
In the present case, 
the doubly degenerate $SO(4)$ spin-coherent states in the upper and lower energy levels provide the 
identical   Wilczek-Zee connection   
\be
{A}  \equiv -i\tilde{\Psi}_1^{\dagger}d\tilde{\Psi}_1=-i\tilde{\Psi}_2^{\dagger}d\tilde{\Psi}_2= -i\frac{1}{2}(U^{\dagger}dU +UdU^{\dagger}) =-\frac{1}{2(1+x_4)}\epsilon_{ijk}x_j dx_i
\sigma_k, \label{tilcoma}
\ee
which exactly coincides with  the $SU(2)$ spin-connection of $S^3$  \cite{Hasebe-2014-2, Nair-Daemi-2004}.

\subsection{Large spin model}

\begin{figure}[tbph]
\center 
\includegraphics*[width=160mm]{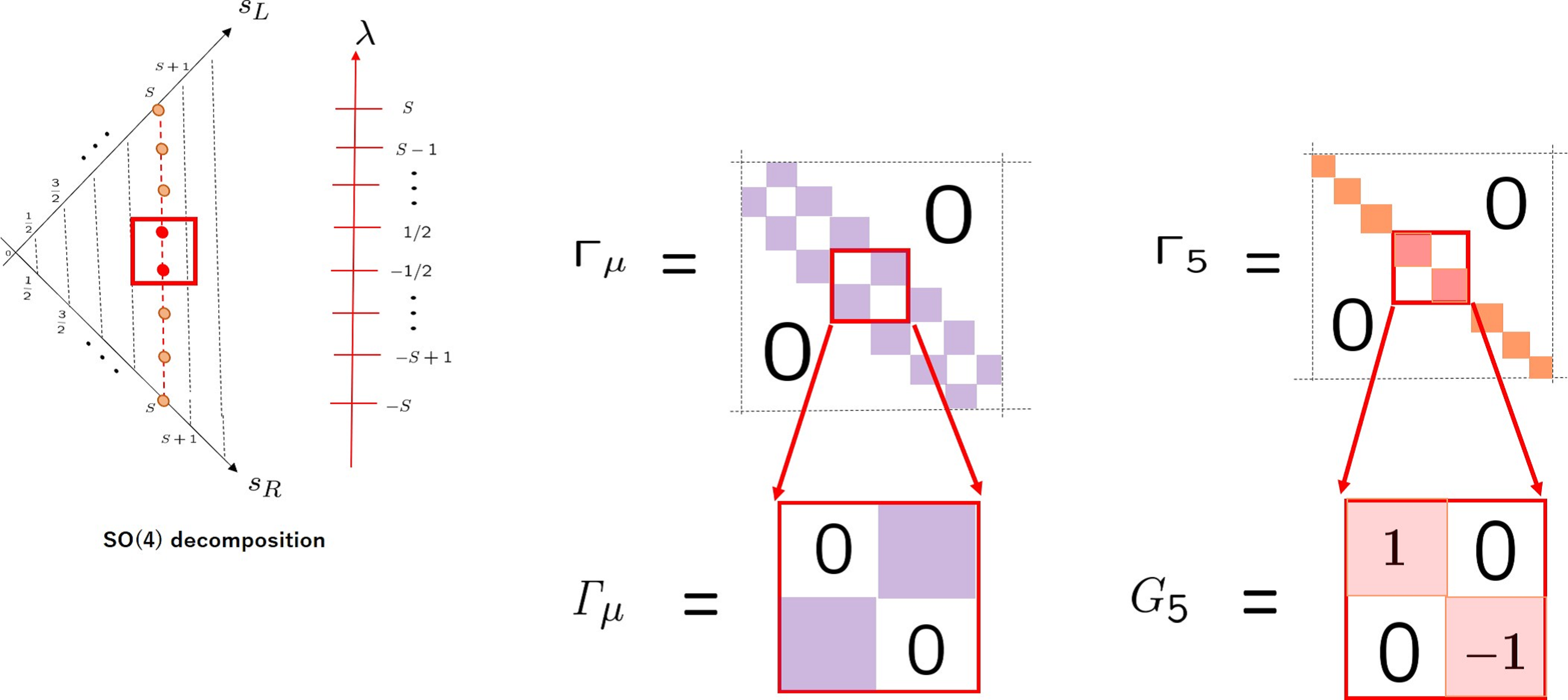}
\caption{The $SO(4)$ subspace of  $(s_L, s_R)=(\frac{2S+1}{4}, \frac{2S-1}{4})\oplus (\frac{2S-1}{4}, \frac{2S+1}{4})$, with dimension, $2\cdot \frac{2S+3}{2}\cdot \frac{2S+1}{2}=\frac{1}{2}(2S+3)(2S+1)$.}
\label{ext5.fig}
\end{figure}

The construction of the $SO(4)$ large-spin Zeeman-Dirac model  is rather tricky. 
One might consider to adopt ${\Gamma}_{\mu=1,2,3,4}$   (\ref{gamma5ge}) as the $SO(4)$ large spin gamma matrices, but 
 $\Gamma_{\mu}$ are not good enough for this purpose. This is because the sum of the squares of $\Gamma_{\mu}$  is not proportional to the unit matrix: 
\be
\sum_{\mu=1}^4 \Gamma_{\mu}\Gamma_{\mu} ~\not\propto~\bs{1}.    \label{sumgmgmne}
\ee
The generalized gamma matrices with the desired  property, $\sum_{\mu=1}^4 \mathit{\Gamma}_{\mu}\mathit{\Gamma}_{\mu} ~\propto~\bs{1}$,  can be constructed  from the $SO(4)$ Landau model \cite{Hasebe-2023-1, Hasebe-2018, Hasebe-2014-2} in the  subspace   \cite{Guralnik&Ramgoolam2001, Ramgoolam2002, Basu-Harvey-2004, JabbariTorabian2005} (Fig.\ref{ext5.fig}): 
\be
(s_L, s_R) =(\frac{S}{2}+\frac{1}{4}, \frac{S}{2}-\frac{1}{4}) \oplus (\frac{S}{2}-\frac{1}{4}, \frac{S}{2}+\frac{1}{4})~~~(2S~:~\text{odd}). \label{subspso4}
\ee
 The subspace  (\ref{subspso4})  geometrically corresponds to the two latitudes adjacent to the equator of the Bloch four-sphere.  The restriction to a sub-space obviously reduces the $SO(5)$ covariance to the $SO(4)$ covariance.   Note that $S$ must be a half-integer value in the $SO(4)$ models, so that $s_{L/R}$ (\ref{subspso4}) takes  integer or half-integer values.  
The matrix elements of $\mathit{\Gamma}_{\mu}$  in the subspace (\ref{subspso4}) are given by 
\be
\mathit{\Gamma}_{\mu}=-(2S+3)\begin{pmatrix}
0 &  Y_{\mu}^{(+,-)}(\frac{2S-1}{4}, \frac{2S+1}{4}) \\
Y_{\mu}^{(-,+)}(\frac{2S+1}{4}, \frac{2S-1}{4}) & 0
\end{pmatrix} ~~~(2S~:~\text{odd}), \label{gammagenso4}
\ee
where $Y_{\mu}^{(+,-)}(\frac{2S-1}{4}, \frac{2S+1}{4})$ are  $\frac{1}{4}(2S+1)(2S+3)\times \frac{1}{4}(2S+1)(2S+3)$  square matrices and $Y_{\mu}^{(-,+)}(\frac{2S+1}{4}, \frac{2S-1}{4})$ are their Hermitian conjugates.\footnote{Explicitly, $Y_{\mu}^{(+,-)}$ are given by \cite{Hasebe-2018}
\begin{align}
&Y_{\mu=1,2}^{(+,-)}(\frac{2S-1}{4}, \frac{2S+1}{4})_{(m'_L, m_R';~ m_L, m_R)} =\frac{1}{2S+3}(-i)^{\mu}\times\nn\\
&~~\biggl(\delta_{m'_L, m_L+\frac{1}{2}}\delta_{m'_R, m_R+\frac{1}{2}} \sqrt{(\frac{2S+3}{4}+m_L)(\frac{2S+1}{4}-m_R)}-(-1)^{\mu}\delta_{m'_L, m_L-\frac{1}{2}}\delta_{m'_R, m_R-\frac{1}{2}} \sqrt{(\frac{2S+3}{4}-m_L)(\frac{2S+1}{4}+m_R)}~\biggr), \nn\\
&Y_{\mu=3,4}^{(+,-)}(\frac{2S-1}{4}, \frac{2S+1}{4})_{(m'_L, m_R'; ~m_L, m_R)} =-\frac{1}{2S+3}(-i)^{\mu}\times\nn\\
&~~\biggl(\delta_{m'_L, m_L+\frac{1}{2}}\delta_{m'_R, m_R-\frac{1}{2}} \sqrt{(\frac{2S+3}{4}+m_L)(\frac{2S+1}{4}+m_R)}+(-1)^{\mu}\delta_{m'_L, m_L-\frac{1}{2}}\delta_{m'_R, m_R+\frac{1}{2}} \sqrt{(\frac{2S+3}{4}-m_L)(\frac{2S+1}{4}-m_R)}~\biggr), \label{expyfuz3}
\end{align}
with $-\frac{2S+1}{4}\le m_L', m_R \le \frac{2S+1}{4}$ and  $-\frac{2S-1}{4}\le m_L, m'_R \le \frac{2S-1}{4}$, 
and 
\be
Y_{\mu}^{(-,+)}(\frac{2S+1}{4}, \frac{2S-1}{4})= Y_{\mu}^{(+,-)}(\frac{2S-1}{4}, \frac{2S+1}{4})^{\dagger}.
\ee 
} 
For $S=1/2$, (\ref{gammagenso4}) is equal to $\gamma_{\mu}$. For $S=3/2$, see Appendix \ref{subsec:applendi3}. 

With (\ref{expyfuz3}), we can explicitly show that  $\mathit{\Gamma}_{\mu}$ (\ref{gammagenso4})  satisfy {\cite{Hasebe-2023-1, Hasebe-2018}\footnote{
Equation (\ref{3bracksph}) realizes a natural generalization of the properties of the $SO(4)$ gamma matrices,   
\be 
\sum_{\mu=1}^4 \gamma_\mu \gamma_\mu =4\cdot \bs{1}_4,~~~~[\gamma_\mu, \gamma_\nu, \gamma_\rho, \gamma_5]=4!\epsilon_{\mu\nu\rho\sigma} \gamma_\sigma. 
\ee
}  
\begin{subequations}
\begin{align}
&\sum_{\mu=1}^4 \mathit{\Gamma}_{\mu}\mathit{\Gamma}_{\mu} =\frac{1}{2}(2S+1)(2S+3)\bs{1}_{\frac{1}{2}(2S+1)(2S+3)},  \label{gammasqso4}  \\
&[\![ \mathit{\Gamma}_{\mu}, \mathit{\Gamma}_{\nu}, \mathit{\Gamma}_{\rho} ]\!]=16(S+1)\epsilon_{\mu\nu\rho\sigma}\mathit{\Gamma}_{\sigma}, \label{3bracksph2}
\end{align} \label{3bracksph}
\end{subequations}
where $[\![~~,~~, ~~]\!]$ signifies the Nambu ``three-bracket'' defined by  
\be
[\![ \mathit{\Gamma}_{\mu}, \mathit{\Gamma}_{\nu}, \mathit{\Gamma}_{\rho} ]\!]\equiv [ \mathit{\Gamma}_{\mu}, \mathit{\Gamma}_{\nu}, \mathit{\Gamma}_{\rho},  G_{5}]=4[\mathit{\Gamma}_{\mu}, \mathit{\Gamma}_{\nu}, \mathit{\Gamma}_{\rho}]G_5
\ee
with 
\be
G_{5} \equiv \begin{pmatrix}
 \bs{1}_{\frac{1}{4}(2S+3)(2S+1)} & 0 \\
 0 & -\bs{1}_{\frac{1}{4}(2S+3)(2S+1)}
 \end{pmatrix}. \label{matg5ex}
\ee
Equations (\ref{3bracksph}) denote the definition of  fuzzy three-sphere  \cite{Ramgoolam2002, Basu-Harvey-2004}. 
The corresponding $SO(4)$ matrix generators  are given by  
\be
\mathit{\Sigma}_{\mu\nu} \equiv   \bigoplus_{\lambda=-\frac{1}{2}}^{\frac{1}{2}}\Sigma_{\mu\nu}^{(\frac{S}{2}+\frac{\lambda}{2}, \frac{S}{2}-\frac{\lambda}{2})}  
=
\begin{pmatrix}
 \Sigma_{\mu\nu}^{ (\frac{2S+1}{4}, \frac{2S-1}{4})} & 0 \\
0  & \Sigma_{\mu\nu}^{ (\frac{2S-1}{4}, \frac{2S+1}{4})} 
\end{pmatrix}. \label{sungeneso4}
\ee
Notice that, while the commutators between $\mathit{\Gamma}_{\mu}$ do not yield $SO(4)$ matrix generators 
(\ref{sungeneso4}) 
(except for $S=1/2$)\footnote{See also Appendix \ref{subsec:applendi3}.}
\be
[\mathit{\Gamma}_{\mu}, \mathit{\Gamma}_{\nu}] ~\neq ~ 4i \mathit{\Sigma}_{\mu\nu}, 
\ee
$\mathit{\Gamma}_{\mu}$ behave as an $SO(4)$ vector under the transformation generated by $\mathit{\Sigma}_{\mu\nu}$: 
\be
[\mathit{\Sigma}_{\mu\nu}, \mathit{\Gamma}_{\rho}] =i\delta_{\mu\rho}\mathit{\Gamma}_{\nu}-i\delta_{\nu\rho}\mathit{\Gamma}_{\mu}.
\ee
The matrix $G_5$ (\ref{matg5ex}) obviously satisfies $[\mathit{\Sigma}_{\mu\nu}, G_5]=0$ and  is immune to  the $SO(4)$ transformations generated by $\mathit{\Sigma}_{\mu\nu}$.  
These properties imply that  (\ref{3bracksph}) are   $SO(4)$ covariant equations. 
Note that any of  $\mathit{\Gamma}_{\mu}$ is diagonalized as 
\begin{align}
\mathit{\Gamma}_\mu ~~~\rightarrow~~\mathit{\Gamma}_{\text{diag}} &\equiv 
\begin{pmatrix}
S\bs{1}_{2S+1} & 0 & 0 & 0 & 0\\
0 & (S-1)\bs{1}_{2S-1} & 0  & 0 & 0 \\
0 & 0 & (S-2)\bs{1}_{2S-3} & 0 & 0 \\
0 & 0 & 0 & \ddots & 0 \\
0 & 0 & 0 & 0 & -S\bs{1}_{2S+1}
\end{pmatrix}
+\frac{1}{2}G_5 \nn\\
&= \bigoplus_{\lambda=-S}^{S} (\lambda+\frac{1}{2}\text{sgn}(\lambda))~\bs{1}_{2|\lambda|+1}. \label{diagdiagammu}
\end{align}
We checked the validity of (\ref{diagdiagammu}) using the explicit form of $\mathit{\Gamma}_{\mu}$ for $S=1/2$, $3/2$, $5/2$ and $7/2$, though we do not have a general proof of (\ref{diagdiagammu}) for arbitrary $S$. 
However,  it may be reasonable to assume  that (\ref{diagdiagammu})  holds for arbitrary $S$, since the parent $SO(5)$ gamma matrices are diagonalized to take the same form $\Gamma_5$ (\ref{gammafivei})  regardless of spin magnitude. 
One may find a resemblance between $\mathit{\Gamma}_{\text{diag}}$ (\ref{diagdiagammu}) and  $\frac{1}{2}\Gamma_5$ (\ref{gammafivei}).  
We now introduce the large-spin $SO(4)$ Zeeman-Dirac Hamiltonian as   
\be
H=\sum_{\mu=1}^4 x_{\mu}\cdot \frac{1}{2}\mathit{\Gamma}_{\mu}. 
\label{so4genehamsp}
\ee
With the $SO(4)$  covariance,  
the  Hamiltonian (\ref{so4genehamsp})  can be  transformed  as  
\be
\mathit{\Psi}^{\dagger}\cdot H \cdot \mathit{\Psi} =\frac{1}{2}\mathit{\Gamma}_4, \label{unitrag4}
\ee
where 
\be
\mathit{\Psi} = e^{i\chi\sum_{i=1}^3 y_i \mathit{\Sigma}_{i4}}
=\begin{pmatrix}
e^{i\chi\sum_{i=1}^3 y_i \Sigma_{i4}^{ (\frac{2S+1}{4}, \frac{2S-1}{4})}} & 0 \\
0 & e^{i\chi\sum_{i=1}^3 y_i\Sigma_{i4}^{ (\frac{2S-1}{4}, \frac{2S+1}{4})}} 
\end{pmatrix}. 
\ee
The matrix $\mathit{\Psi}$ is factorized as 
\be
\mathit{\Psi}(\chi,\theta,\phi)
= {\mathcal{N}(\theta,\phi)}^{\dagger} ~e^{ i\chi \mathit{\Sigma}_{3 4}}~ \mathcal{N}(\theta, \phi) , \label{lllnonre4}
\ee 
with 
\be
\mathcal{N}(\theta,\phi) = e^{ i\theta \mathit{\Sigma}_{3 1}} e^{i\phi \mathit{\Sigma}_{1 2}}.
\ee
Equation (\ref{unitrag4}) obviously has the $SO(3)$ symmetry generated by 
$\mathit{\Sigma}_{ij}$, and so each energy level 
accommodates the $2|\lambda|+1$ degeneracy,  accordingly.  
\begin{figure}[tbph]
\center 
\includegraphics*[width=180mm]{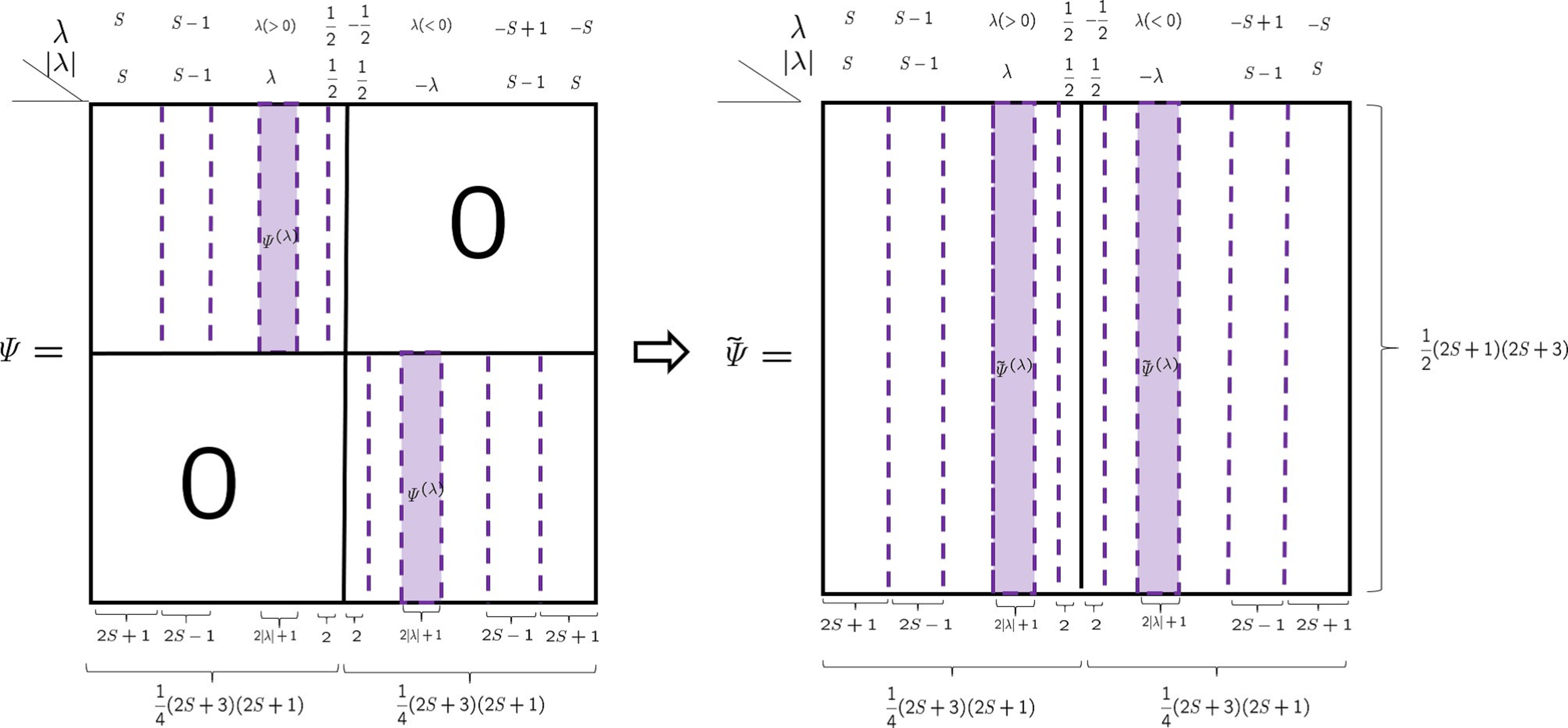}
\caption{(Left) $\mathit{\Psi}$ (\ref{lllnonre4}). For positive/negative $\lambda$, $\mathit{\Psi}^{(\lambda)}$ appears in the up-left/down-right block of $\mathit{\Psi}$. (Right)   $\tilde{\mathit{\Psi}}$ (\ref{tildemmatso4}).  For positive/negative $\lambda$,  $\tilde{\mathit{\Psi}}^{(\lambda)}$ appears in the left/right block of $\tilde{\mathit{\Psi}}$.}
\label{mat4.fig}
\end{figure}

The rectangular  matrices   
 $\Psi^{(\lambda)}$ 
 in Fig.\ref{mat4.fig}     are  made of the $SO(4)$ monopole harmonics  $\boldsymbol{\phi}_{\alpha}^{(\lambda)}\equiv \boldsymbol{\phi}_{\alpha}^{(s_L, s_R)=(\frac{S}{2}+\frac{1}{4}\text{sgn}(\lambda), \frac{S}{2}-\frac{1}{4}\text{sgn}(\lambda))}$ (\ref{so4monopolehar})  as\footnote{See Appendix \ref{appendix:so4monopoleharm} for more details on the $SO(4)$ monopole harmonics. Here, $\boldsymbol{\phi}_{\alpha}^ {(\lambda)}$ denote the lowest sub-band eigenstates of  $S-|\lambda|$th Landau level with the chirality $\text{sgn}(\lambda)$ in the background of the $SU(2)$ monopole with the spin index $|\lambda|$.} 
\be
\mathit{\Psi}^{(\lambda)} \equiv 
\begin{pmatrix}
\mathit{\Psi}^{(\lambda)}_1 & \mathit{\Psi}^{(\lambda)}_2 & \cdots & \mathit{\Psi}^{(\lambda)}_{2|\lambda|+1}   
\end{pmatrix}
=
\begin{pmatrix}
{\boldsymbol{\phi}_{1}^{(\lambda)}}^{\dagger}\\
{\boldsymbol{\phi}_{2}^{(\lambda)}}^{\dagger}\\
{\boldsymbol{\phi}_{3}^{(\lambda)}}^{\dagger}\\
\vdots \\
{\boldsymbol{\phi}_{(S+\frac{3}{2})(S+\frac{1}{2})}^{(\lambda)}}^{\dagger}\\
\end{pmatrix}.
\ee
See Fig.\ref{so4ll.fig} also. 
\begin{figure}[tbph]
\center 
\includegraphics*[width=140mm]{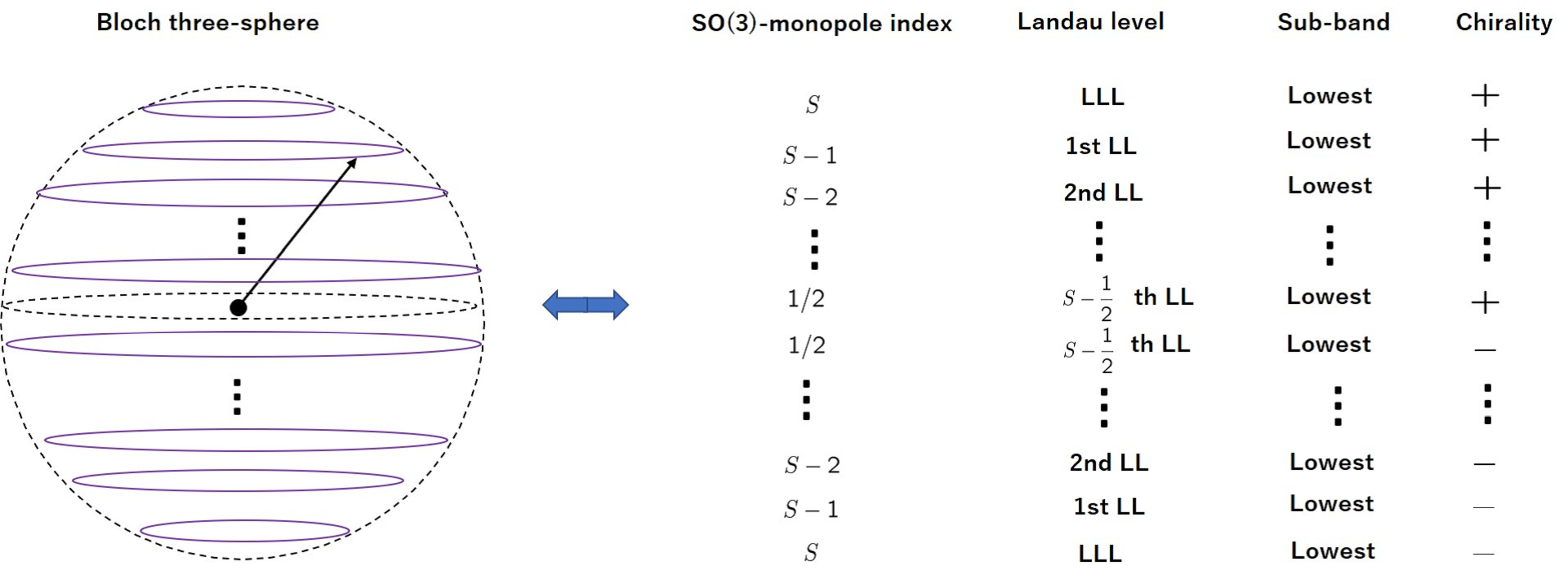}
\caption{Bloch three-sphere and the $SO(4)$ Landau level eigenstates  }
\label{so4ll.fig}
\end{figure}

With an appropriate unitary matrix $\mathcal{V}$, $\mathit{\Gamma}_4$ is diagonalized as in (\ref{diagdiagammu}):  
\be
\mathcal{V}^{\dagger}\mathit{\Gamma}_4 \mathcal{V}=\mathit{\Gamma}_{\text{diag}}. 
\ee
Therefore, with  
\be
\tilde{\mathit{\Psi}} \equiv \mathit{\Psi}\mathcal{V}, \label{tildemmatso4}
\ee
we can diagonalize $H$ (\ref{so4genehamsp}) as 
\be
\tilde{\mathit{\Psi}}^{\dagger} H \tilde{\mathit{\Psi}} =\frac{1}{2}\mathit{\Gamma}_{\text{diag}} 
= 
\begin{pmatrix}
\frac{S}{2}\bs{1}_{2S+1} & 0 & 0 & 0 & 0\\
0 & (\frac{S}{2}-\frac{1}{2})\bs{1}_{2S-1} & 0  & 0 & 0 \\
0 & 0 & (\frac{S}{2}-1)\bs{1}_{2S-3} & 0 & 0 \\
0 & 0 & 0 & \ddots & 0 \\
0 & 0 & 0 & 0 & -\frac{S}{2}\bs{1}_{2S+1}
\end{pmatrix}
+\frac{1}{4}G_5.
\ee
The eigenvalues rang from $-(\frac{S}{2}+\frac{1}{4})$ to $+\frac{S}{2}+\frac{1}{4}$,   equally spaced by $1/2$, except for the spacing $1$ between  $1/2$ and $-1/2$ (Fig.\ref{4I.fig}).  
\begin{figure}[tbph]
\center 
\includegraphics*[width=150mm]{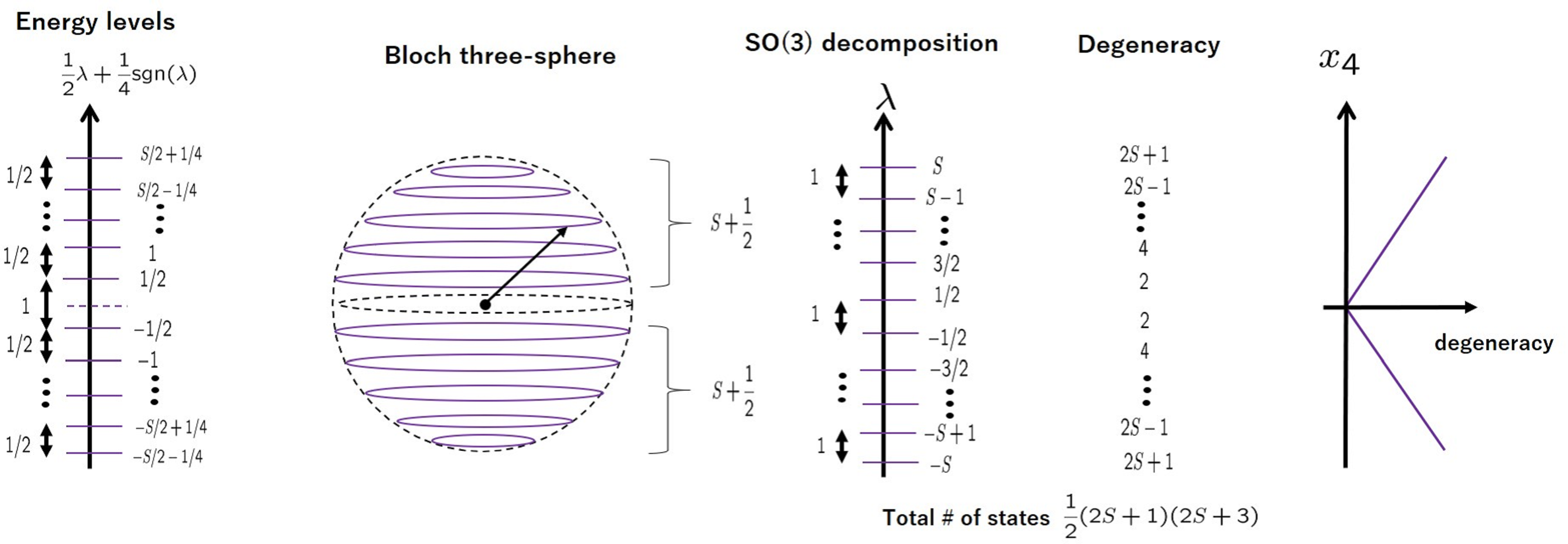}
\caption{For odd $2S$, there are $2S+1$ energy levels.  Note that zero-energy state is void.   }
\label{4I.fig}
\end{figure}

The $SO(4)$ spin-coherent states ${\tilde{\mathit{\Psi}}^{(\lambda)}_{\sigma}}$ are  realized in $\tilde{\mathit{\Psi}}$ as (Fig.\ref{mat4.fig}): 
\footnotesize
\begin{align}
\tilde{\mathit{\Psi}}
&=\biggl(\tilde{\mathit{\Psi}}^{(S)} ~\vdots ~\tilde{\mathit{\Psi}}^{(S-1)})  ~\vdots~  \cdots~ \vdots~ \tilde{\mathit{\Psi}}^{(1/2)} ~\vdots~ \tilde{\mathit{\Psi}}^{(-1/2)} ~\vdots~ \cdots ~\vdots~ \tilde{\mathit{\Psi}}^{(-S+1)} ~\vdots~ \mathit{\Psi}^{(-S)} \biggr)\nn\\
&=\biggl( 
\tilde{\mathit{\Psi}}^{(S)}_S    \cdots   \tilde{\mathit{\Psi}}^{(S)}_{-S} ~\vdots~  \tilde{\mathit{\Psi}}^{(S-1)}_{S-1}  \cdots   \tilde{\mathit{\Psi}}^{(S-1)}_{-(S-1)} 
  ~\vdots~  \cdots  ~\vdots~  \tilde{\mathit{\Psi}}^{(1/2)}_{1/2}  ~ \tilde{\mathit{\Psi}}^{(1/2)}_{-1/2} ~\vdots~ \tilde{\mathit{\Psi}}^{(-1/2)}_{1/2} ~\tilde{\mathit{\Psi}}^{(1/2)}_{-1/2} ~\vdots~ \cdots   ~\vdots~  \tilde{\mathit{\Psi}}^{(-S+1)}_{S-1}  \cdots   \tilde{\mathit{\Psi}}^{(-S+1)}_{-(S-1)} ~\vdots~ \tilde{\mathit{\Psi}}^{(-S)}_S   \cdots    \tilde{\mathit{\Psi}}^{(-S)}_{-S} 
\biggr), \label{eigeninmtil}
\end{align}
\normalsize 
and they satisfy  
\be
H\tilde{\mathit{\Psi}}^{(\lambda)}_{\sigma} =\frac{1}{2}(\lambda+\frac{1}{2}\text{sgn}(\lambda)) \cdot \tilde{\mathit{\Psi}}^{(\lambda)}_{\sigma} ~~~~~~(\sigma=\overbrace{|\lambda|, |\lambda|-1, |\lambda|-2, \cdots, -|\lambda|}^{= 2|\lambda|+1}).
\ee
Their ortho-normal relations are given by 
\be
({\tilde{\mathit{\Psi}}^{(\lambda)}_{\sigma}})^{\dagger}~{\tilde{\mathit{\Psi}}^{(\lambda')}_{\tau}}=\delta_{\sigma\tau}\delta_{\lambda\lambda'}. 
\ee
Note that the energy levels of the $SO(5)$ and $SO(4)$ Zeeman-Dirac models are equal only for the $S=1/2$ case, but are generally different. (Compare Fig.\ref{4I.fig} with 
Fig.\ref{lll.fig}). 
As the energy level approaches zero by $1/2$, the corresponding degeneracy decreases by 2, leading to the absence of a zero-energy state (Fig.\ref{4I.fig}). 
As usual, an $SO(3)$ gauge invariant quantity is given by the  Bloch vector:  
\be
({\tilde{\mathit{\Psi}}}^{(\lambda)})^{\dagger}
\mathit{\Gamma}_\mu{\tilde{\mathit{\Psi}}}^{(\lambda)} =(\lambda +\frac{1}{2}\text{sgn}(\lambda)) \cdot  x_\mu \bs{1}_{2|\lambda|+1}. 
\ee
The quantum geometric tensor is given by a matrix-valued $SO(3)$ covariant quantity, 
\be
\chi^{(\lambda)}_{\theta_i \theta_j}\equiv \partial_{\theta_i}(\tilde{\mathit{\Psi}}^{(\lambda)})^{\dagger}  \partial_{\theta_j}{\tilde{\mathit{\Psi}}^{(\lambda)}} -     \partial_{\theta_i}(\tilde{\mathit{\Psi}}^{(\lambda)})^{\dagger}  {\tilde{\mathit{\Psi}}^{(\lambda)}}~ (\tilde{\mathit{\Psi}}^{(\lambda)})^{\dagger}\partial_{\theta_j}{\tilde{\mathit{\Psi}}^{(\lambda)}}~~~~~(\theta_i, \theta_j=\chi, \theta, \phi), \label{qgtso4g}
\ee
and the trace of its symmetric part gives rise to the  three-sphere metric,
\be
g^{(\lambda)}_{\theta_i \theta_j}
=\frac{1}{2}\text{tr}(\chi^{(\lambda)}_{\theta_i \theta_j}+\chi^{(\lambda)}_{\theta_j \theta_i}) ~\propto~g_{\theta_i \theta_j}^{(S^3)} =
\text{diag}(1,  ~\sin^2\chi, ~\sin^2\chi \sin^2\theta). 
 \label{fishermetge3}
\ee
We checked (\ref{fishermetge3}) for $S=1/2, 3/2$ and $5/2$.  
The proportionality coefficients omitted in (\ref{fishermetge3}) depend on both $S$ and $|\lambda|$. 
The Wilczek-Zee connection  is derived as  
\be
-i\tilde{\mathit{\Psi}}^{\dagger}d\tilde{\mathit{\Psi}} ={\mathcal{V}}^{\dagger}(-i\mathit{\Psi}^{\dagger} d\mathit{\Psi}){\mathcal{V}}=
\left(
\begin{array}{cccc:cccc}
A^{(S)} & * & * & * & * & * & * & *\\
* & A^{(S -1)} & *  & * & * & * & * & *  \\
*  & * & \ddots  & * & * & * & * & * \\
*  & * & *  & A^{(1/2)} & * & * & * & * \\ 
\hdashline
*  & * & * & * & A^{(-1/2)}  & * & * & * \\
*  & * & * & * & * & \ddots & *  & * \\ 
*  & * & * & * & * & * & A^{(-S+1)}  & * \\    
*  & * & * & * & * & * & * &  A^{(-S)}
\end{array}
\right), \label{tildeberryco}
\ee
where 
\be
A^{(\lambda)} = -i(\tilde{\mathit{\Psi}}^{(\lambda)})^{\dagger}d\tilde{\mathit{\Psi}}^{(\lambda)}=-\frac{1}{1+x_4}\epsilon_{ijk}x_j S_k^{(|\lambda|)}dx_i=-i\frac{1}{2}({U^{(|\lambda|)}}^{\dagger}dU^{(|\lambda|)} +U^{(|\lambda|)}d{U^{(|\lambda|)}}^{\dagger})=A^{(-\lambda)}, \label{su2monogau}
\ee
with 
\be
U^{(|\lambda|)}\equiv e^{i\chi \sum_{i=1}^3 y_i S_i^{(|\lambda|)}}.
\ee
We explicitly evaluated  (\ref{su2monogau}) for $S=1/2$, $3/2$, $5/2$ and $7/2$ to have 
\be
A^{(\lambda)} =\frac{1}{2}\omega_{ij\theta_k}\epsilon_{ijk'}S^{(|\lambda|)}_{k'} d\theta_k, \label{alambdas3}
\ee
where $\omega_{ij \theta_{k}}$ denote the spin-connection of $S^3$. 
The corresponding curvature $F_{\theta_i \theta_j}=\partial_{\theta_i} A_{\theta_j} -\partial_{\theta_j} A_{\theta_i} +i[A_{\theta_i}, A_{\theta_j}]$ is  the antisymmetric part of (\ref{qgtso4g}): 
\be
F_{\theta_i \theta_j}^{(\lambda)}  =-i(\chi_{\theta_i\theta_j}^{(\lambda)}-\chi_{\theta_j\theta_i}^{(\lambda)})
=\frac{1}{2}e^{i'}_{~\theta_{i}}\wedge e^{j'}_{~\theta_{j}} \epsilon_{i'j'k'}S_{k'}^{(S)},
\ee
where $e^{i'}_{~\theta_{i}}$ denote the dreibein of $S^3$ \cite{Hasebe-2018}.

\section{Bloch hyper-spheres in even higher dimensions}\label{sec:quansphigh}

This section discusses how the previous discussions are generalized in arbitrary dimensions. 
  While $SO(d+1)$ large-spin gamma matrices can in principle be derived  using the Landau level eigenstates of  higher dimensional Landau models \cite{Hasebe-Kimura-2003, Hasebe-2014-1, Hasebe-2017}, the evaluation of their explicit matrix forms is a formidable task. We therefore deduce general results from a group theoretical analysis. 

\subsection{General properties }

As discussed in the previous sections, the $SO(5)$ Zeeman-Dirac model and the $SO(4)$  model respect the $SO(4)$ symmetry and the $SO(3)$ symmetry, respectively. These symmetries  introduce  degeneracies and the  Wilczek-Zee connections equivalent to the  $SO(4)$ and $SO(3)$ monopole gauge fields.   We will discuss how these features can be understood geometrically and extended to arbitrary dimensions.  
Let us consider the $SO(d+1)$ Zeeman-Dirac model
\be
H=\sum_{a=1}^{d+1}x_a\cdot \frac{1}{2}\Gamma_a ~~~~(\sum_{a=1}^{d+1}x_a x_a =1),\label{sod+1hamge}
\ee
where $x_a$ are given parameters that denote the Bloch vector. In general, the $SO(d+1)$ Hamiltonian (\ref{sod+1hamge}) has an $SO(d)$ symmetry,\footnote{Meanwhile, the $SO(d+1)$ Landau model has the $SO(d+1)$ symmetry and each of the  Landau levels exhibits degeneracy due to the $SO(d+1)$ symmetry. The degenerate Landau level eigenstates constitute an irreducible representation of $SO(d+1)$. } 
\be
U^{\dagger} HU =H ~~~~(U \in SO(d)).
\ee
Each of the energy levels accommodates the degeneracy attributed to the $SO(d)$ symmetry.  The geometric origin of this  $SO(d)$ symmetry is  explained as follows. Suppose that $S_{ab} \in SO(d)$ denotes the rotation about the direction of the Bloch vector (Fig.\ref{stab.fig}). 
\begin{figure}[tbph]
\center 
\includegraphics*[width=50mm]{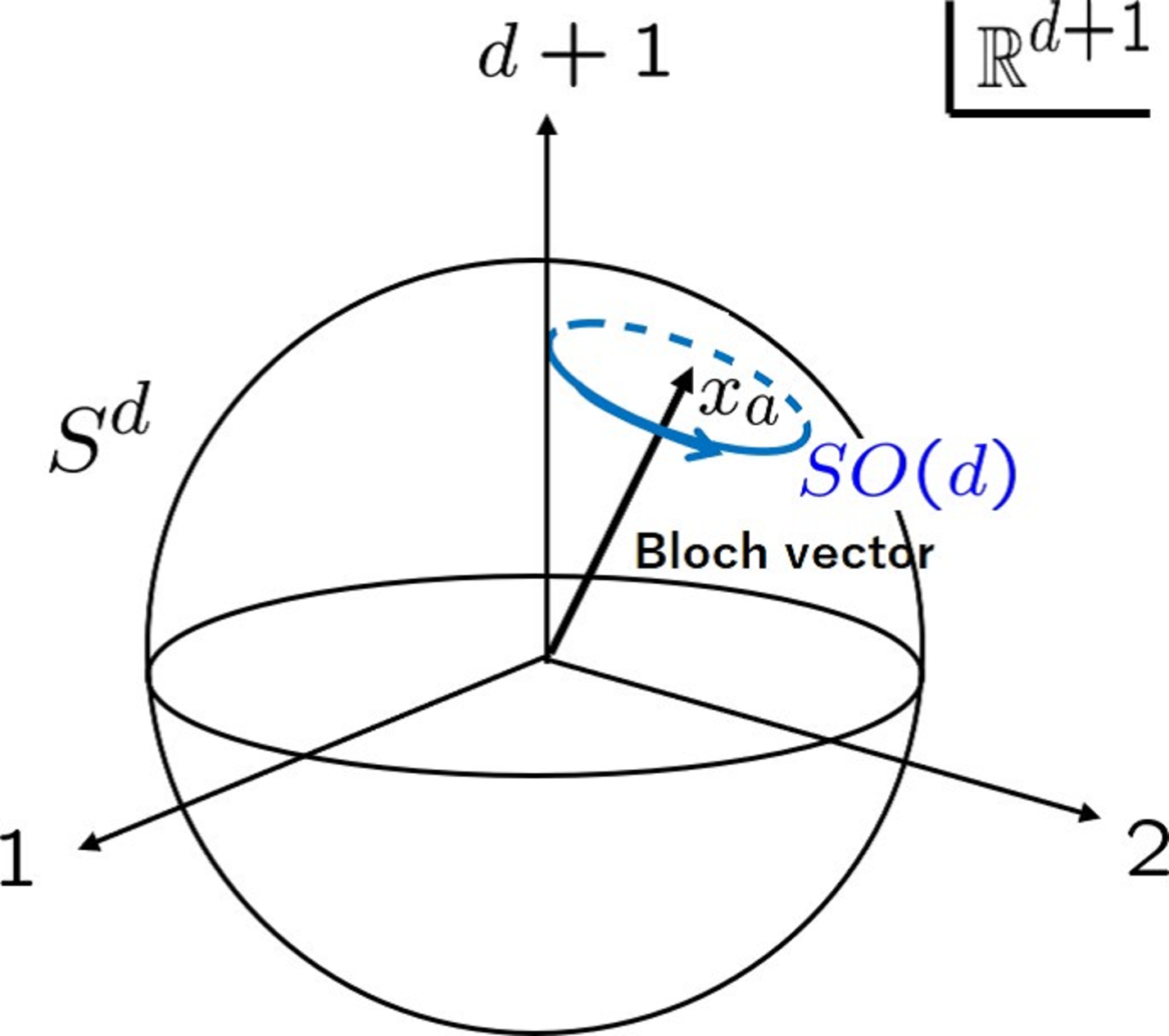}
\caption{$SO(d)$ stabilizer group   that does not transform the point $x_a$ (Bloch vector) on $S^d$.    }
\label{stab.fig}
\end{figure}
Under such a transformation,   the Bloch vector is apparently invariant  
\be
x_a ~~~\rightarrow~~S_{ab}x_b =x_a~~~~~(S_{ab} \in SO(d)).
\ee
This transformation that does not change a point on manifold is known as the  stabilizer group.   
The $SO(d)$ stabilizer group appears as the denominator  of the coset  $S^d \simeq SO(d+1)/SO(d)$. 
The $SO(d)$ invariance of the Bloch vector can be reinterpreted as a symmetry of the Hamiltonian (\ref{sod+1hamge}):
\be
H=\sum_{a=1}^{d+1}x_a\cdot \frac{1}{2}\Gamma_a~~\rightarrow ~~\sum_{a}(\sum_b S_{ab}x_b)\cdot \frac{1}{2}\Gamma_a=\sum_a   x_a \cdot \frac{1}{2} \overbrace{(\sum_b S_{ba}\Gamma_b)}^{=U^{\dagger} \Gamma_a U} =U^{\dagger} HU. 
\ee
Thus,  the $SO(d)$ symmetry of the $SO(d+1)$ Zeeman-Dirac Hamiltonian originates from the stabilizer group of the Bloch hyper-sphere.  This $SO(d)$ symmetry necessarily introduces a corresponding degeneracy to each energy level.  
Next, we clarify the geometric origin of the $SO(d)$ monopole gauge field.  Through adiabatic evolution, an $SO(d+1)$ spin-coherent state experiences transitions among the degenerate states within each energy level, giving rise to  the Wilczek-Zee connection.    This  Wilczek-Zee connection is attributed to the $SO(d)$ holonomy of $S^d$, which is identical to the gauge field  of  the $SO(d)$ non-Abelian monopole. The above mechanics is summarized  in Fig.\ref{genf.fig}.  
\begin{figure}[tbph]
\includegraphics*[width=150mm]{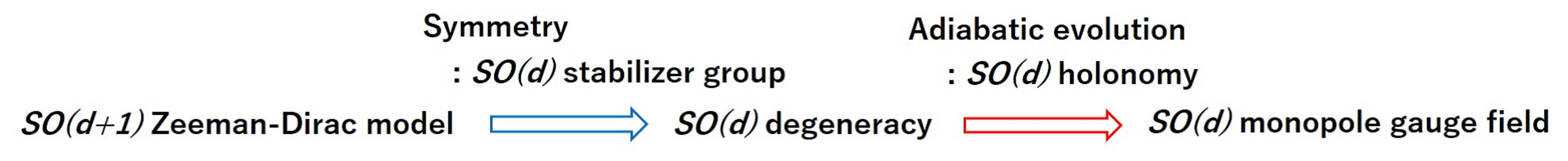}
\caption{Emergence of the $SO(d)$ monopole gauge field from the $SO(d+1)$ Zeeman-Dirac model  }
\label{genf.fig}
\end{figure}
In the following, we confirm these speculations through more concrete analyses. 

\subsection{Irreducible representations}

Before going to details,  we present a general argument about the representations of the orthogonal groups. Suppose that 
$[l_1, l_2, \cdots, l_{k}]_{SO(2k+1)}$ and $[l_1, l_2, \cdots, l_{k}]_{SO(2k)}$ signify the Young tableaux of the groups $SO(2k+1)$ and $SO(2k)$, respectively \cite{IachelloBook}.\footnote{For $SO(5)$, the index $(p,q)$ in Appendix \ref{subsec:higherlls} is related to $[l_1, l_2]_{SO(5)}$ as 
\be
p=l_1+l_2, ~~q=l_1-l_2.
\ee
For $SO(4)$, the bi-spin index $(s_L, s_R)$  is related to $[l_1, l_2]_{SO(4)}$ as 
\be
s_L=\frac{l_1+l_2}{2}, ~~s_R=\frac{l_1-l_2}{2} . 
\ee
}  
The representations of our interest are designated as 
\begin{align}
&[ \lambda]_{SO(2k+1)}\equiv [S,  \lambda]_{SO(2k+1)}\equiv [\overbrace{S, S, \cdots, S, \lambda}^{k}]_{SO(2k+1)}~~~~(0\le \lambda\le S), \\
&[ \lambda]_{SO(2k)}\equiv [ S,\lambda]_{SO(2k)}\equiv [\overbrace{S, S, \cdots, S, \lambda}^{k}]_{SO(2k)}~~~~(-S\le \lambda \le S), 
\label{so2ksdimtext} 
\end{align}
with dimensions 
\begin{subequations}
\begin{align}
&D_{SO(2k+1)}(\lambda) \equiv D_{SO(2k+1)} ( S, \lambda)\equiv \frac{2\lambda+1}{2S+1}\prod_{j=1}^{k-1}\frac{S-\lambda+k-j}{k-j}\frac{S+\lambda+k-j+1}{2S+k-j+1}\cdot \prod_{l=1}^{k}\prod_{i=1}^l \frac{2S+l+i-1}{l+i-1}, \label{so2k+1dimr} \\
&D_{SO(2k)}(\lambda) \equiv D_{SO(2k)}(S, \lambda)\equiv \prod_{j=1}^{k-1}\frac{(S+j)^2-\lambda^2}{j^2}~\cdot ~\prod_{l=1}^{k-2}\prod_{i=1}^{k-l-1}\frac{2S+2l+i}{2l+i}=D_{SO(2k)}( -\lambda). \label{so2ksdim} 
\end{align}\label{twosodeg}
\end{subequations}
In particular,\footnote{For instance, 
\begin{align}
&D_{SO(3)}(S)=2S+1, ~~~D_{SO(5)}(S)=\frac{1}{3}(S+1)(2S+1)(2S+3), ~~D_{SO(7)}(S)=\frac{1}{90}(S+1)(S+2)(2S+1)(2S+3)^2(2S+5), \nn\\
&D_{SO(9)}(S)=\frac{1}{18900}(S+1)(S+2)^2(S+3)(2S+1)(2S+3)^2(2S+5)^2(2S+7),
\end{align}
and 
\begin{align}
&D_{SO(2)}(1/2)=1,~~~~~D_{SO(4)}(1/2)=\frac{1}{4}(2S+1)(2S+3),~~~~D_{SO(6)}(1/2)=\frac{1}{192}(2S+1)(2S+3)^3(2S+5),\nn\\
&D_{SO(8)}(1/2)=\frac{1}{69120}(S+2)(2S+1)(2S+3)^3(2S+5)^3(2S+7).
\end{align}
} 
\begin{subequations}
\begin{align}
&D_{SO(2k+1)}(S)=D_{SO(2k+2)}(\pm  S) =\prod_{l=1}^k \prod_{i=1}^l \frac{2S+l+i-1}{l+i-1}  ~~\sim~~S^{\frac{1}{2}k(k+1)}~\sim~S \cdot D_{SO(2k)(1/2)}=S^1, S^3, S^6, S^{10}, \cdots, \label{so2k+1fuldim} \\
&D_{SO(2k)}(1/2)=D_{SO(2k)}(-1/2)=\prod_{j=1}^{k-1}\frac{(2S+2j)^2-1}{(2j)^2}~\cdot ~\prod_{l=1}^{k-2}\prod_{i=1}^{k-l-1}\frac{2S+2l+i}{2l+i} ~~\sim~~S^{\frac{1}{2}(k+2)(k-1)}=S^0, S^2, S^5, S^9, \cdots .
\end{align}
\end{subequations}
As we will see in Secs.\ref{subsec:arbeven} and \ref{subsec:arbodd},  $D_{SO(2k+1)}(\lambda)/D_{SO(2k)}(\lambda)$  indicates the degeneracy of the energy level indexed by $\lambda$ of the $SO(2k+2)/SO(2k+1)$  model.   
The degeneracies (\ref{twosodeg}) are shown in Fig.\ref{dis.fig}. 
\begin{figure}[tbph]
\hspace{-0.4cm}
\includegraphics*[width=180mm]{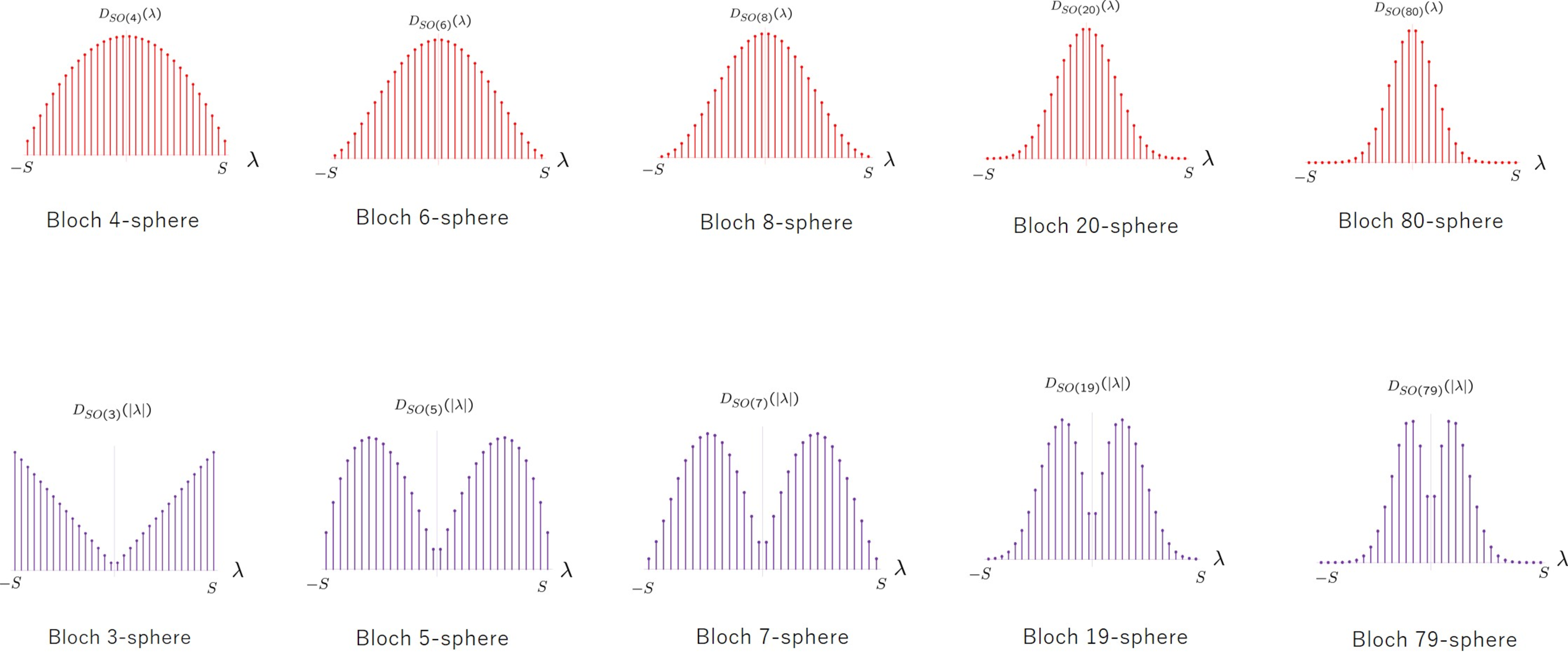}
\caption{The upper/lower figure represents the distributions of  the degeneracies of the $SO(2k+1)$/$SO(2k)$ Zeeman-Dirac model for $2S=31$.  }
\label{dis.fig}
\end{figure}
There are interesting relations between  adjacent dimensions:  
\begin{subequations}
\begin{align}
&D_{SO(2k+1)} (S) =\sum_{\lambda=-S}^{S} D_{SO(2k)}(\lambda), \label{reladimhie1} \\
&D_{SO(2k)} ( 1/2) =\sum_{\lambda=\frac{1}{2}}^{S} D_{SO(2k-1)}( \lambda) =\sum_{\lambda=-S}^{-\frac{1}{2}} D_{SO(2k-1)}( -\lambda)=D_{SO(2k)} ( -1/2)
~~~~(2S~:~\text{odd}), \label{reladimhie2}
\end{align}\label{hier}
\end{subequations}
which imply 
\begin{subequations}
\begin{align}
&\Sigma_{\mu\nu}^{[S]_{SO(2k+1)}} =\bigoplus_{\lambda=-S}^S \Sigma_{\mu\nu}^{[\lambda]_{SO(2k)}}  ~~~(\mu,\nu =1,2,\cdots, 2k), \\
&\Sigma_{ij}^{[1/2]_{SO(2k)}} =\bigoplus_{\lambda=1/2}^S \Sigma_{ij}^{[\lambda]_{SO(2k-1)}} ~~(i,j =1,2,\cdots, 2k-1). 
\end{align}
\end{subequations}
Notice that (\ref{reladimhie2}) holds $\it{only}$ for odd $2S$, not for even $2S$.   (Recall that odd dimensional Bloch hyper-spheres are defined only for half-integer $S$.) 
Equation (\ref{hier}) implies the dimensional hierarchies between  even  and odd dimensions.\footnote{Such a dimensional hierarchy has been observed in the  Landau models \cite{Hasebe-2017, Hasebe-2014-1, Hasebe-Kimura-2003} and also in  the non-linear sigma models \cite{Hasebe-2020-3}. }

\subsection{$SO(2k+1)$ Zeeman-Dirac model}\label{subsec:arbeven}

 As in the $SO(5)$ case, there exist large spin gamma matrices  for arbitrary $SO(2k+1)$ groups (see Refs.\cite{Hasebe-2010, DeBellisSS2010} for reviews  and references therein).  
 Using such  gamma matrices,  we can construct the large spin $SO(2k+1)$ Zeeman-Dirac model.  
For a better understanding, we analyze the $SO(2k+1)$ minimal model in Appendix \ref{append:dimeven}. 

The $SO(2k+1)$ large-spin gamma matrices satisfy two basic equations\footnote{The matrices ${\Gamma}_{a}$ satisfy the orthonormal relations: 
\be
\tr({\Gamma}_{a}{\Gamma}_{b}) =4\frac{S(S+k)}{2k+1}D_{SO(2k+1)}(S)~\delta_{ab}. \label{gamso2k+1ortho}
\ee
}
\begin{subequations}
\begin{align}
&\sum_{a=1}^{2k+1} \Gamma_a \Gamma_a =  4S(S+k)\bs{1}_{D_{SO(2k+1)}(S)} , \\
&[\Gamma_{a_1}, \Gamma_{a_2}, \cdots, \Gamma_{a_{2k}}] = i^k \frac{(2k)! !~(2S+2k-2)!!}{(2S)!!} ~\epsilon_{a_1 a_2 \cdots a_{2k+1}}\Gamma_{a_{2k+1}},
\label{propgammasggen} 
\end{align}\label{propgammasggenboth} 
\end{subequations}
where $[~,~,\cdots , ~]$ is called the $2k$-bracket that signifies totally antisymmetric combination of the $2k$ quantities inside the bracket.  The matrices  $\Gamma_a$ thus   satisfy the quantum Nambu geometry \cite{Azuma-Bagnoud-2003, Azuma-thesis-2004} and act as the coordinates of a fuzzy $2k$-sphere.   
The commutators between $\Gamma_a$s yield the $SO(2k+1)$ generators of symmetric representation\footnote{The sum of the squares  of (\ref{sigso2k+1reps}) is given by  
\be
\sum_{a<b=1}^{2k+1}\Sigma_{ab}^{[S]_{SO(2k+1)}}\Sigma_{ab}^{[S]_{SO(2k+1)}} =kS(S+k)\bs{1}_{D_{SO(2k+1)}(S)}. 
\ee
} 
\be
\Sigma_{ab}^{[S]_{SO(2k+1)}} =-i\frac{1}{4}[\Gamma_a, \Gamma_b]. \label{sigso2k+1reps}
\ee
The $SO(2k+1)$ covariance of $\Gamma_a$ is represented as  $[\Sigma_{ab}^{[S]_{SO(2k+1)}}, \Gamma_c]=i\delta_{ac}\Gamma_b-i\delta_{bc}\Gamma_a$. 
The $SO(2k+1)$ Zeeman-Dirac Hamiltonian 
\be
H=\sum_{a=1}^{2k+1}x_a \cdot \frac{1}{2}\Gamma_a ~~~~~~(\sum_{a=1}^{2k+1}x_ax_a=1) \label{sumgamevenhamil}
\ee
is diagonalized as 
\be
\Psi^{\dagger}H \Psi =\frac{1}{2}\Gamma_{2k+1} =\bigoplus_{\lambda=-S}^{S}~\lambda~\bs{1}_{D_{SO(2k)}(\lambda)} , \label{diagdiraczee2k+1}
\ee
where  
\be
\Psi=e^{i\theta_{2k}\sum_{\mu=1}^{2k}y_{\mu}\Sigma_{\mu, {2k+1}}^{[S]_{SO(2k+1)}}}=N^{\dagger}\cdot e^{i\theta_{2k}\Sigma_{2k,2k+1}^{[S]_{SO(2k+1)}}}\cdot  N~~~~~(y_{\mu=1,2,\cdots, 2k}=\frac{1}{\sin\theta_{2k}} x_{\mu},~~~x_{2k+1}=\cos\theta_{2k}),  
\ee 
with 
\be
N=e^{i\theta_{2k-1}\Sigma_{2k, 2k-1}^{[S]_{SO(2k+1)}}}~e^{i\theta_{2k-2}\Sigma_{2k-1, 2k-2}^{[S]_{SO(2k+1)}}}~\cdots~e^{i\theta_4\Sigma_{54}^{[S]_{SO(2k+1)}}} ~e^{i\theta_3\Sigma_{43}^{[S]_{SO(2k+1)}}}~e^{i\theta \Sigma_{31}^{[S]_{SO(2k+1)}}}~e^{i\phi\Sigma_{12}^{[S]_{SO(2k+1)}}}.  \label{factorizeh}
\ee
As shown in (\ref{diagdiraczee2k+1}), the $SO(2k+1)$  Hamiltonian  exhibits  $2S+1$ energy levels 
\be
\lambda =S,~S-1, ~S-2, \cdots, -S, \label{speso2k+1} 
\ee
with degeneracies $D_{SO(2k)}(\lambda)$ (\ref{so2ksdim}).  The spectrum (\ref{speso2k+1}) is symmetric with respect to the origin, and the geometric picture of the Bloch $2k$-sphere is similar to that of the Bloch four-sphere (Fig.\ref{lll.fig}), up to the energy level degeneracy.  

This $SO(2k)$ degeneracy  comes from the $SO(2k)$ symmetry of (\ref{diagdiraczee2k+1}) 
\be
\Psi ~\rightarrow~\Psi\cdot e^{i\frac{1}{2}\omega_{\mu\nu}\Sigma^{[S]_{SO(2k+1)}}_{\mu\nu}}. 
\ee
The $SO(2k)$ decomposition (\ref{reladimhie1}) and the analyses of  Appendix \ref{append:dimeven} 
suggest that the Wilczek-Zee $SO(2k)$ connection is given by the $SO(2k)$ monopole gauge field, 
\be
A^{(\lambda)}=-\frac{1}{1+x_{2k+1}}\Sigma_{\mu\nu}^{[ \lambda]_{SO(2k)}}x_{\nu}dx_{\mu}, 
\label{so2kmonogau}
\ee
where $\Sigma_{\mu\nu}^{[ \lambda]_{SO(2k)}}$ denote the $SO(2k)$ generators of 
$[\lambda]_{SO(2k)}$.  For $k=2$, (\ref{so2kmonogau}) reproduces  the $SO(4)$ monopole connection (\ref{so4bwzg}).     We   also checked the validity of (\ref{so2kmonogau}) using  generalized $SO(7)$ gamma matrices for $S=1/2$, $1$ and $3/2$.     The non-trivial topology of the $SO(2k)$ monopole field configuration is specified by the   $k$th Chern number 
\be
\text{ch}_k = \frac{1}{k!(2\pi)^k}\int_{S^{2k}}\tr(F^k), \label{kthcs}
\ee
which is equivalent to the homotopy map from the equator of $S^{2k}$ to the $SO(2k)$ transition function,  
\be
\pi_{2k-1}(SO(2k)) ~\simeq ~\mathbb{Z}. 
\ee
For the monopole field configuration (\ref{so2kmonogau}),  the $k$th Chern number is evaluated as   
\be
\text{ch}_k^{[ \lambda ]_{SO(2k)}} =\text{sgn}(\lambda) \cdot D_{SO(2k+1)}
(S-\frac{1}{2}, |\lambda|-\frac{1}{2}) =-\text{ch}_k^{[-\lambda ]_{SO(2k)}} \label{geneechenexp}
\ee
with $\text{sgn}(0)\equiv 0$.  Equation (\ref{geneechenexp}) is an apparent generalization of  the previous $k=2$ case  (\ref{chernumbsec}). 
Two opposite energy levels with respect to the zero-energy have the same  magnitude of Chern numbers with opposite signs.

\subsection{$SO(2k)$ Zeeman-Dirac model}\label{subsec:arbodd}

 The $SO(2k)$ large-spin gamma matrices  are realized in the subspace $\lambda=(+1/2) \oplus  (-1/2$)  of the $SO(2k+1)$ large-spin gamma matrices \cite{Ramgoolam2002, JabbariTorabian2005}.  The spin magnitude  $S$ should be a half-integer for the same reason as in the $SO(4)$ model.  
 Analysis of the $SO(2k)$ minimal model is presented in Appendix \ref{append:dimodd}. 

The $SO(2k)$ large spin gamma matrices are given by the following off-diagonal block matrices, 
\be
\mathit{\Gamma}_{\mu} =\begin{pmatrix}
0 & \mathcal{Y}_{\mu}^{\dagger} \\
\mathcal{Y}_{\mu} & 0 
\end{pmatrix}~~~~~~(\mu=1,2,\cdots, 2k).
\ee
They  satisfy the two equations:\footnote{Together with
\be
\mathit{\Gamma}_{2k+1} = \sqrt{\frac{(2S+1)(2S+2k-1)}{4k}} ~  G_{2k+1},
\ee
$\mathit{\Gamma}_{a=1,2,\cdots, 2k+1}$ satisfy the orthonormal relations, 
\be
\tr(\mathit{\Gamma}_a \mathit{\Gamma}_b)=\frac{(2S+1)(2S+2k-1)}{2k} D_{SO(2k)}(1/2)~\delta_{ab}, \label{mathgamso2k+1ortho}
\ee
and  the quantum Nambu algebra, 
\be
[\mathit{\Gamma}_{a_1}, \mathit{\Gamma}_{a_2}, \cdots, \mathit{\Gamma}_{a_{2k}}] = i^k \sqrt{\frac{(2S+1)(2S+2k-1)}{4k}} ~\frac{(2k)! !~(2S+2k-2)!!}{(2S)!!} ~\epsilon_{a_1 a_2 \cdots a_{2k+1}}\mathit{\Gamma}_{a_{2k+1}}. 
\ee
} 
\begin{subequations}
\begin{align}
&\sum_{\mu=1}^{2k} \mathit{\Gamma}_\mu \mathit{\Gamma}_\mu =  \frac{1}{2}(2S+1)(2S+2k-1)\bs{1}_{2D_{SO(2k)}(1/2)} ,  \label{propgammasggen2k1}\\ 
&[\![ \mathit{\Gamma}_{\mu_1}, \mathit{\Gamma}_{\mu_2}, \cdots, \mathit{\Gamma}_{\mu_{2k-1}}]\!] = -i^k ~\frac{(2k)! !~(2S+2k-2)!!}{(2S)!!}  ~\epsilon_{\mu_1 \mu_2 \cdots \mu_{2k}}\mathit{\Gamma}_{\mu_{2k}},  
\label{propgammasggen2k}
\end{align}\label{sq3eq2}
\end{subequations}
where 
\be
[\![ \mathit{\Gamma}_{\mu_1}, \mathit{\Gamma}_{\mu_2}, \cdots, \mathit{\Gamma}_{\mu_{2k-1}} ]\!]\equiv [\mathit{\Gamma}_{\mu_1}, \mathit{\Gamma}_{\mu_2}, \cdots, \mathit{\Gamma}_{\mu_{2k-1}}, G_{{2k+1}}]=2k~[\mathit{\Gamma}_{\mu_1}, \mathit{\Gamma}_{\mu_2}, \cdots, \mathit{\Gamma}_{\mu_{2k-1}}] G_{{2k+1}}. 
\ee 
Equation (\ref{propgammasggen2k1}) was derived in Ref.\cite{Ramgoolam2002}.  
The matrix $G_{2k+1}$ is a diagonal matrix 
\be
G_{2k+1} =    \begin{pmatrix}
\bs{1}_{D_{SO(2k)} (1/2)} & 0 \\
0 & -\bs{1}_{D_{SO(2k)} (1/2)}
\end{pmatrix},
\ee 
which anti-commutes with  all $\mathit{\Gamma}_{\mu}$s: 
\be
\{\mathit{\Gamma}_{\mu} , G_{2k+1}\}=0. 
\ee
With such $\mathit{\Gamma}_{\mu}$s, 
we construct the $SO(2k)$ Zeeman-Dirac Hamiltonian as 
\be
{H}=\sum_{\mu=1}^{2k}x_\mu \cdot \frac{1}{2}\mathit{\Gamma}_\mu  =\frac{1}{2}\begin{pmatrix}
0 & \mathit{Q}^{(-)} \\
\mathit{Q}^{(+)} & 0 
\end{pmatrix}
~~~~~~(\sum_{\mu=1}^{2k}x_\mu x_\mu =1), \label{so2klargespinham}
\ee
where 
\be
\mathit{Q}^{(+)}\equiv \sum_{\mu=1}^{2k} x_{\mu}\mathcal{Y}_{\mu}, ~~~~\mathit{Q}^{(-)}\equiv {\mathit{Q}^{(+)}}^{\dagger}=\sum_{\mu=1}^{2k} x_{\mu}\mathcal{Y}_{\mu}^{\dagger}. 
\ee
The Hamiltonian (\ref{so2klargespinham}) obviously respects the chiral symmetry: 
\be
\{\mathit{H} , G_{2k+1}\}=0.
\ee
While the commutators between $\mathit{\Gamma}_{\mu}$s  do not realize $SO(2k)$ generators,  $\mathit{\Gamma}_{\mu}$  transform as a vector under the $SO(2k)$ transformations generated by the following $SO(2k)$ generators  
\cite{Hasebe-2023-1},
\be
\mathit{\Sigma}_{\mu\nu} \equiv \begin{pmatrix}
\Sigma_{\mu\nu}^{[+1/2]_{SO(2k)}} & 0 \\
0 & \Sigma_{\mu\nu}^{[-1/2]_{SO(2k)}}
\end{pmatrix}. 
\ee
The non-linear realization matrix is constructed as 
\be
\mathit{\Psi} =e^{i\theta_{2k-1}\sum_{i=1}^{2k-1}y_{i}\mathit{\Sigma}_{i, 
{2k}}}=\mathcal{N}^{\dagger}~e^{i\theta_{2k-1}\mathit{\Sigma}_{2k-1, 2k}}~\mathcal{N} =\begin{pmatrix}
\mathcal{U}^{[+1/2]} & 0 \\
0 & \mathcal{U}^{[-1/2]}
\end{pmatrix},
\ee
where 
\begin{subequations}
\begin{align}
&y_{i=1,2,\cdots, 2k-1}=\frac{1}{\sin(\theta_{2k-1})} x_{i} ~~~x_{2k}=\cos(\theta_{2k-1}), \\
&\mathcal{N} =e^{i\theta_{2k-2}\mathit{\Sigma}_{2k-1, 2k-2}}e^{i\theta_{2k-3}\mathit{\Sigma}_{2k-2, 2k-3}} \cdots e^{i\theta_{4}\mathit{\Sigma}_{54}}e^{i\theta_{3}\mathit{\Sigma}_{4 3}}e^{i\theta\mathit{\Sigma}_{3 1}}e^{i\phi\mathit{\Sigma}_{12}}, \\
&\mathcal{U}^{[\pm 1/2]} \equiv e^{i\theta_{2k-1}\sum_{i=1}^{2k-1}y_{i}{\Sigma}^{[\pm 1/2]_{SO(2k)}}_{i, 
{2k}}}. \label{ulambda}
\end{align}
\end{subequations}
The matrix $\mathit{\Psi}$  transforms the $SO(2k)$ Hamiltonian (\ref{so2klargespinham}) into the form 
\be
\mathit{\Psi}^{\dagger}{H}\mathit{\Psi} =\frac{1}{2}\mathit{\Gamma}_{2k}. \label{diagham2k}
\ee
With an appropriate unitary matrix $\mathcal{V}$,  $\mathit{\Gamma}_{2k}$ is diagonalized as\footnote{
We can check the validity of (\ref{gamma2kdiag}) using the explicit matrix form of   $\mathit{\Gamma}_{2k}$. } 
\be
\mathcal{V}^{\dagger}\mathit{\Gamma}_{2k} \mathcal{V} =\mathit{\Gamma}_{\text{diag}}\equiv {\bigoplus_{\lambda=-S}^{S}}~ (\lambda+\frac{1}{2}\text{sgn}(\lambda))~ \bs{1}_{D_{SO(2k-1)}(|\lambda|)}= {\bigoplus_{\lambda=-S}^{S}}~ \lambda~ \bs{1}_{D_{SO(2k-1)}(|\lambda|)}+\frac{1}{2}G_{2k+1}. \label{gamma2kdiag}
\ee
In other words, 
 we can diagonalize the Hamiltonian  with $\tilde{\mathit{\Psi}}=\mathit{\Psi}\mathcal{V}$ as 
\be
\tilde{\mathit{\Psi}}^{\dagger} {H} \tilde{\mathit{\Psi}}=\frac{1}{2}\mathit{\Gamma}_{\text{diag}}={\bigoplus_{\lambda=-S}^{S}}~\frac{1}{2} (\lambda+\frac{1}{2}\text{sgn}(\lambda)) ~\bs{1}_{D_{SO(2k-1)}(|\lambda|)}. 
\ee
Apparently there are $SO(2k-1)$ degrees of freedom in (\ref{diagham2k}): 
\be
\mathit{\Psi} ~~\rightarrow ~~\mathit{\Psi}\cdot e^{i\frac{1}{2}\sum_{i,j=1}^{2k-1}\omega_{ij}\mathit{\Sigma}_{ij}}
\ee
or 
\be
\tilde{\mathit{\Psi}} ~~\rightarrow ~~\tilde{\mathit{\Psi}}\cdot e^{i\frac{1}{2}\sum_{i,j=1}^{2k-1}\omega_{ij}\tilde{\mathit{\Sigma}}_{ij}}~~~~~~~(\tilde{\mathit{\Sigma}}_{ij} \equiv  \mathcal{V}^{\dagger}\mathit{\Sigma}_{ij} \mathcal{V}).
\ee
For a Hamiltonian with chiral symmetry,  we can define  the winding number \cite{Ryu-S-F-L-2010} 
\be 
\nu^{(\pm)} \equiv (-i)^{k-1}\frac{1}{(2\pi)^k}\frac{(k-1)!}{(2k-1)!}\int_{S^{2k-1}}\tr( (-i\mathit{Q}^{(\mp)}d\mathit{Q}^{(\pm)})^{2k-1})=\pm D_{SO(2k+1)}(S-\frac{1}{2}, 0)=\text{ch}_k^{[ \pm \frac{1}{2}]_{SO(2k)}}, 
\label{winding2k-1}
\ee
which corresponds to  the homotopy map
\be
\pi_{2k-1}(SO(2k)) ~\simeq ~\mathbb{Z}. 
\ee
The analysis of the $SO(2k)$ spinor representation (Appendix \ref{append:dimodd}) and the $SO(2k-1)$ decomposition (\ref{reladimhie2}) suggest that the diagonal blocks of $-i\tilde{\mathit{\Psi}}^{\dagger}d\tilde{\mathit{\Psi}}$ may yield the $SO(2k-1)$ Wilczek-Zee  connection in a similar fashion to (\ref{tildeberryco}): 
\be
A^{(\lambda)}
=-\frac{1}{1+x_{2k}}\Sigma_{ij}^{[|\lambda|]_{SO(2k-1)}}x_{j}dx_{i}. \label{so2k-1conn}
\ee
For $k=2$, (\ref{so2k-1conn}) is reduced to the $SO(3)$ monopole connection (\ref{alambdas3}). 

The obtained results are illustrated in Fig.\ref{highs.fig}.
\begin{figure}[tbph]
\hspace{-0.4cm}
\includegraphics*[width=180mm]{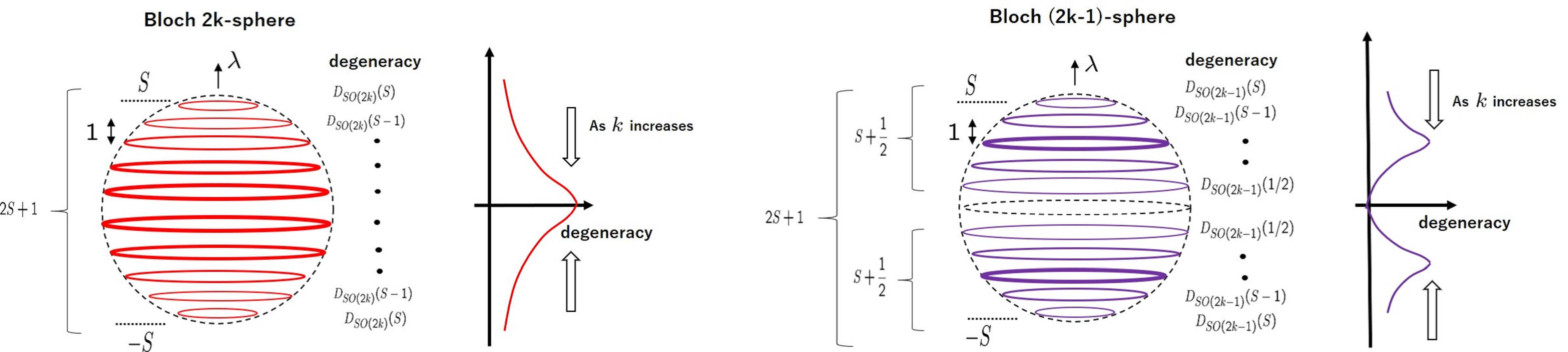}
\caption{ The Bloch $2k$-sphere (left) and the Bloch $(2k-1)$-sphere (right). 
In both cases, there are $2S+1$ energy levels.   
 For the Bloch $2k$-spheres, the degeneracies increase toward the equator: As $k$ increases, the peak at the equator becomes sharper (see also the upper figures of Fig.\ref{dis.fig}). For the Bloch $(2k-1)$-spheres, the degeneracies 
have two peaks in the northern  and  southern hemispheres:    As $k$ increases, the two peaks  approach  the equator (see also the lower figures of Fig.\ref{dis.fig}.) }
\label{highs.fig}
\end{figure}

\section{Bloch hyper-balls and quantum statistics }\label{sec:blochball}

We refer to the $d+1$ dimensional hyper-volume region surrounded by the Bloch hyper-sphere $S^d$ as    
the Bloch hyper-ball, $B^{d+1}$. 
Here, we consider  $2S+1$-level density matrices whose parameters are given by the coordinates of  $B^{d+1}$  and investigate the corresponding von Neumann entropies  and the Bures information metrics.

\subsection{Bloch hyper-balls and density matrices}

Arbitrary $2\times 2$ density matrix is represented as  
\be
\rho=\frac{1}{2}(\bs{1}_2+r\sum_{i=1}^3 x_i \sigma_i) ~~~~(0 \le r \le 1,~~~{\sum_{i=1}^3 x_ix_i}=1), \label{density2times2}
\ee
which is formally equivalent to  
\be
\rho =\frac{1}{2}\bs{1}_2 +r H. 
\ee
Here, $H$ denotes  the $SO(3)$ Zeeman-Dirac Hamiltonian (\ref{so3spinmodemin}). 
The parameters $rx_i$ indicate a position inside the Bloch three-ball to specify the density matrix (\ref{density2times2}).

In the following, we explore the density matrix made of  the $SO(d+1)$ Zeeman-Dirac Hamiltonian $H$:  
\be
\rho =\alpha \bs{1} + \beta H, \label{generho}
\ee
where $\alpha$ and $\beta$ are quantities to be determined so that $\rho$ satisfies the necessary conditions for density matrix: 
\be
 1.~\text{$\rho$ is  Hermitian} ~~~~~~
2.~\text{ $\tr (\rho)=1$}  ~~~~~~~
3.~\text{The eigenvalues of $\rho$ are non-negative}. \nn
\ee
The first condition implies that $\alpha$ and $\beta$ should be real parameters. The second condition gives $\alpha=\frac{1}{\tr \bs{1}}$, provided that $H$ is a traceless matrix as in the case of the Zeeman-Dirac Hamiltonian. The third condition yields $0\le \beta \le \alpha/(h_1\equiv \text{Max}(\text{eigenvalues of} ~H))$ when the spectrum of $H$ is symmetric  with respect to the zero-energy, as in the present case. 
Consequently, we have  
\be
\alpha=\frac{1}{\tr \bs{1}}, ~~~~~~~\beta =\frac{\alpha}{h_1}r, ~~~~~(0\le r \le 1)
\ee
and (\ref{generho}) becomes  
\be
\rho =\frac{1}{\tr \bs{1}}(\bs{1}+\frac{1}{h_1}r H). 
\ee
The present density matrix represents a  special multi-level density matrix. See Refs.\cite{Kimura-2003, Byrd-Khaneja-2003} for a general multi-level density matrix. In such a general model, the geometry of the allowed parameter region is  much more intricate than the simple volume region of a hyper-ball.


For the case of the $SO(2k+1)$  model, the parameters are identified as  
$\alpha=D_{SO(2k+1)}(S)$ and $h_1=S$. Therefore, the density matrix  becomes   
\be
\rho=\frac{1}{D_{SO(2k+1)}(S)} \biggl(\bs{1}_{D_{SO(2k+1)}(S)} + r\frac{1}{S}\sum_{a=1}^{2k+1} x_a \cdot \frac{1}{2}\Gamma_a\biggr)~~~~~~~~(0\le r\le 1,~~~{\sum_{a=1}^{2k+1}x_ax_a}=1 ). 
\label{den2k+1mat}
\ee
The inequality $0\le r \le 1$ indicates the region occupied by the Bloch $2k+1$-ball, and the density matrix is defined at each point inside the $B^{2k+1}$.

Similarly,  for the   $SO(2k)$  model, the parameters are identified as  
$\alpha=2D_{SO(2k)}(1/2)$ and $h_1=\frac{1}{2}(S+\frac{1}{2})$.    
The density matrix is  then given by  
\be
\rho=\frac{1}{2D_{SO(2k)}(1/2)} \biggl(\bs{1}_{2D_{SO(2k)}(1/2)} + r\frac{4}{2S+1}\sum_{\mu=1}^{2k} x_\mu \cdot \frac{1}{2} \mathit{\Gamma}_\mu\biggr) ~~~~~(2S:\text{odd}, ~~~0 \le r\le 1 , ~~~{\sum_{\mu=1}^{2k}x_\mu x_\mu}= 1). \label{den2kmat}
\ee

\subsection{von Neumann entropies}

With a given density matrix $\rho$,  the von Neumann entropy is defined as 
\be
S_{vN} =-\tr(\rho\ln \rho) 
=-\sum_{\lambda} D(\lambda) ~\rho_{\lambda} \ln \rho_\lambda ~~~~(\tr \rho=
\sum_{\lambda} D(\lambda) ~\rho_{\lambda}=1) , \label{vnentgen}
\ee
where $\rho_{\lambda}$ denote the eigenvalues of  $\rho$ with  degeneracy $D(\lambda)$. 
For the present models,  they are given by 
\begin{subequations}
\begin{align}
&B^{2k+1}~~:~\rho_\lambda(r) =\frac{1}{D_{SO(2k+1)}(S)} \biggl(1+\frac{\lambda}{S} r\biggr),~~ ~~D(\lambda) =D_{SO(2k)}(\lambda)
\label{probabs1},\\
&B^{2k}~~:~\rho_\lambda(r) =\frac{1}{2D_{SO(2k)}(1/2)} \biggl(1+\frac{2\lambda +\text{sgn}(\lambda)}{2S+1} r\biggr),~~ ~~D(\lambda) =D_{SO(2k-1)}(|\lambda|). \label{probabs1}
\end{align}\label{probabsbb}
\end{subequations}
Using (\ref{hier}), we can readily confirm that (\ref{probabsbb}) satisfies $\tr \rho=\sum_{\lambda=-S}^S D(\lambda)\rho_{\lambda}(r)=1$.  
Their von Neumann entropies (\ref{vnentgen}) are evaluated as 
\begin{subequations}
\begin{align}
B^{2k+1}:~S_{vN}(r)& =\ln (D_{SO(2k+1)}(S) )-\frac{1}{D_{SO(2k+1)}(S) }\sum_{\lambda=-S}^{S}D_{SO(2k)}(\lambda) \cdot(1+\frac{\lambda }{S} r)\cdot \ln (1+\frac{\lambda }{S}r),  \label{vonneu2k1} \\
B^{2k}:~S_{vN}(r)&=\ln (2 D_{SO(2k)}(1/2) )\nn\\
&-\frac{1}{2 D_{SO(2k)}(1/2) }\sum_{\lambda=-S}^{S}D_{SO(2k-1)}(|\lambda|) \cdot (1+\frac{2\lambda+\text{sgn}(\lambda) }{2S+1} r)\cdot \ln (1+\frac{2\lambda+\text{sgn}(\lambda) }{2S+1} r), \label{vonneu2k}
\end{align}\label{vnnn}
\end{subequations}
where we used (\ref{hier}) again.  The core of the Bloch hyper-ball 
($r=0$) signifies the maximally mixed ensemble: 
\be
\rho =\frac{1}{N}\bs{1}_N ,~~\text{Max}(S_{vN})=\ln N ~~~(N=D_{SO(2k+1)(S)}, ~~2D_{SO(2k)(1/2)}).
\ee
The von Neumann entropy (\ref{vnnn}) decreases monotonically as $r$ increases regardless of the parity of dimensions (see the left of Fig.\ref{vn.fig}). 
 
\begin{figure}[tbph]
\center 
\includegraphics*[width=140mm]{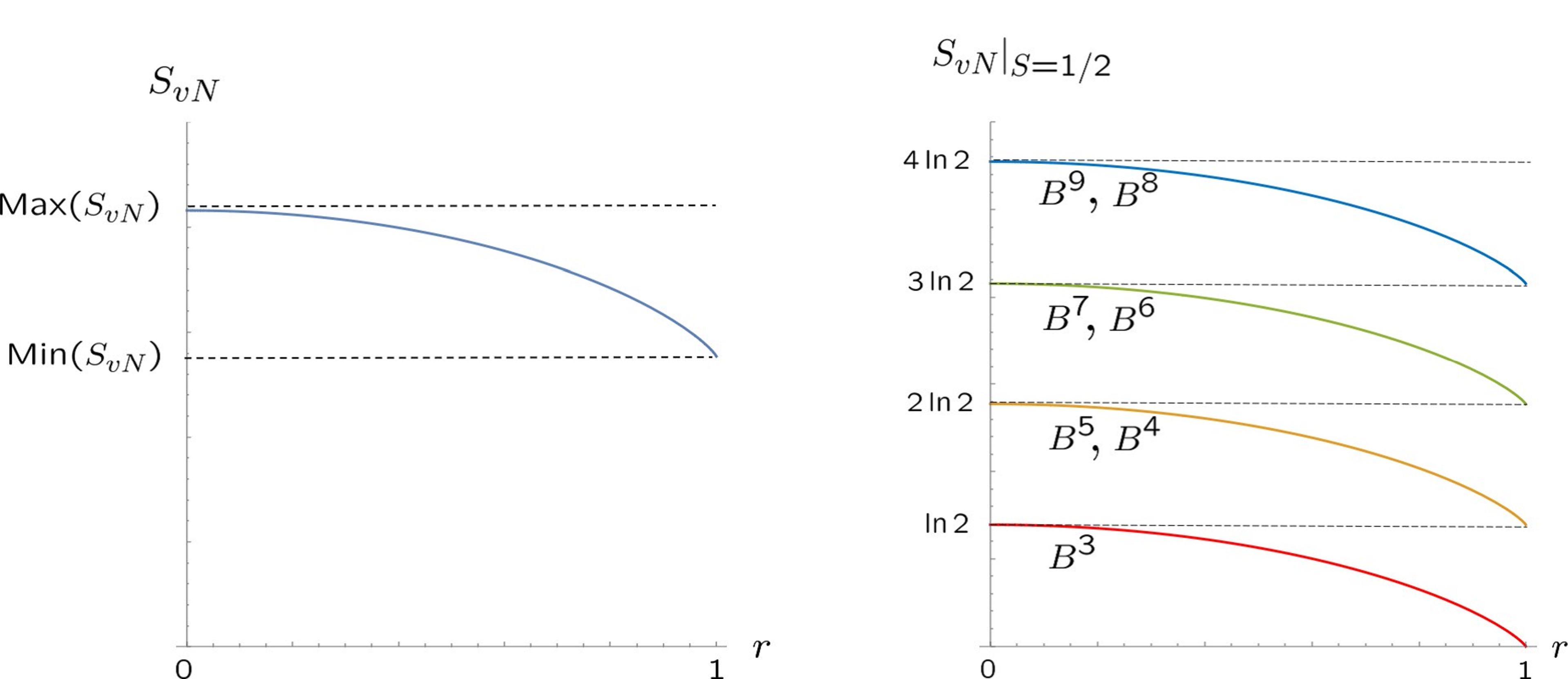}
\caption{(Left) General behavior of the von Neumann entropy  for a Bloch hyper-ball. (Right) The von Neumann entropies for the minimal Bloch $d+1$-balls $(k=[(d+1)/2]).$}
\label{vn.fig}
\end{figure}

For the  Bloch balls with minimal spin $S=1/2$, the density matrices are given by 
\be
B^{2k+1}~:~\rho|_{S=1/2}=\frac{1}{2^k} (\bs{1}_{2^k} +r\sum_{a=1}^{2k+1} x_a \gamma_a), ~~~~~~~~~~~B^{2k}~:~\rho|_{S=1/2}=\frac{1}{2^k} (\bs{1}_{2^k} +r\sum_{\mu=1}^{2k} x_\mu \gamma_\mu), 
\ee
where $\gamma_a$ and $\gamma_{\mu}$ denote the $SO(2k+1)$ and the $SO(2k)$ gamma matrices (\ref{so2k+1gammafun}), respectively. 
Both of them are  diagonalized as 
\be
\rho|_{S=1/2} \rightarrow \frac{1}{2^k}\begin{pmatrix}
(1+r)\bs{1}_{2^{k-1}} & 0 \\
0 & (1-r)\bs{1}_{2^{k-1}} 
\end{pmatrix}, \label{diagrho1/2}
\ee
and so the  von Neumann entropies for $B^{2k+1}$ and $B^{2k}$  take the same value (see the right of Fig.\ref{vn.fig}),    
\be
S_{vN}(r)|_{S=1/2} = k\ln 2 -\frac{1}{2}(1+r)\ln (1+r) -\frac{1}{2}(1-r)\ln(1-r). 
\ee
Their maximum  and minimum values are respectively given by  
\be
\text{Max}(S_{vN})|_{S=1/2}=S_{vN}(0)|_{S=1/2} =k\ln2, ~~~~~\text{Min}(S_{vN})|_{S=1/2} =S_{vN}(1)|_{S=1/2} =(k-1)\ln 2. 
\ee
The maximum value 
$\ln(2^k)$ is due to the $2^{k}$ matrix dimension of the  minimal  Hamiltonian, while the minimum value  
$\ln(2^{k-1})$  comes from the  $2^{k-1}$  degeneracy of the upper eigenvalue of the density matrix (\ref{diagrho1/2}). 

\subsection{Quantum statistical geometry }

We discuss quantum statistical geometries.  
First let us examine the trace distance $L\equiv \frac{1}{2}\tr(\sqrt{(\rho-\rho')^2})$ between two density matrices $\rho$ and $\rho'$. From (\ref{probabsbb}), the trace distance is  easily derived as\footnote{In the derivation of (\ref{tradis}), we used the formula, $\tr (\sqrt{H^2}) =\sum_h |h| \cdot D(h)$, which holds for arbitrary Hermitian matrix $H$ with eigenvalues $h$ of degeneracy $D(h)$.
} 
\be
L
 =
  c(S,d+1) \cdot \sqrt{\sum_{\alpha=1}^{d+1}(rx_\alpha-r'x'_\alpha)^2} , \label{tradis}
\ee
where 
\be
c(S,2k+1) \equiv \sum_{\lambda  =-S}^S \frac{|\lambda|}{4S}
\frac{D_{SO(2k)}(\lambda)}{ D_{SO(2k+1)}(S)} , 
~~~~~c(S,2k) \equiv \sum_{\lambda=1/2}^S \frac{2\lambda +1}{2S+1}
\frac{D_{SO(2k-1)}(\lambda)}{D_{SO(2k)}(1/2)}. \label{tracedens2}
\ee 
 In particular, for $S=1/2$, $c$s (\ref{tracedens2}) do not depend on $k$,  $c(1/2, 2k+1) =1/4$ and $c'(1/2,2k)=1$. In general, $c$s decrease monotonically as $S$ and $k$ increase. 
The trace distance (\ref{tradis}) is proportional to the distance between the vectors $rx_{\alpha}$ and $r'x'_{\alpha}$ in the $d+1$ dimensional flat Euclidean space.  

Next, let us discuss the Bures metric \cite{Bures-1969, Uhlmann-1992}. We will see that various curved spaces with $SO(d+1)$ rotational symmetry emerge as Bures geometries for  $B^{d+1}$, corresponding to the gamma matrices with different  spins.  From the formula of \cite{Hubner-1992}, 
we can evaluate the Bures metrics 
\begin{subequations}
\begin{align}
&B^{2k+1}~:~B_{a b}
=\sum_{\lambda,\lambda'=-S}^S \frac{1}{2(\rho_{\lambda} +\rho_{\lambda'})}  \tr \biggl({\Psi^{(\lambda)
}}^{\dagger}  \frac{\partial \rho}{\partial  X_a} \Psi^{(\lambda')}
~  {\Psi^{(\lambda')}}^{\dagger}  \frac{\partial \rho}{\partial X_b} {\Psi^{(\lambda)}} \biggr), \label{burespresentodd} \\ 
&B^{2k}~~~:~B_{\mu\nu}
=\sum_{\lambda,\lambda'=-S}^S \frac{1}{2(\rho_{\lambda} +\rho_{\lambda'})}  \tr \biggl((\tilde{\mathit{\Psi}}^{(\lambda)})^{\dagger} \frac{\partial \rho}{\partial  X_\mu} \tilde{\mathit{\Psi}}^{(\lambda')}
~ (\tilde{\mathit{\Psi}}^{(\lambda')})^{\dagger}  \frac{\partial \rho}{\partial  X_\nu} {\tilde{\mathit{\Psi}}^{(\lambda)}}\biggr). 
\label{burespresenteven}
\end{align}\label{burespresent}
\end{subequations}
While the Bures metrics (\ref{burespresent}) may take various forms depending on the functional forms of the spin-coherent states,    
 they  generally take the   $SO(d+1)$ spherically symmetric form
\be
B_{\alpha\beta} =f(r) \delta_{\alpha\beta} +g(r) x_{\alpha}x_{\beta} ,
\ee
or\footnote{Utilizing the re-parametrization of the radial coordinate, 
\be
r'=\sqrt{f(r)}~r, 
\ee
we can further transform (\ref{metricsphere}) into the standard form \cite{Weinberg-book}
\be
ds^2= h(r')dr'^2 +r'^2{dl_{S^d}}^2, 
\ee
where 
\be
h(r') \equiv \biggl(1+\frac{g(r)}{f(r)}\biggr){\biggl(1+\frac{f'(r)}{2f(r)}r\biggr)^{-2}}\biggr|_{r=r(r')}.
\ee
Information of the spherical space metric can be incorporated in the single function $h$. 
} 
\be
\sum_{\alpha, \beta=1}^{d+1}B_{\alpha\beta}~d(rx_{\alpha})~d(rx_{\beta}) =(f(r)+g(r))dr^2 +f(r)r^2 {dl_{S^d}}^2~~~~~~({dl_{S^d}}^2\equiv \sum_{\alpha=1}^{d+1}dx_{\alpha}dx_{\alpha}), \label{metricsphere}
\ee
where $dl_{S^d}$ denotes the line element of  $S^d$, and  $f(r)$ and $g(r)$ are some functions that depend on both $S$ and $d$. 
(Several examples are shown in Table \ref{table:fgs}.) We find that  various $SO(d+1)$ symmetric curved geometries emerge  for different values of $S$ and $k$. The  (1/4 of)  Ricci scalar curvatures are shown in Fig.\ref{scurvature.fig} and  they exhibit qualitatively distinct behavior depending on the parity of the dimensions. These Ricci scalars do not have any singularities.  We also evaluated the Kretschmann scalars   
$R_{\mu\nu\rho\sigma}R^{\mu\nu\rho\sigma}$ and confirmed that they have no singularities either.   

\begin{table}
\begin{center}
   \begin{tabular}{|c||c|c||c|c||c|c|}\hline
   &  \multicolumn{2}{|c||}{$S=1/2$} &  \multicolumn{2}{|c||}{$S=1$} &  \multicolumn{2}{|c|}{$S=3/2$} \\ \hline
        &  $f(r)+g(r)$ &    $f(r)$ & $f(r)+g(r)$ &   $f(r)$ &   $f(r)+g(r)$ &   $f(r)$  \\ \hline \hline%
 $B^3$  &  $\frac{1}{4(1-r^2)}$   & $\frac{1}{4}$   & $\frac{1}{6(1-r^2)}$ & $\frac{2}{3(4-r^2)}$ & $\frac{5-r^2}{4(1-r^2)(9-r^2)}$  & $\frac{45-8r^2}{36(9-4r^2)}$            \\ \hline %
$B^4$ &  $\frac{1}{4(1-r^2)}$   & $\frac{1}{4}$  & $\slash$ &  $\slash$ & $\frac{27-4r^2}{4(9-r^2)(9-4r^2)}$ & $\frac{324-65r^2}{972(4-r^2)}$        \\ \hline  
 $B^5$ & $\frac{1}{4(1-r^2)}$    &  $\frac{1}{4}$ & $\frac{3}{20(1-r^2)}$& $\frac{3}{5(4-r^2)}$ & ~~ $\frac{21-5r^2}{20(1-r^2)(9-r^2)}$ ~~ & ~~ $\frac{21-4r^2}{20(9-4r^2)}$     ~~      \\ \hline   
    \end{tabular}       
\end{center}
\caption{Explicit functional forms of $f(r)$ and $g(r)$ for the low dimensional Bloch balls and the small spin magnitudes.   For $S=1/2$, $f(r)$ and $g(r)$ are universal regardless of the dimensions, but for other $S$s, they are not.   
}
\label{table:fgs}
\end{table}

\begin{figure}[tbph]
\center 
\includegraphics*[width=160mm]{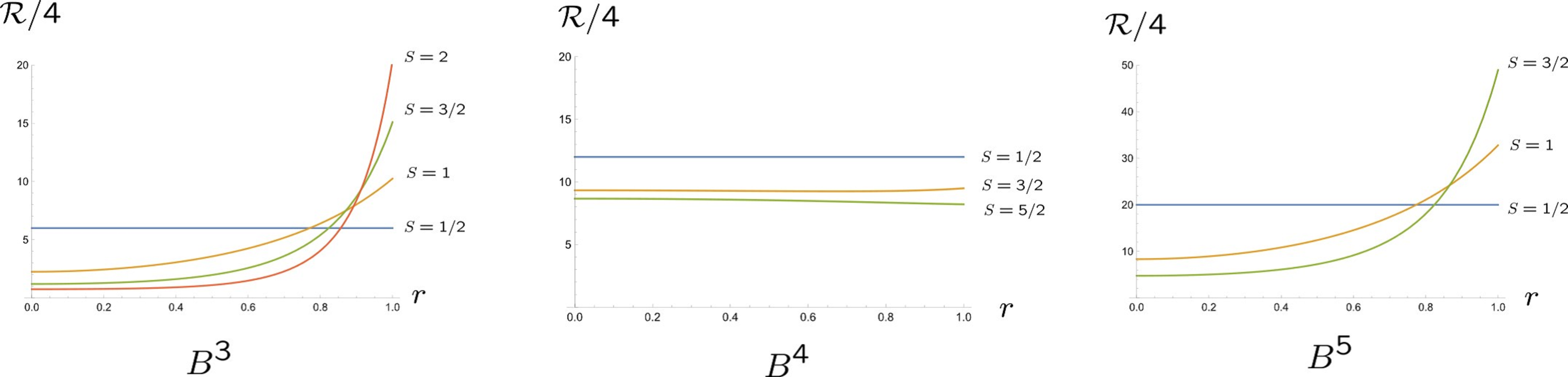}
\caption{The Ricci scalar curvatures  $\mathcal{R}$ for the low dimensional Bures geometries for the metrics  in Table \ref{table:fgs}. There is no singularity in the Ricci scalar curvatures.   As  $r$ increases, the scalar curvatures $(S\neq 1/2)$  increase monotonically and  grow rapidly near the surfaces ($r=1$) for $d+1=3$ and $5$, but not for $d+1=4$. In the case of $S=1/2$, we find  $\mathcal{R}/4 = d(d+1)$, $i.e.$, the  Ricci scalar curvature of $S^{d+1}$. (See also the discussions around Eq.(\ref{metricbures1/2hyperhem}).) }
\label{scurvature.fig}
\end{figure}

In the case $S=1/2$,  the Bures geometry of $B^{d+1}$ is simply  illustrated  by  the hemi-sphere geometry of  the hyper-sphere $S^{d+1}$. 
It is not difficult to  calculate (\ref{burespresent}) explicitly,  using the results of Appendix \ref{sec:fundarb}. Either (\ref{burespresentodd}) or (\ref{burespresenteven}) yields    
\be
B_{\alpha\beta} |_{S=1/2}=\frac{1}{4}\biggl(\delta_{\alpha\beta}+\frac{r^2}{1-r^2}x_{\alpha}x_{\beta}\biggr)~~~~~~~~(\alpha,\beta=1,2,\cdots, d+1)\label{brmetr}
\ee
or 
\be
B_{\alpha\beta} |_{S=1/2}~d(rx_{\alpha})~d(rx_{\beta}) = \frac{1}{4}\biggl(  \frac{1}{1-r^2}dr^2 +r^2{dl_{S^d}}^2 \biggr). \label{buremini}
\ee
The corresponding Bures volume  is evaluated as 
\be
V
|_{S=1/2}\equiv \int_{S^{d}} d\Omega_d ~\int_{0}^1 dr ~r^d \overbrace{\sqrt{\det(B_{\alpha\beta} |_{S=1/2})}}^{=\frac{1}{2^{d+1}}\frac{1}{\sqrt{1-r^2}}}=(\frac{\pi}{2})^{[\frac{d}{2}]+1} \frac{1}{d!!} \label{buresvol}
\ee
where we used $\int_0^1 dr ~r^d \frac{1}{\sqrt{1-r^2}} =  (\frac{\pi}{2})^{\frac{1+(-1)^d}{2}} \frac{(d-1)!!}{d!!}$ and 
\be 
A(S^d)\equiv \int_{S^d} d\Omega_d=\frac{2}{(d-1)!!}(2\pi)^{[\frac{d}{2}]} ~\pi^{\frac{1-(-1)^d}{2}}. \label{sarea} 
\ee
The Bures metric  (\ref{brmetr}) is exactly equal to the metric of the $(d+1)$-sphere of radius $1/2$: 
\be
\sum_{\alpha, \beta=1}^{d+1} B_{\alpha\beta} |_{S=1/2}~~dX_{\alpha}~dX_{\beta} =\sum_{A=1}^{d+2}dX_{A}dX_A, \label{metricbures1/2hyperhem}
\ee
where  
\be
X_{\alpha=1,2,\cdots, d+1}\equiv \frac{1}{2}rx_{\alpha},~~~~X_{d+2} \equiv \frac{1}{2}\sqrt{1-r^2} ~~~~~~(\sum_{\alpha=1}^{d+1}X_{\alpha}X_{\alpha} +X_{d+2}X_{d+2}=(\frac{1}{2})^2). 
\ee
Due to $0\le  r \le 1$,  the present Bures geometry is equal to the north hemisphere of the $(d+1)$-sphere with radius $1/2$ (Fig.\ref{bbs.fig}).\footnote{One can confirm that the scalar curvature $\mathcal{R}$ for the metric (\ref{buremini})   
and the Bures volume (\ref{buresvol}) are equal to those of the $d+1$-hemisphere of radius $1/2$: 
\be
\mathcal{R} ={4} d(d+1), ~~~~
 V
 |_{S=1/2} =  \frac{1}{2^{d+1}} \cdot ~\frac{ A(S^{d+1})}{2}.  
\ee
} The  $SO(d+1)$ symmetry of the Bures geometry corresponds to  the  rotational symmetry around the $X_{d+2}$ axis for the northern hemisphere.  
This is a natural generalization of the known result for  $d=2$  \cite{Hubner-1992}.
The Bures distance between $\rho(X)|_{S=1/2}$ and $\rho(X')|_{S=1/2}$ coincides with the  length of the geodesic curve connecting $X_A$ and $X'_A$ on the 
$(d+1)$-hemisphere (Fig.\ref{bbs.fig}):    
\be
D_{X,X'}=\frac{1}{2}\arccos\biggl(4\sum_{A=1}^{d+2} X_A X'_A\biggr) =\frac{1}{2}\arccos\biggl(4\sum_{\alpha=1}^{d+1} X_{\alpha}X_{\alpha}'+\sqrt{(1-r^2)(1-{r'}^2)}\biggr),
\ee
where $r^2=4\sum_{\alpha=1}^{d+1} X_{\alpha} X_{\alpha}$ and ${r'}^2=4\sum_{\alpha=1}^{d+1} X'_{\alpha}X'_{\alpha}$.
 
\begin{figure}[tbph]
\center 
\includegraphics*[width=70mm]{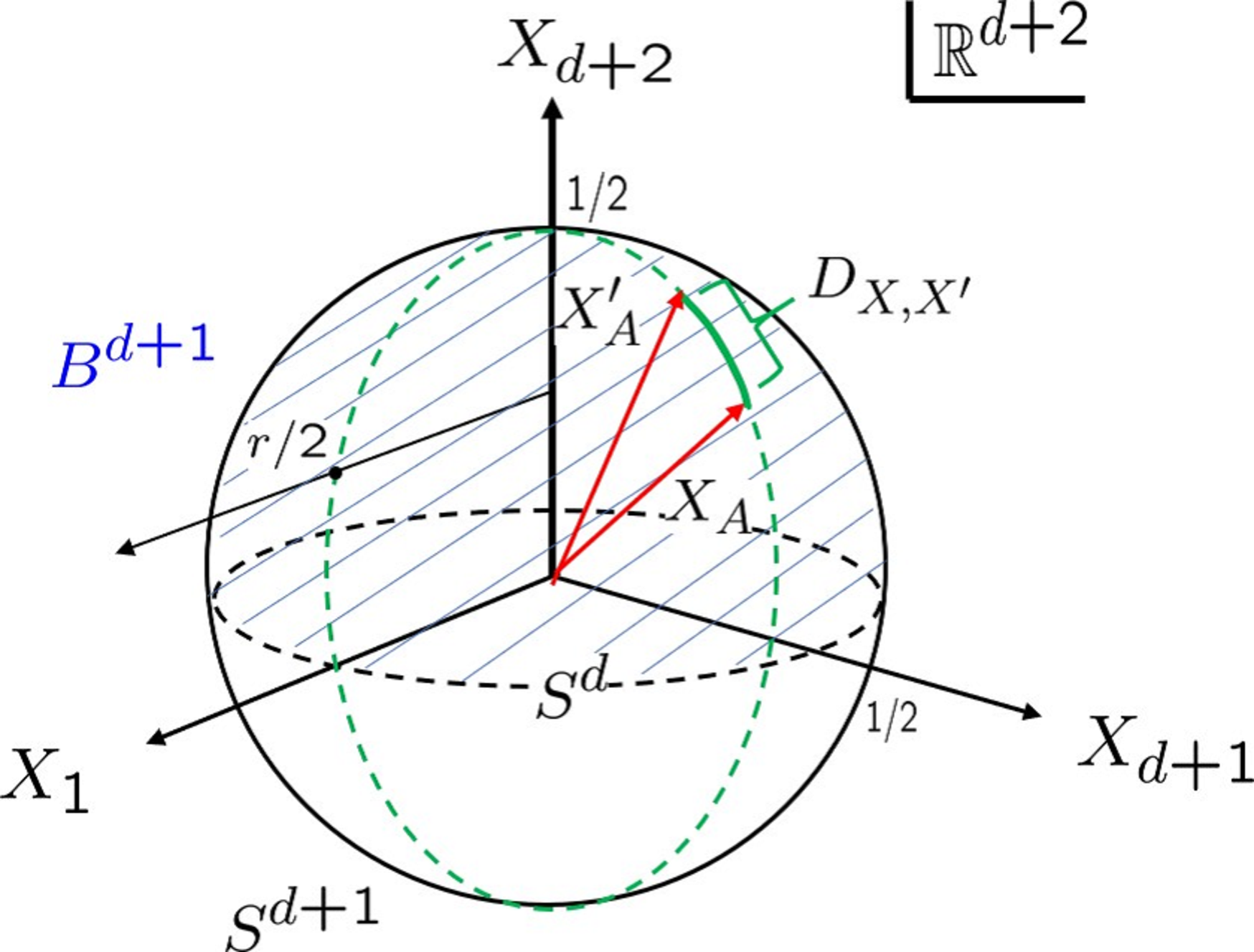}
\caption{ Bures geometry of  
$S=1/2$ is  equal to the hyper-hemisphere.}
\label{bbs.fig}
\end{figure}

\section{Summary}\label{sec:summary}

Taking advantage of the analogies between the Landau model and the precessing spin system, we explored a higher dimensional generalization of the Zeeman-Dirac model and Bloch sphere with large spins. The $SO(3)$ Zeeman-Dirac model has $2S+1$ eigenvalues ranging from $-S$ to $+S$ with interval $1$. Through the analyses using the concrete matrix realization, we showed that the $SO(5)$ Zeeman-Dirac model has the same spectrum as the $SO(3)$ model and that each level accommodates the $SO(4)$ degeneracy.  The $SO(4)$ Zeeman-Dirac model was similarly analyzed to have $2S+1$ energy levels, each of which accommodates the degeneracy attributed to the $SO(3)$ symmetry. 
These properties are naturally generalized in higher dimensions: 
\begin{itemize}
\item  The $SO(2k+1)$ Zeeman-Dirac model is defined for any non-negative integer $2S$. The $SO(2k+1)$ Zeeman-Dirac Hamiltonian has the spectrum ranging from $-S$ to $+S$ with  interval $1$. There are $2S+1$  energy levels with $SO(2k)$ degeneracies. The distribution of the degeneracies has a peak at the equator of the Bloch $2k$-sphere.  This peak becomes sharper as the dimension increases.    
\item The $SO(2k)$ Zeeman-Dirac model is defined only for odd non-negative integer $2S$. The $SO(2k)$ Zeeman-Dirac Hamiltonian exhibits the spectrum ranging from $-\frac{S}{2}-\frac{1}{4}$ to 
$+\frac{S}{2}+\frac{1}{4}$ with interval $1/2$ excluding the zero energy level. There are $2S+1$ energy levels with $SO(2k-1)$ degeneracies.  The distribution  of the degeneracies  has two peaks at the opposite latitudes of the two hemispheres of the Bloch $2k-1$-sphere. These two peaks become sharper and  approach the equator as the dimension increases.    
\end{itemize}
The $d$ dimensional Bloch hyper-sphere geometry exists behind the $SO(d+1)$ Zeeman-Dirac model and  accounts for the particular properties of this model: The $SO(d)$ stabilizer group symmetry of this Bloch hypersphere endows the energy levels with the $SO(2k)$ degeneracies. 
 The $SO(d)$ holonomy group of the Bloch hyper-sphere induces the Wilczek-Zee connection, which is identical to the $SO(d)$ non-Abelian monopole. 
We  investigated the density matrices described by the Bloch hyper-balls and the corresponding von Neumann entropies and Bures metrics. As one moves from the core of the Bloch hyper-ball to its hyper-sphere surface, the von Neumann entropy  decreases monotonically and reaches its minimum value at the  surface.  
 The Bures statistical geometries of these density matrices represent various curved spherical geometries depending on the dimensions and  spin magnitudes.  These Bures geometries are rotationally spherical like the corresponding Bloch hyper-balls and have no singularities. They also show qualitatively different behavior depending on the parity of the dimensions. 
 The Bures geometries for $S=1/2$ were  explicitly evaluated and  identified as the simple hyper-hemispheres with the same dimensions as the Bloch hyper-balls.

It may be worthwhile to emphasize that the quantum Nambu matrix geometry serves as the underlying geometry of M(atrix) theory \cite{Ho-Matsuo-2016, Baggeretal2013}.  
This line of research thus provides an intersection where the exotic concept of non-commutative geometry and string theory meets the quantum information and quantum matter. It is also highly anticipated that the ingenuity of upcoming experiments with artificial gauge fields and synthetic dimensions will further facilitate access to novel physical phenomena of higher dimensional physics.

\section*{Acknowledgments}

 This work was supported by JSPS KAKENHI Grant No.~21K03542.

\appendix

\section{Examples of the generalized gamma matrices}\label{appendix:genegamm}

For a better understanding,  we present a concrete matrix realization of the $SO(5)$ generalized gamma matrices for $S=1$ and the  $SO(4)$ generalized gamma matrices for $S=3/2$.

\subsection{$SO(5)$ $\Gamma_a$ for $S=1$ }\label{subsecI2so5gamma}

The $SO(5)$ gamma matrices with $S=1$ are given by the following $10\times 10$ matrices,   
\footnotesize
\begin{align}
&\!\!\!\!\!\!\!\!\!\hspace{-0.6cm}\Gamma_1\!=\!\!
 \left(
\begin{array}{ccc:cccc:ccc}
 0& 0& 0& 0& \sqrt{2} i& 0& 0& 0& 0& 0 \\
 0& 0& 0& i& 0& 0& i& 0& 0& 0 \\
 0& 0& 0& 0& 0& \sqrt{2} i& 0& 0& 0& 0 \\
  \hdashline
 0& -i& 0& 0& 0& 0& 0& 0& i& 0 \\
 -\sqrt{2} i& 0& 0& 0& 0& 0& 0& 0& 0& \sqrt{2} i \\
 0& 0& -\sqrt{2} i& 0& 0& 0& 0& \sqrt{2} i& 0& 0 \\
 0& -i& 0& 0& 0& 0& 0& 0& i& 0 \\
  \hdashline
 0& 0& 0& 0& 0& -\sqrt{2} i& 0& 0& 0& 0 \\
 0& 0& 0& -i& 0& 0& -i& 0& 0& 0 \\
 0& 0& 0& 0& -\sqrt{2} i& 0& 0& 0& 0& 0
\end{array}
\right)\!\!,~~\Gamma_2\!=\!\!
 \left(
\begin{array}{ccc:cccc:ccc}
 0& 0& 0& 0& \sqrt{2} & 0& 0& 0& 0& 0 \\
 0& 0& 0& -1& 0& 0& 1& 0& 0& 0 \\
 0& 0& 0& 0& 0& -\sqrt{2} & 0& 0& 0& 0 \\
  \hdashline
 0& -1& 0& 0& 0& 0& 0& 0& 1& 0 \\
 \sqrt{2} & 0& 0& 0& 0& 0& 0& 0& 0& \sqrt{2} \\
 0& 0& -\sqrt{2} & 0& 0& 0& 0& -\sqrt{2} & 0& 0 \\
 0& 1& 0& 0& 0& 0& 0& 0& -1& 0 \\
  \hdashline
 0& 0& 0& 0& 0& -\sqrt{2} & 0& 0& 0& 0 \\
 0& 0& 0& 1& 0& 0& -1& 0& 0& 0 \\
 0& 0& 0& 0& \sqrt{2} & 0& 0& 0& 0& 0
\end{array}
\right)\!\!,\nonumber\\
&\hspace{-0.6cm}\!\!\!\Gamma_3=
 \left(
\begin{array}{ccc:cccc:ccc}
 0& 0& 0& \sqrt{2} i& 0& 0& 0& 0& 0& 0 \\
 0& 0& 0& 0& -i& i& 0& 0& 0& 0 \\
 0& 0& 0& 0& 0& 0& -\sqrt{2} i& 0& 0& 0 \\
  \hdashline
 -\sqrt{2} i& 0& 0& 0& 0& 0& 0& \sqrt{2} i& 0& 0 \\
 0& i& 0& 0& 0& 0& 0& 0& i& 0 \\
 0& -i& 0& 0& 0& 0& 0& 0& -i& 0 \\
 0& 0& \sqrt{2} i& 0& 0& 0& 0& 0& 0& -\sqrt{2} i\\ \hdashline
 0& 0& 0& -\sqrt{2} i& 0& 0& 0& 0& 0& 0 \\
 0& 0& 0& 0& -i& i& 0& 0& 0& 0 \\
 0& 0& 0& 0& 0& 0& \sqrt{2} i& 0& 0& 0
\end{array}
\right)\!\! , ~~
\Gamma_4= 
 \left(
\begin{array}{ccc:cccc:ccc}
 0& 0& 0& \sqrt{2} & 0& 0& 0& 0& 0& 0 \\
 0& 0& 0& 0& 1& 1& 0& 0& 0& 0 \\
 0& 0& 0& 0& 0& 0& \sqrt{2} & 0& 0& 0 \\
 \hdashline
 \sqrt{2} & 0& 0& 0& 0& 0& 0& \sqrt{2} & 0& 0 \\
 0& 1& 0& 0& 0& 0& 0& 0& 1& 0 \\
 0& 1& 0& 0& 0& 0& 0& 0& 1& 0 \\
 0& 0& \sqrt{2} & 0& 0& 0& 0& 0& 0& \sqrt{2} \\
 \hdashline
 0& 0& 0& \sqrt{2} & 0& 0& 0& 0& 0& 0 \\
 0& 0& 0& 0& 1& 1& 0& 0& 0& 0 \\
 0& 0& 0& 0& 0& 0& \sqrt{2} & 0& 0& 0
\end{array}
\right), \nn\\
&\hspace{-0.6cm}\Gamma_5=
 \left(
\begin{array}{ccc:cccc:ccc}
2 & 0 & 0 & 0 & 0 & 0 & 0 & 0 & 0 & 0  \\
0 & 2 & 0 & 0 & 0 & 0 & 0 & 0 & 0 & 0   \\
0 & 0 & 2 & 0 & 0 & 0 & 0 & 0 & 0 & 0 \\
\hdashline
0 & 0 & 0 & 0 & 0 & 0 & 0 & 0 & 0 & 0 \\
0 & 0 & 0 & 0 & 0 & 0 & 0 & 0 & 0 & 0 \\
0 & 0 & 0 & 0 & 0 & 0 & 0 & 0 & 0 & 0 \\
0 & 0 & 0 & 0 & 0 & 0 & 0 & 0 & 0 & 0 \\
\hdashline
0 & 0 & 0 & 0 & 0 & 0 & 0 & -2 & 0 & 0 \\
0 & 0 & 0 & 0 & 0 & 0 & 0 & 0 & -2 & 0 \\
0 & 0 & 0 & 0 & 0 & 0 & 0 & 0 & 0 & -2
\end{array}
\right),
\label{so5rep10gammarest}
\end{align}
\normalsize
which satisfy 
\be
\sum_{a=1}^5 \Gamma_a\Gamma_a=12\cdot \bs{1}_{10}, ~~~~~[\Gamma_a, \Gamma_b, \Gamma_c, \Gamma_d]=- 32~\sum_{e=1}^5\epsilon_{abcde} \Gamma_e. 
\ee
The corresponding $SO(5)$ generators, $\Sigma_{ab}=-i\frac{1}{4}[\Gamma_a, \Gamma_b]$,  
satisfy  $\sum_{a<b=1}^5 {\Sigma_{ab}}\Sigma_{ab}=6\cdot \bs{1}_6$. The $SO(4)$ decomposition 
\be
(p,q)=(2,0)~~\rightarrow~~(s_L, s_R)=(1,0)\oplus (1/2, 1/2) \oplus (0,1)  
\ee
 implies that the  $SO(4)$  matrices of $\Sigma_{ab}$ 
 take the following form  
\be
\Sigma_{\mu\nu}=\begin{pmatrix}
\Sigma_{\mu\nu}^{(1,0)} & 0 & 0 \\
0 & \Sigma_{\mu\nu}^{(1/2,1/2)} & 0 \\
0 & 0 & \Sigma_{\mu\nu}^{(0, 1)}
\end{pmatrix}, \label{sigmatso4de}
\ee
where 
\be
\Sigma_{\mu\nu}^{(1,0)} =\eta_{\mu\nu}^iS_i^{(1)}, ~~~~\Sigma_{\mu\nu}^{(1/2,1/2)} =\frac{1}{2}\eta_{\mu\nu}^{(+)i}\sigma_i\otimes \bs{1}_2 +\bs{1}_2\otimes \frac{1}{2}{\eta}_{\mu\nu}^{(-)i}\sigma_i, ~~~~\Sigma_{\mu\nu}^{(0,1)} ={\eta}_{\mu\nu}^{(-)i}S_i^{(1)}.
\ee 
We can confirm (\ref{sigmatso4de}) using (\ref{so5rep10gammarest}) explicitly.

\subsection{$SO(4)$ $\mathit{\Gamma}_{\mu}$ for $S=3/2$}\label{subsec:applendi3}

The $SO(5)$ gamma matrices with $S=3/2$ are given by $20 \times 20$  matrices. According to the $SO(4)$ subgroup decomposition 
\be
(p,q) =(3, 0) ~~\longrightarrow~~(s_L, s_R) =(3/2,0)\oplus (1, 1/2) \oplus (1/2, 1) \oplus (0, 3/2) \label{3decompso4}
\ee
or 
\be
\bs{20}  ~~\longrightarrow~~\bs{4}\oplus \bs{6}\oplus \bs{6}\oplus \bs{4},  \label{3decompso4rep}
\ee
the $SO(4)$ subspace of our interest   $(1, 1/2)\bigoplus (1/2,1)$  corresponds to   $\bs{6}\oplus \bs{6}$ in (\ref{3decompso4rep}). Thus, the $SO(4)$ gamma matrices with  $S=3/2$ are given by  the following $12 \times 12$ matrices: 
\be
\mathit{\Gamma}_{\mu} =\begin{pmatrix}
0 & Y_{\mu} \\
{Y_{\mu}}^{\dagger} & 0 
\end{pmatrix}, ~~~~~\mathit{\Gamma}_5 =\begin{pmatrix}
\bs{1}_6 & 0 \\
0 & -\bs{1}_6
\end{pmatrix}, \label{genso4spin32}
\ee
where 
\footnotesize
\begin{align}
&Y_1 \equiv 
\begin{pmatrix}
 0 & \sqrt{2}i & 0 & 0 & 0 & 0 \\ 
 0 & 0 & 2i & 0 & 0 & 0 \\ 
 \sqrt{2}i & 0 & 0 & 0 & i & 0 \\ 
   0 & i & 0 & 0 & 0 & \sqrt{2}i \\ 
 0 & 0 & 0 & 2i & 0 & 0 \\
 0 & 0 & 0 & 0 & \sqrt{2}i & 0 
\end{pmatrix},~~Y_2 \equiv 
\begin{pmatrix}
 0 & \sqrt{2} & 0 & 0 & 0 & 0 \\ 
 0 & 0 & 2 & 0 & 0 & 0 \\ 
 -\sqrt{2} & 0 & 0 & 0 & 1 & 0 \\ 
   0 & -1 & 0 & 0 & 0 & \sqrt{2} \\ 
 0 & 0 & 0 & -2 & 0 & 0 \\
 0 & 0 & 0 & 0 & -\sqrt{2} & 0 
\end{pmatrix}, \nn\\
&Y_3\equiv  
\begin{pmatrix}
 {2}i & 0 & 0 & 0 & 0 & 0 \\ 
 0 & \sqrt{2}i & 0 & 0 & 0 & 0 \\ 
 0 & -i & 0 & \sqrt{2}i & 0 & 0 \\ 
 0 & 0 & -\sqrt{2}i & 0 & i & 0 \\ 
  0 & 0 & 0 & 0 & -\sqrt{2}i & 0 \\
  0 & 0 & 0 & 0 & 0 & -2i 
\end{pmatrix}, ~~~Y_4\equiv  
\begin{pmatrix}
 {2} & 0 & 0 & 0 & 0 & 0 \\ 
 0 & \sqrt{2} & 0 & 0 & 0 & 0 \\ 
 0 & 1 & 0 & \sqrt{2} & 0 & 0 \\ 
 0 & 0 & \sqrt{2} & 0 & 1 & 0 \\ 
  0 & 0 & 0 & 0 & \sqrt{2} & 0 \\
  0 & 0 & 0 & 0 & 0 & 2 
\end{pmatrix}.
\end{align}
\normalsize
The matrices (\ref{genso4spin32}) satisfy 
\be
\sum_{\mu=1}^4\mathit{\Gamma}_{\mu}\mathit{\Gamma}_{\mu} =12\cdot \bs{1}_{12},~~~~~[\![\mathit{\Gamma}_\mu, \mathit{\Gamma}_\nu, \mathit{\Gamma}_\rho]\!]= 40~\sum_{\sigma=1}^4\epsilon_{\mu\nu\rho\sigma} \mathit{\Gamma}_\sigma. . 
\ee
We can diagonalize $\mathit{\Gamma}_4$   as 
\be
\mathcal{V}^{\dagger}\mathit{\Gamma}_4 \mathcal{V} = \begin{pmatrix}
2\cdot \bs{1}_4 & 0 & 0 & 0 \\
0 &1\cdot \bs{1}_2 & 0 & 0 \\
0 & 0 & -1\cdot \bs{1}_2 & 0 \\
0 & 0 & 0 & -2\cdot \bs{1}_4 
\end{pmatrix},
\ee
where  
\footnotesize
\be 
\mathcal{V}=\frac{1}{\sqrt{6}}
\left(
\begin{array}{cccccc:cccccc}
 \sqrt{3} & 0 & 0 & 0 & 0  & 0 & 0 & 0 & -\sqrt{3} & 0 & 0 & 0 \\ 
 0 & 1 & 0 & 0 & -\sqrt{2} & 0 & \sqrt{2} & 0 & 0  & -1 & 0  & 0 \\ 
  0 & \sqrt{2} & 0 & 0 & 1 & 0 & -1 & 0 & 0  & -\sqrt{2} & 0  & 0 \\  
 0 & 0 & \sqrt{2}   & 0 & 0 & -1 & 0  & 1 & 0 & 0 & -\sqrt{2} & 0 \\ 
0 & 0 & 1   & 0 & 0 & \sqrt{2} & 0  & -\sqrt{2} & 0 & 0 & -1 & 0 \\ 
 0 & 0 & 0 & \sqrt{3}  & 0 & 0 & 0 & 0  & 0 & 0 & 0 & -\sqrt{3} \\
 \hdashline  
 \sqrt{3}  & 0 & 0 & 0 & 0  & 0 & 0 & 0 & \sqrt{3} & 0 & 0 & 0 \\ 
 0 & \sqrt{2}  & 0 & 0 & -1 & 0  & -1 & 0 & 0 & \sqrt{2} & 0 & 0 \\ 
 0 & 0 & 1 & 0 & 0 & -\sqrt{2} & 0 & -\sqrt{2} & 0 & 0  & 1 & 0  \\ 
 0 & 1 & 0 & 0 & \sqrt{2} & 0 & \sqrt{2} & 0 & 0  & 1 & 0  & 0 \\ 
 0 & 0 & \sqrt{2}  & 0 & 0 & 1 & 0  & 1 & 0 & 0 & \sqrt{2} & 0 \\ 
 0 & 0 & 0 & \sqrt{3}  & 0 & 0 & 0 & 0  & 0 & 0 & 0 & \sqrt{3} 
\end{array}
\right). 
\label{vi3}
\ee
\normalsize
The $SO(4)$ matrix generators,  $\mathit{\Sigma}_{\mu\nu}$, are represented as  
\be
\mathit{\Sigma}_{\mu\nu} =\begin{pmatrix}
\Sigma_{\mu\nu}^{(1,\frac{1}{2})} & 0 \\
0 & \Sigma_{\mu\nu}^{(\frac{1}{2}, 1)} 
\end{pmatrix} =
\begin{pmatrix}
\eta_{\mu\nu}^{(+)i} S_i^{(1)}\otimes \bs{1}_2
+\bs{1}_3\otimes  {\eta}_{\mu\nu}^{(-)i} \frac{1}{2}\sigma_i & 0 \\
0 & \eta_{\mu\nu}^i \frac{1}{2}\sigma_i\otimes \bs{1}_3 + \bs{1}_2\otimes {\eta}^{(-)i}_{\mu\nu} S_i^{(1)} 
\end{pmatrix}.
\ee
Note that $\mathit{\Sigma}_{\mu\nu}  \neq -i\frac{1}{4}[\mathit{\Gamma}_{\mu}, \mathit{\Gamma}_{\nu}]$.

\section{Matrix-valued quantum geometric tensor }\label{append:matrixtensgeo}

Here, we consider  $N$-fold degenerate quantum states represented by a $M\times N$ rectangular matrix $\Psi$. This subject has been addressed in 
\cite{Levay-1992,  Nowakowski-Trautman-1978}. We assume that $\Psi$ satisfies the normalization condition,  
\be
\Psi^{\dagger}\Psi=\bs{1}_N. 
\ee
In terms of the rectangular matrix $\Psi$,  the quantum geometric tensor \cite{Provost-Vallee-1980} may be  generalized as  a  matrix-valued  quantity  
\be
\chi_{\mu\nu} =\partial_{\mu}\Psi^{\dagger}~\partial_{\nu}\Psi -\partial_{\mu}\Psi^{\dagger}\Psi \cdot \Psi^{\dagger}\partial_{\nu}\Psi,  \label{matqgt}
\ee
which satisfies 
\be
{\chi_{\mu\nu}}^{\dagger}=\chi_{\nu\mu}. \label{chidagger}
\ee
It is straightforward to show that the matrix quantum geometric tensor (\ref{matqgt}) is covariant under the gauge transformation: 
\be
\Psi ~~\rightarrow~~\Psi \cdot g ~~~(g^{\dagger}g =\bs{1}_N), ~~~~\chi_{\mu\nu} ~~\rightarrow~~g^{\dagger}\chi_{\mu\nu}g. \label{vovtrachi}
\ee
A field theoretical model of rectangular matrix-valued field with gauge symmetry is discussed in \cite{Macfarlane-1979}. The target space of this model is  the Grassmannian manifold, $Gr(M,N)\simeq U(M)/(U(N)\otimes U(M-N))$, which naturally realizes a matrix extension of the $\mathbb{C}P^{N-1}=Gr(N,1)$ with the Fubini-Study metric.    
We adopt the same procedure to explore the matrix version of the quantum geometric tensor.     
We introduce the auxiliary gauge field and  the covariant derivative as 
\be
A_{\mu} =-i\Psi^{\dagger}\partial_{\mu} \Psi =A_{\mu}^{\dagger}, ~~~~D_{\mu}\Psi\equiv \partial_{\mu}\Psi -i\Psi A_{\mu} , ~~~(D_{\mu}\Psi)^{\dagger}= \partial_{\mu}\Psi^{\dagger} +i A_{\mu} \Psi^{\dagger}, \label{dmupsi}
\ee
which transform as 
\be
A_{\mu} ~~\rightarrow ~~g^{\dagger}A_{\mu}g -ig^{\dagger}\partial_{\mu}g, ~~~D_{\mu}\Psi ~~\rightarrow~~(D_{\mu}\Psi) \cdot g , ~~~~(D_{\mu}\Psi)^{\dagger} ~~\rightarrow~~g^{\dagger}\cdot (D_{\mu}\Psi)^{\dagger}. 
\ee
The matrix $\chi_{\mu\nu}$ is simply represented as  
\be
\chi_{\mu\nu} =(D_{\mu}\Psi)^{\dagger}~D_{\nu}\Psi. \label{dpsidpsi}
\ee
Equation (\ref{dpsidpsi}) manifestly shows that $\chi_{\mu\nu}$ is not generally gauge invariant, but rather covariant under the transformation (\ref{vovtrachi}). The matrix-valued quantum geometric tensor is decomposed into  its symmetric (Hermitian) part and its antisymmetric (anti-Hermitian) part as 
\be
\chi_{\mu\nu} =G_{\mu\nu}+i\frac{1}{2}F_{\mu\nu},
\ee
where 
\begin{subequations}
\begin{align}
&G_{\mu\nu} \equiv \frac{1}{2}(\chi_{\mu\nu}+\chi_{\nu\mu}) =\frac{1}{2}((D_{\mu}\Psi)^{\dagger}~D_{\nu}\Psi+(D_{\nu}\Psi)^{\dagger}~D_{\mu}\Psi),\\
&F_{\mu\nu} \equiv -i(\chi_{\mu\nu}-\chi_{\nu\mu}) =-i((D_{\mu}\Psi)^{\dagger}~D_{\nu}\Psi-(D_{\nu}\Psi)^{\dagger}~D_{\mu}\Psi).
\end{align}
\end{subequations}
Equation (\ref{chidagger}) implies that both $G_{\mu\nu}$ and $F_{\mu\nu}$ are Hermitian:  
\be
{G_{\mu\nu}}^{\dagger}=G_{\mu\nu}, ~~~{F_{\mu\nu}}^{\dagger}=F_{\mu\nu}.
\ee
It is obvious that both $G_{\mu\nu}$ and $F_{\mu\nu}$ transform as covariantly
\be
G_{\mu\nu} ~~\rightarrow~~g^{\dagger}G_{\mu\nu}g, ~~~~~~F_{\mu\nu} ~~\rightarrow~~g^{\dagger}F_{\mu\nu}g.
\ee
Using $A_{\mu}$, we can represent $G_{\mu\nu}$ and $F_{\mu\nu}$  as 
\begin{subequations}
\begin{align}
&G_{\mu\nu} =\frac{1}{2}(\partial_{\mu}\Psi^{\dagger}\partial_{\nu}\Psi+\partial_{\nu}\Psi^{\dagger}\partial_{\mu}\Psi )-\frac{1}{2}(A_{\mu}A_{\nu}+A_{\nu}A_{\mu}), \label{fmunuofg}\\
&F_{\mu\nu}=\partial_{\mu}A_{\nu}-\partial_{\nu}A_{\mu}+i[A_{\mu}, A_{\nu}]. \label{fmunuofa} 
\end{align}
\end{subequations}
Note that $F_{\mu\nu}$ (\ref{fmunuofa}) stand for the field strength of the gauge field $A_{\mu}$,\footnote{From (\ref{dmupsi}), we obtain the field strength (\ref{fmunuofa}) as 
\be
-i[D_{\mu}, D_{\nu}]\Psi =\Psi F_{\mu\nu}.
\ee
} while $G_{\mu\nu}$  (\ref{fmunuofg}) cannot be expressed only in terms of $A_{\mu}$. 
The matrix $G_{\mu\nu}$ may be considered as  a matrix-valued version of the quantum metric, because its trace gives rise to the quantum metric, 
\be
g_{\mu\nu} \equiv \tr(G_{\mu\nu}). 
\ee
 For groups with traceless generators, such as a special unitary group or  a special orthogonal group, 
the trace of the quantum geometric tensor directly yields the quantum metric,  
\be
\tr(\chi_{\mu\nu}) =\tr(G_{\mu\nu})+i\frac{1}{2}\overbrace{\tr(F_{\mu\nu})}^{=0}= g_{\mu\nu} ~~\text{for}~~SU(N), ~SO(N),~\text{etc}.
\ee

\section{$SO(4)$ monopole harmonics from the $SO(4)$ non-linear realization }\label{appendix:so4monopoleharm}

We revisit the analysis of the $SO(4)$ Landau model \cite{Hasebe-2018, Nair-Daemi-2004} from the perspective of non-linear realization.

\subsection{$SO(3)$ decomposition of the $SO(4)$ matrix generators}

Due to $SO(4)\simeq SU(2)_L\otimes SU(2)_R$, the $SO(4)$ irreducible representation is indexed by  $SU(2)$ bi-spin, $s_L$ and $s_R$. 
The $SO(4)$ matrix generators for an irreducible representation are generally given by  
\be
\Sigma_{\mu\nu}^{(s_L, s_R)} =\eta^{(+)i}_{\mu\nu} S_i^{(s_L)}\otimes \bs{1}_{2s_R+1} +{\eta}^{(-)i}_{\mu\nu}  \bs{1}_{2s_L+1}\otimes S_i^{(s_R)}, \label{so4genematexp}
\ee
where $\eta_{\mu\nu}^{(\pm) i}$ are the `t Hoof tensors (\ref{defthoof}) and   $S_i^{(s_L)}$ and $S_i^{(s_R)}$ signify the $SU(2)$ matrices with spins $s_L$ and $s_R$, respectively    
 $(\sum_{i=1}^3 {S_i^{(s_{L/R})}}S_i^{(s_{L/R})}=s_{L/R}(s_{L/R}+1)\bs{1}_{2s_{L/R}+1})$. 
In detail, 
\begin{subequations}
\begin{align}
&\Sigma_{ij}^{(s_L, s_R)}=\epsilon_{ijk} (S_k^{(s_L)}\otimes \bs{1}_{2s_R+1} + \bs{1}_{2s_L+1}\otimes S_k^{(s_R)}), \label{sigmajk} \\
&\Sigma_{i4}^{(s_L, s_R)} =-\Sigma_{4i}^{(s_L,s_R)}=S_i^{(s_L)}\otimes \bs{1}_{2s_R+1} - \bs{1}_{2s_L+1}\otimes S_i^{(s_R)}.
\end{align}
\end{subequations}
The sum of their squares  gives 
\be
\sum_{\mu > \nu=1}^4 {\Sigma_{\mu\nu}^{(s_L, s_R)}}~{\Sigma_{\mu\nu}^{(s_L, s_R)}} = 2(s_L(s_L+1)+s_R(s_R+1))\bs{1}_{(2s_L+1)(2s_R+1)}. 
\ee
Note that $\Sigma_{ij}^{(s_L,s_R)}$ (\ref{sigmajk}) represents the tensor product of two $SU(2)$ spins, which is irreducibly decomposed by the $SU(2)$ group as  
\be
O~\Sigma_{ij}^{(s_L, s_R)}~O^t=\epsilon_{ijk} \bigoplus_{J=|s_L-s_R|}^{s_L+s_R} S_k^{(J)}, 
 \label{diagsig}
\ee
where $O$ denotes an orthogonal matrix of the Clebsch-Gordan coefficients,   
\be
O_{\alpha\beta}\equiv C^{(JM)}_{s_L,m_L; ~s_R, m_R}~~~~(\alpha,\beta=1,2,\cdots, (2s_L+1)(2s_R+1)),
\ee
with identification 
\begin{align}
&\alpha\equiv (J, M) ~~~~~~~~~(J=s_L+s_R, s_L+s_R-1, \cdots, |s_L-s_R|, ~~M=J, J-1, \cdots, -J),\nn\\
&\beta\equiv (m_L, m_R) ~~~~~~(m_L=s_L, s_L-1, \cdots, -s_L, ~~~m_R=s_R, s_R-1, \cdots, -s_R). 
\end{align}

\subsection{$SO(4)$ monopole harmonics}

Using the parameterization of  $x_{\mu}$  (\ref{xsso4}), 
we introduce  the non-linear realization matrix 
\be
\Psi^{(s_L, s_R)} \equiv e^{-i\sum_{i=1}^3\chi y_i \Sigma_{i 4}^{(s_L, s_R)}}=e^{(-i\chi \sum_{i=1}^3 y_i S_i^{(s_L)})\otimes \bs{1}_{2s_R+1} +  \bs{1}_{2s_L+1} \otimes (i\chi\sum_{i=1}^3 y_i S_i^{(s_R)}) }, 
\label{ddpsipre}
\ee
or 
\be
\Psi^{(s_L, s_R)}=D^{(s_L)}(\bs{\chi})\otimes D^{(s_R)}(-\bs{\chi}) \label{ddpsi}
\ee
where 
\be
D^{(s_L)}(\bs{\chi}) \equiv e^{-i\chi\sum_{i=1}^3 y_i S_i^{(s_L)}},~~~~~D^{(s_R)}(-\bs{\chi}) \equiv e^{i\chi \sum_{i=1}^3y_i S_i^{(s_R)}}. 
\ee
The covariant derivative is defined as 
\be
D_{\mu}^{(s_L, s_R)} =\partial_{\mu}+iA_{\mu}^{(s_L, s_R)}
\ee
where 
\be
A_{\mu}^{(s_L, s_R)}dx_{\mu} =-\frac{1}{1+x_4}\Sigma_{ij}^{(s_L, s_R)}x_{j}dx_i=-\frac{1}{1+x_4}\epsilon_{ijk}(S_k^{(s_L)}\otimes \bs{1}_{2s_R+1} +\bs{1}_{2s_L+1}\otimes S_k^{(s_R)} )x_j dx_i. 
\ee
The matrix  $\Psi^{(s_L, s_R)}$ satisfies 
\be
L_{\mu\nu}^{(s_L, s_R)}  \Psi^{(s_L, s_R)}=\Psi^{(s_L, s_R)} \Sigma_{\mu\nu}^{(s_L, s_R)}, \label{lmunusphi}
\ee
where 
\be
L_{\mu\nu} ^{(s_L, s_R)} =-ix_{\mu}D_{\nu}^{(s_L, s_R)} +ix_{\nu}D_{\mu}^{(s_L, s_R)} +F_{\mu\nu}^{(s_L, s_R)}. 
\ee
Therefore, with 
\be
\Psi^{(s_L, s_R)} =\begin{pmatrix}
\bs{\Psi}^{(s_L, s_R)}_1 & \bs{\Psi}^{(s_L, s_R)}_2 & \bs{\Psi}^{(s_L, s_R)}_3 & \cdots & \bs{\Psi}^{(s_L, s_R)}_{(2s_L+1)(2s_R+1)} 
\end{pmatrix}, 
\ee
we have 
\be
L_{\mu\nu}^{(s_L, s_R)}  \bs{\Psi}^{(s_L, s_R)}_{\alpha}=\bs{\Psi}^{(s_L, s_R)}_{\beta} (\Sigma_{\mu\nu}^{(s_L, s_R)})_{\beta\alpha}. \label{transso4psi}
\ee
From (\ref{diagsig}),  we can obtain the $SU(2)$ irreducible decomposition of Eq.(\ref{transso4psi}) 
\be
O L_{\mu\nu}^{(s_L, s_R)}O^t \cdot O\bs{\Psi}^{(s_L, s_R)}_{\alpha} =  O\bs{\Psi}^{(s_L, s_R)}_{\beta} (\Sigma_{\mu\nu}^{(s_L, s_R)})_{\beta\alpha} 
\ee
as 
\be
(\bigoplus_{J=|s_L-s_R|}^{s_L+s_R} L_{\mu\nu}^{(J)})\bs{\Phi}^{(s_L, s_R)}_{\alpha}
=\bs{\Phi}^{(s_L, s_R)}_{\beta} (\Sigma_{\mu\nu}^{(s_L, s_R)})_{\beta\alpha},
\ee
where 
\be
\bs{\Phi}^{(s_L, s_R)}_{\alpha} \equiv O \bs{\Psi}^{(s_L, s_R)}_{\alpha}, ~~~~~
L_{\mu\nu}^{(J)} \equiv -ix_{\mu}D^{(J)}_{\nu}+ix_{\nu}D^{(J)}_{\mu} +F^{(J)}_{\mu\nu}, 
\ee
with 
\be
A^{(J)}_{\mu}dx_{\mu} \equiv -\frac{1}{1+x_4}\epsilon_{ijk}x_jS^{(J)}_k dx_i. 
\ee
Assume that $J$ includes $S$, 
\be
J=s_L+s_R, s_L+s_R-1, \cdots, S, \cdots, |s_L-s_R|. 
\ee
We introduce the $(2S+1)$ component ``vector'' $\bs{\phi}_{\alpha}^{(s_L, s_R)}$ with  its $A$th component being  
\be
(\bs{\phi}_{\alpha}^{(s_L, s_R)})_A\equiv C^{S, A}_{s_L, m_L;~s_R, m_R} (\bs{\Psi}_{\alpha}^{(s_L, s_R)}) _{m_L, m_R}~~~~~~(\alpha=1,2,\cdots, (2s_L+1)(2s_R+1), ~~~A=S, S-1,\cdots, -S), 
\ee
or 
\be
(\bs{\phi}_{m_L, m_R}^{(s_L, s_R)})_A\equiv C^{S, A}_{s_L, m'_L;~s_R,m'_R} D^{(s_L)}(\bs{\chi})_{m'_L, m_L} D^{(s_R)}(-\bs{\chi})_{m'_R, m_R} ~~~~~(-s_L\le m_L\le s_L,~~-s_R \le m_R\le s_R), \label{so4monopolehar}
\ee
which is consistent with the expression 
 in Refs.\cite{Hasebe-2018, Nair-Daemi-2004}. 
These $SO(4)$ monopole harmonics satisfy
\be
L^{(S)}_{\mu\nu} \bs{\phi}_{m_L, m_R}^{(s_L, s_R)} =\bs{\phi}_{m_L, m_R}^{(s_L, s_R)} \Sigma_{\mu\nu}^{(s_L, s_R)}
\ee
where 
\be
L_{\mu\nu}^{(S)} =-ix_{\mu}D^{(S)}_{\nu}+ix_{\nu}D^{(S)}_{\mu} +F^{(S)}_{\mu\nu}, 
\ee
with 
\be
A^{(S)}_{\mu}dx_{\mu} =-\frac{1}{1+x_4}\epsilon_{ijk}x_jS^{(S)}_k dx_i. 
\ee
Consequently, 
\be
\sum_{\mu >\nu}{L^{(S)}_{\mu\nu}}^2 ~\bs{\phi}_{m_j, m_k}^{(s_L, s_R)}=2(s_L(s_L+1)+s_R(s_R+1)) \bs{\phi}_{m_j, m_k}^{(s_L, s_R)}.
\ee
The ortho-normal relations of the $SO(4)$  monopole harmonics are  given by 
\be
\int_{S^3} d\Omega_3 ~{\bs{\phi}_{\alpha}^{(s_L, s_R)}}^{\dagger}\bs{\phi}_{\beta}^{(s_L, s_R)} =A(S^3)\frac{D_{SO(3)}(S)}{D_{SO(4)}(s_L, s_R)}\delta_{\alpha\beta}=2\pi^2\frac{ 2S+1}{(2s_L+1)(2s_R+1)}\delta_{\alpha\beta},
\ee
where $d\Omega_3=\sin^2\chi\sin\theta d\chi d\theta d\phi$, $A(S^3)=\int_{S^3}d\Omega_3=2\pi^2$ and $D_{SO(4)}(s_L, s_R)=(2s_L+1)(2s_R+1)$.

\section{Nested Bloch four-spheres from higher Landau levels}\label{subsec:higherlls}

Here, we extend the  analysis of the $SO(5)$ lowest Landau level  (Sec.\ref{subsec:so5largespin}) to  higher Landau levels. 
Since the quantum matrix geometry exhibits a nested structure in higher Landau levels \cite{Hasebe-2023-1, Hasebe-2020-1},  the corresponding Zeeman-Dirac model also exhibits a nested structure.  The Landau level $N$ and the spin  index $S$ of the $SU(2)$ monopole are identified with the $SO(5)$ Casimir indices as 
\be
(p,q) = (N+2S, N) \label{pqgene}
\ee
or $[l_1,l_2]=[\frac{1}{2}(p+q), \frac{1}{2}(p-q)]=[N+S, S]$. The degeneracy of the $N$th Landau level is  given by 
\be
D(N, S) \equiv \frac{1}{6}(N+1)(2S+1)(N+2S+2)(2N+2S+3). 
\ee
Evaluating the matrix coordinates with the $N$th Landau level eigenstates
\be
(\Gamma_a)_{\alpha\beta} ~\propto~\langle \psi_{\alpha}|x_a|\psi_{\beta}\rangle, 
\ee
we can derive   $D(N, S)\times D(N, S)$ generalized  gamma matrices $\Gamma_{a=1,2,3,4,5}$ \cite{Hasebe-2023-1},
which  satisfy 
\be
\sum_{a=1}^5 \Gamma_a \Gamma_a = 4\frac{(N+S+2) S(S+1)}{N+S+1} \bs{1}_{D(N, S)}~\propto~\bs{1}.  
\ee
The matrix $\Gamma_5$ is a diagonal matrix (see Fig.\ref{higherll.fig} also)
\be
\Gamma_5 =\frac{2}{N+S+1} \bigoplus_{n=0}^N (n+S+1) 
~\biggl(\bigoplus_{\lambda=-S}^{S} \lambda  \cdot \bs{1}_{(n+S+1+\lambda)(n+S+1-\lambda)}\biggr). 
\ee
\begin{figure}[tbph]
\hspace{-0.4cm} 
\includegraphics*[width=150mm]{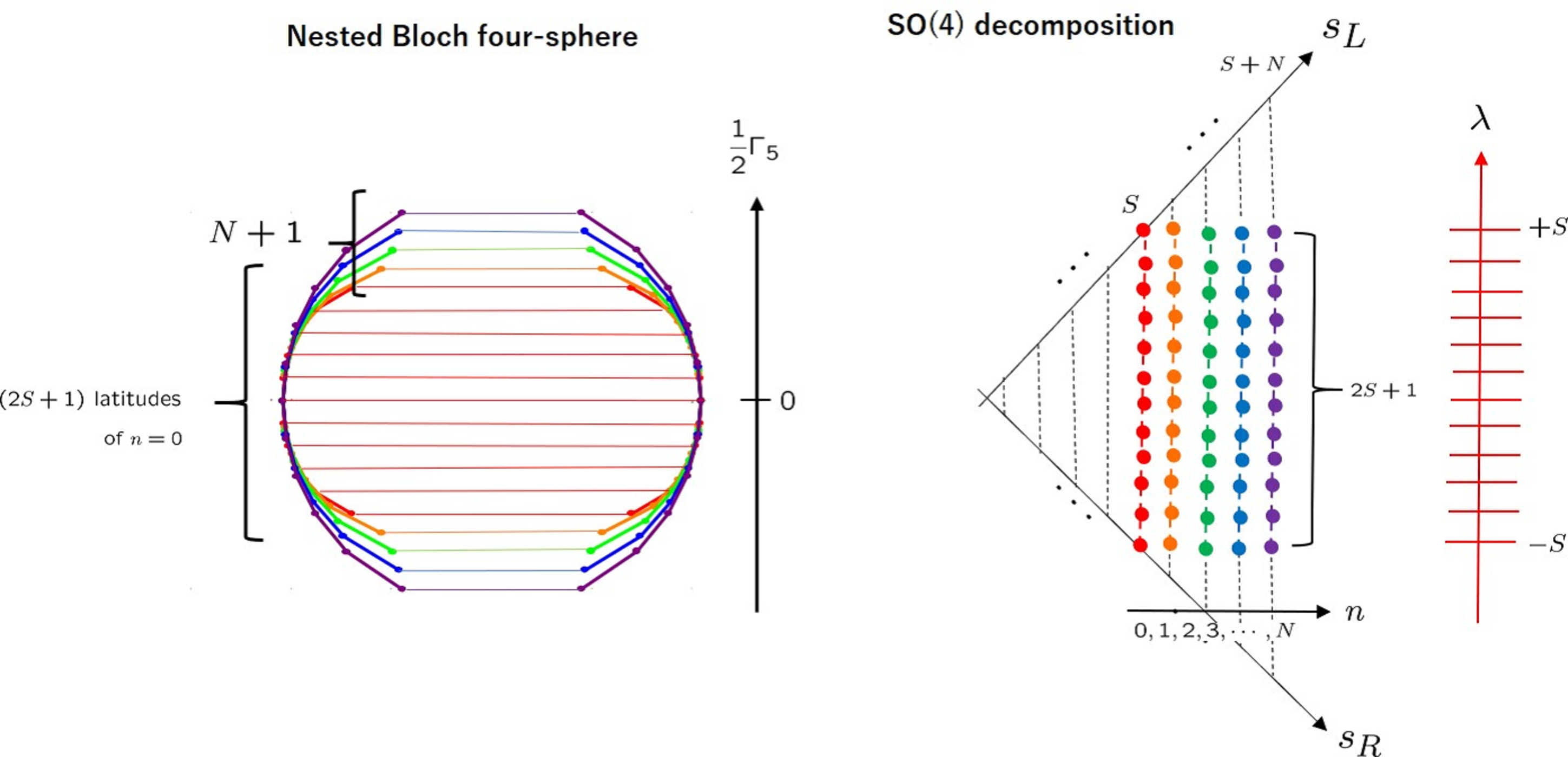}
\caption{The $SO(5)$ Zeeman-Dirac model for $(p,q)=(2S+N,N)$. Taken from \cite{Hasebe-2023-1}.   }
\label{higherll.fig}
\end{figure}
With the $SO(5)$ matrix generators $\Sigma_{ab}$ of the representation (\ref{pqgene}),   $\Gamma_a$ transform as an $SO(5)$ vector \cite{Hasebe-2023-1}\footnote{
Unlike the generalized $SO(5)$ gamma matrices in Sec.\ref{subsec:so5largespin},  the  commutators of the present $\Gamma_a$ $(N\ge 1)$ do not yield the $SO(5)$ matrix generators:  
\be
[\Gamma_a, \Gamma_b] ~\not\propto ~4i\Sigma_{ab}. 
\ee
} 
\be
[\Sigma_{ab}, \Gamma_c] =i\delta_{ac}\Gamma_b-i\delta_{bc}\Gamma_a. \label{so5covgam}
\ee
The $SO(5)$ Zeeman-Dirac Hamiltonian is constructed as  
\be
H=\sum_{a=1}^5 x_a \cdot \frac{1}{2}\Gamma_a .  ~~~~(\sum_{a=1}^5 x_ax_a =1) \label{highllzd}
\ee
Since $\Gamma_a$ transform as an $SO(5)$ vector (\ref{so5covgam}), 
$\Psi=e^{i\xi \sum_{\mu =1}^4 y_{\mu}\Sigma_{\mu 5}}$ diagonalizes this Hamiltonian:    
\be
\Psi^{\dagger}~\sum_{a=1}^5 (x_a \cdot \frac{1}{2}\Gamma_a)~ \Psi =  \frac{1}{2}\Gamma_5.\label{spinsso5cohmat}
\ee
Therefore, the eigenvalues of the Hamiltonian (\ref{highllzd})  are given by  
\be
\frac{n+S+1}{N+S+1}\lambda ~~~~(n=0,1,2,\cdots, N,~~~\lambda=S,S-1, S-2, \cdots, -S). \label{enelevgenso5}
\ee
Notice that the energy levels are indexed by two quantities, $n$ and $\lambda$, and the  degeneracies are given by 
$(n+S+\lambda+1)(n+S-\lambda+1)$.  
Consequently, there are $(N+1)(2S+1)$ energy levels (Fig.\ref{higherll.fig}).  
The Wilczek-Zee connection for the energy level (\ref{enelevgenso5}) is equal to  the $SO(4)$ monopole gauge field: 
\be
A^{(s_L, s_R)} =-\frac{1}{1+x_5}\Sigma_{\mu\nu}^{(s_L, s_R)} x_{\nu}dx_{\mu},
\ee
where $\Sigma_{\mu\nu}^{(s_L, s_R)}$ are given by (\ref{so4genematexp}) with 
\be 
(s_L, s_R) \equiv (\frac{n}{2}+\frac{S}{2}+\frac{\lambda}{2}, \frac{n}{2}+\frac{S}{2}-\frac{\lambda}{2}). \label{so4biiden}
\ee 

The correspondence to the Landau model eigenstates is given as follows. For the $SO(5)$ Landau model in the $SO(4)$ monopole background with bi-spin index $(I_+/2, I_-/2)$, the Landau level $L$ and the $l$th sector are related to the $SO(5)$ and $SO(4)$ Casimir indices as  
\begin{subequations}
\begin{align}
&(p,q) =(L+I_++I_--l, L+l), \\
&(s_L, s_R) =(\frac{I_+}{2}, \frac{I_-}{2}). 
\end{align}
\end{subequations}
For the $SO(5)$ Zeeman-Dirac model,  the relations are given by (\ref{pqgene}) and (\ref{so4biiden}). 
Consequently, their  identification is given by    
\begin{subequations}
\begin{align}
&L=N-n, ~l=n,  \\
&\frac{I_+}{2} =\frac{n}{2}+\frac{S}{2}+\frac{\lambda}{2},~~~~~\frac{I_-}{2}=\frac{n}{2}+\frac{S}{2}-\frac{\lambda}{2}.
\end{align}
\end{subequations}
Let ${\Psi_{\sigma}}$ denote the degenerate $SO(5)$ spin-coherent states, all of which are aligned to the direction of the $\lambda$-latitude  on the $n$th shell of the nested Bloch four-sphere (see the left of Fig.\ref{higherll.fig}), and $\bs{\psi}^{(n)}_{\alpha,N-n}$ stand for the $(N-n)$th Landau level eigenstates of the $n$-sector  in the  $SO(4)$ monopole background with the bi-spin index, $(\frac{n}{2}+\frac{S}{2}+\frac{\lambda}{2}, \frac{n}{2}+\frac{S}{2}-\frac{\lambda}{2})$. They are related as 
\be
\begin{pmatrix} 
\Psi_1 & \Psi_2 & \cdots & \Psi_{(n+S+\lambda+1)(n+S-\lambda+1)}
\end{pmatrix} =\begin{pmatrix}
{\bs{\psi}^{(n)}_{1,N-n}}^{\dagger} \\
{\bs{\psi}^{(n)}_{2,N-n}}^{\dagger} \\
{\bs{\psi}^{(n)}_{3,N-n}}^{\dagger} \\
\vdots \\
{\bs{\psi}^{(n)}_{D(N, S),N-n}}^{\dagger} 
\end{pmatrix}.
\ee

\section{$SO(d+1)$ minimal Zeeman-Dirac models}\label{sec:fundarb}

We investigate the $SO(d+1)$ Zeeman-Dirac models made of  the spinor representation gamma matrices.  This minimal version of the $SO(2k+1)$ Zeeman-Dirac models  has been analyzed in \cite{Benedict-Feher-Horvath-1989, Levay-1992,Levay-1994}. 

\subsection{$SO(d+1)$ spinor representation matrices}\label{subsec:fundarb}

The $SO(2k+1)$ gamma matrices $\gamma_{a=1,2,\cdots, 2k+1}$ are given by 
\be
\gamma_{\mu=1,2,\cdots, 2k} =\begin{pmatrix}
0 & \bar{g}_{\mu} \\
g_{\mu} & 0 
\end{pmatrix}, ~~~\gamma_{2k+1} =\begin{pmatrix}
\bs{1}_{2^{k-1}} & 0 \\
0 & -\bs{1}_{2^{k-1}}
\end{pmatrix}, \label{so2k+1gammafun}
\ee
where 
\be
g_{\mu}\equiv\{-i\gamma'_i, \bs{1}_{2^{k-1}}\}, ~~\bar{g}_{\mu}\equiv\{i\gamma'_i, \bs{1}_{2^{k-1}}\},
\ee
with $SO(2k-1)$ gamma matrices $\gamma'_{i=1,2,\cdots, 2k-1}$. The matrices  (\ref{so2k+1gammafun}) satisfy 
\be
\{\gamma_a, \gamma_b\}=2\delta_{ab}\bs{1}_{2^k}, 
\ee
and their commutators provide 
the $SO(2k+1)$ matrix 
generators, 
\be
\sigma_{ab} \equiv -i\frac{1}{4}[\gamma_a, \gamma_b].
\ee
The matrices $\sigma_{\mu\nu}$ are the  matrix generators  of the $SO(2k)$ group:  
\be
\sigma_{\mu\nu} =\begin{pmatrix}
\sigma_{\mu\nu}^{[+ 1/2]} & 0 \\
0 & \sigma_{\mu\nu}^{[-1/2]}
\end{pmatrix},
\ee
where  
\be
\sigma_{ij}^{[+1/2]} =\sigma_{ij}^{[-1/2]} \equiv \sigma'_{ij}\equiv 
-i\frac{1}{4}[\gamma'_i, \gamma'_j], ~~~~~\sigma_{i,{2k}}^{[+1/2]} =-\sigma_{i,2k}^{[-1/2]}\equiv \frac{1}{2}\gamma'_i.  
\ee

\subsection{$SO(2k+1)$ minimal  spin  model}\label{append:dimeven}

The spinor representation  of the $SO(2k+1)$ is specified by 
\be
[1/2, 1/2, \cdots, 1/2]_{SO(2k+1)}.
\ee
We construct the $SO(2k+1)$  minimal  Zeeman-Dirac Hamiltonian as 
\be
H=\sum_{a=1}^{2k+1}x_a\cdot \frac{1}{2}\gamma_a.~~~~~(\sum_{a=1}^{2k+1}x_ax_a=1) \label{so2k+1zdham}, \\ 
\ee
Using the non-linear realization matrix  
\begin{align}
&\Psi=e^{i\theta_{2k}\sum_{\mu=1}^{2k}y_{\mu}\sigma_{\mu, {2k+1}}} = \cos(\frac{\theta_{2k}}{2})~\bs{1}_{2^k} + 2i\sin (\frac{\theta_{2k}}{2}) \sum_{\mu=1}^{2k}y_{\mu}\sigma_{\mu, {2k+1}}\nn\\
&~~=\frac{1}{\sqrt{2(1+x_{2k+1})}}
\begin{pmatrix}
(1+x_{2k+1}) \bs{1}_{2^{k-1}} & -\sum_{\mu=1}^{2k}x_{\mu}\bar{g}_{\mu} \\
\sum_{\mu=1}^{2k}x_{\mu}g_{\mu} & (1+x_{2k+1})\bs{1}_{2^{k-1}}
\end{pmatrix}=(\Psi^{(1/2)}~~\Psi^{(-1/2)}), \label{so2k+1nlrm}
\end{align}
we can diagonalize the Hamiltonian (\ref{so2k+1zdham}) as 
\be
\Psi^{\dagger} H\Psi =\frac{1}{2}\gamma_{2k+1}.  \label{diagsimpso2k+1}
\ee
The energy levels are  $\pm 1/2$ with degeneracy $2^{k-1}$ for each. 
Equation (\ref{diagsimpso2k+1}) is invariant under the $SO(2k)$ transformation, 
\be
\Psi~~\rightarrow~~\Psi\cdot e^{i\frac{1}{2}\omega_{\mu\nu}\sigma_{\mu\nu}}. 
\ee
We can derive the Bloch vector  as 
\be
{\Psi^{(\pm 1/2)}}^{\dagger}\gamma_a \Psi^{(\pm 1/2)} =\pm x_a \bs{1}_{2^{k-1}}.  
\ee
The matrix-valued quantum geometric tensor is given by 
\be
\chi^{(\pm 1/2)}_{\theta_{\mu} \theta_{\nu}}
=\partial_{\theta_\mu}{\Psi^{(\pm 1/2)}}^{\dagger}  \partial_{\theta_\nu}{\Psi^{(\pm 1/2)}} -   \partial_{\theta_\mu}{\Psi^{(\pm 1/2)}}^{\dagger}  {\Psi^{(\pm 1/2)}}~ {\Psi^{(\pm 1/2)}}^{\dagger}\partial_{\theta_\nu}{\Psi^{(\pm 1/2)}}
~~~~~(\theta_\mu, \theta_\nu=\theta_{2k}, \theta_{2k-1},\cdots,  \theta, \phi). \label{fishermetge2k}
\ee
The trace of (\ref{fishermetge2k}) provides the metric of the $2k$-sphere: 
\be
g_{\theta_\mu \theta_\nu}^{(\pm 1/2)} =\frac{1}{2}\tr(\chi^{(\pm 1/2)}_{\theta_{\mu} \theta_{\nu}}+\chi^{(\pm 1/2)}_{\theta_{\nu} \theta_{\mu}}) 
=2^{k-3} ~\text{diag}(1, ~\sin^2\theta_{2k}, ~\sin^2\theta_{2k}\sin^2\theta_{2k-1}, ~\cdots, ~\prod_{i=3}^{2k}\sin^2\theta_i,~\sin^2\theta \prod_{i=3}^{2k}\sin^2\theta_i). 
\ee 
The Wilczek-Zee  connections  are derived  as 
\be
A^{(\pm 1/2)} =-i{\Psi^{(\pm 1/2)}}^{\dagger}d\Psi^{(\pm 1/2)}=-\frac{1}{1+x_{2k+1}}\sigma_{\mu\nu}^{[\pm 1/2]}x_{\nu}dx_{\mu}, \label{so2kmonofun}
\ee
which  coincide with the gauge fields of the $SO(2k)$ monopoles for $\text{ch}_k^{(\pm 1/2)} = 
\pm 1$.  
The corresponding curvature $F_{\theta_\mu \theta_\nu}=\partial_{\theta_\mu} A_{\theta_{\nu}} -\partial_{\theta_{\nu}} A_{\theta_{\mu}} +i[A_{\theta_{\mu}}, A_{\theta_{\nu}}]$ represents the antisymmetric part of the matrix-valued quantum geometric tensor: 
\be 
F_{\theta_\mu \theta_\nu}^{(\pm 1/2)} =-i(\chi^{(\pm 1/2)}_{\theta_{\mu} \theta_{\nu}}-\chi^{(\pm 1/2)}_{\theta_{\nu} \theta_{\mu}})
=\frac{1}{2}e^{\mu'}_{~\theta_{\mu}}\wedge e^{\nu'}_{~\theta_{\nu}} \sigma_{\mu'\nu'}^{[\pm 1/2]},
\ee
where $e^{\mu'}_{~\theta_{\mu}}$ denote the vielbein of $S^{2k}$. 

\subsection{$SO(2k)$ minimal spin  model}\label{append:dimodd}

The spinor representation  of the $SO(2k)$ is designated by    
\be
[1/2, 1/2, \cdots, \pm 1/2]_{SO(2k)}. 
\ee
We introduce the $SO(2k)$ minimal Zeeman-Dirac Hamiltonian as 
\be
H=\sum_{\mu=1}^{2k}x_{\mu}\cdot \frac{1}{2}\gamma_{\mu}~~~~~(\sum_{\mu=1}^{2k}x_{\mu}x_{\mu}=1), \label{so2k-1funham} \\
\ee
and the non-linear realization matrix as  
\be
{\Psi}=e^{i\theta_{2k-1}\sum_{i=1}^{2k-1}y_i {\sigma}_{i,2k}}=\cos(\frac{\theta_{2k-1}}{2})~\bs{1}_{2^k} +2i\sin(\frac{\theta_{2k-1}}{2}) ~\sum_{i=1}^{2k-1}y_i{\sigma}_{i,2k}=\begin{pmatrix}
{U} & 0 \\
0 & {U}^{\dagger}
\end{pmatrix},
\ee
where 
\begin{subequations}
\begin{align}
&{\sigma}_{i, 2k} \equiv \begin{pmatrix}
\sigma_{i,2k}^{[+1/2]} & 0 \\
0 & \sigma_{i,2k}^{[-1/2]}
\end{pmatrix} =\frac{1}{2}\begin{pmatrix}
\gamma'_{i} & 0 \\
0 & -\gamma'_{i}
\end{pmatrix}, \\
&{U}=e^{i{\theta_{2k-1}}\sum_{i=1}^{2k-1}y_i\sigma_{i, 2k}^{[+1/2]}} 
=\frac{1}{\sqrt{2(1+x_{2k})}} ((1+x_{2k})\bs{1}_{2^{k-1}}+ix_i\gamma'_i). 
\end{align}
\end{subequations}
The Hamiltonian (\ref{so2k-1funham}) is  diagonalized as 
\be
\tilde{{\Psi}}^{\dagger}H\tilde{{\Psi}} =\frac{1}{2}\gamma_{2k+1}   \label{diagsimpso2k1} 
\ee
where 
\be
\tilde{{\Psi}} ={\Psi}{V} =\frac{1}{\sqrt{2}}\begin{pmatrix}
{U} & -{U} \\
{U}^{\dagger} & {U}^{\dagger}
\end{pmatrix} =(\tilde{{\Psi}}^{(1/2)} ~\tilde{{\Psi}}^{(-1/2)})
\ee
with 
\be
{V} =\frac{1}{\sqrt{2}}\begin{pmatrix}
\bs{1}_{2^{k-1}} & -\bs{1}_{2^{k-1}} \\
\bs{1}_{2^{k-1}} & \bs{1}_{2^{k-1}} 
\end{pmatrix}  .
\ee
The energy levels are  $\pm 1/2$ with degeneracy $2^{k-1}$ for each. 
Equation (\ref{diagsimpso2k1}) is invariant under the $SO(2k-1)$ transformation, 
\be
\tilde{{\Psi}}~~\rightarrow~~\tilde{{\Psi}}\cdot e^{i\frac{1}{2}\omega_{ij}\tilde{\sigma}_{ij}} ~~~~~~(\tilde{\sigma}_{ij} \equiv \mathcal{V}^{\dagger}{\sigma}_{ij}\mathcal{V}).  
\ee
We can derive the Bloch vector  as 
\be
(\tilde{{\Psi}}^{(\pm 1/2)})^{\dagger}\gamma_\mu \tilde{{\Psi}}^{(\pm 1/2)} =\pm x_\mu \bs{1}_{2^{k-1}}. 
\ee
The matrix-valued quantum geometric tensor is given by 
\be
\chi^{(\pm 1/2)}_{\theta_i \theta_j}
= \partial_{\theta_i}(\tilde{{\Psi}}^{(\pm 1/2)})^{\dagger}  \partial_{\theta_j}{\tilde{{\Psi}}}^{(\pm 1/2)}
-  
   \partial_{\theta_i}(\tilde{{\Psi}}^{(\pm 1/2)})^{\dagger}  {\tilde{{\Psi}}}^{(\pm 1/2)}~ 
(\tilde{{\Psi}}^{(\pm 1/2)})^{\dagger}\partial_{\theta_j}\tilde{{\Psi}}^{(\pm 1/2)}
~~~~~(\theta_i, \theta_j=\theta_{2k-1},\cdots,  \theta_3, \theta, \phi). \label{fishermetge2kchi} 
\ee
Its symmetric part of $\chi^{(\lambda)}_{\theta_i \theta_j}$ provides    the metric of $(2k-1)$-sphere: 
\be
g^{(\pm 1/2)}_{\theta_i \theta_j}=\frac{1}{2}\tr(\chi^{(\pm 1/2)}_{\theta_i \theta_j}+\chi^{(\pm 1/2)}_{\theta_j \theta_i})
=2^{k-3}~\text{diag}(1, ~\sin^2\theta_{2k-1}, ~\sin^2\theta_{2k-1}\sin^2\theta_{2k-2}, ~\cdots, ~\prod_{i=3}^{2k-1}\sin^2\theta_i,~\sin^2\theta \prod_{i=3}^{2k-1}\sin^2\theta_i). 
\ee
The Wilczek-Zee connections are derived as 
\be
-i\tilde{{\Psi}}^{\dagger} d\tilde{{\Psi} }  ={{V}}^{\dagger}(-i\Psi^{\dagger} d\Psi){{V}}=
\begin{pmatrix}
{A}^{(+1/2)} & * \\
* & {A}^{(-1/2)}
\end{pmatrix}, 
\ee
where $A^{(+1/2)}=A^{(-1/2)}$ is equal to  the gauge field of the $SO(2k-1)$ monopole: 
\begin{align}
{A}^{(+1/2)}&=-i(\tilde{{\Psi}}^{(1/2)})^{\dagger}d\tilde{{\Psi}}^{(1/2)}=-i\frac{1}{2}({U}^{\dagger}d{U} +{U}d{U}^{\dagger})=-\frac{1}{1+x_{2k}}\sigma'_{ij}x_jdx_i\nn\\
&=-i(\tilde{{\Psi}}^{(-1/2)})^{\dagger}d\tilde{{\Psi}}^{(-1/2)}={A}^{(-1/2)}. \label{so2k-1gaugefifun}
\end{align}
The corresponding curvature $F_{\theta_i \theta_j}=\partial_{\theta_i} A_{\theta_j} -\partial_{\theta_j} A_{\theta_i} +i[A_{\theta_i}, A_{\theta_j}]$ represents the antisymmetric part of (\ref{fishermetge2kchi}): 
\be
F_{\theta_i \theta_j}^{( 1/2)} =-i(\chi^{(1/2)}_{\theta_i \theta_j}-\chi^{(1/2)}_{\theta_j \theta_i})
=\frac{1}{2}e^{i'}_{~\theta_{i}}\wedge e^{j'}_{~\theta_{j}} \sigma'_{i'j'}=F_{\theta_i \theta_j}^{(- 1/2)},
\ee
where $e^{i'}_{~\theta_{i}}$ denote the vielbein of $S^{2k-1}$.



\end{document}